\documentclass[12pt,a4paper]{article}

\usepackage{multirow}
\usepackage{ifthen} 
\newboolean{pdflatex}
\setboolean{pdflatex}{true} 
\newboolean{articletitles}
\setboolean{articletitles}{true} 
\newboolean{uprightparticles}
\setboolean{uprightparticles}{false} 
\newboolean{inbibliography}
\setboolean{inbibliography}{false} 
\usepackage{longtable} 
\usepackage{dcolumn}   
\newcolumntype{.}{D{.}{.}{-1}}

\usepackage{microtype}
\usepackage{lineno}  
\usepackage{xspace} 
                    
\usepackage{caption}

\usepackage{graphicx}  
\usepackage{color}
\usepackage{colortbl}
\graphicspath{{./figs/}}

\usepackage{amsmath} 
\usepackage{amssymb}
\usepackage{amsfonts}
\usepackage{upgreek} 

\newcommand*\patchAmsMathEnvironmentForLineno[1]{
\expandafter\let\csname old#1\expandafter\endcsname\csname #1\endcsname
\expandafter\let\csname oldend#1\expandafter\endcsname\csname
end#1\endcsname
 \renewenvironment{#1}
   {\linenomath\csname old#1\endcsname}
   {\csname oldend#1\endcsname\endlinenomath}
}
\newcommand*\patchBothAmsMathEnvironmentsForLineno[1]{
  \patchAmsMathEnvironmentForLineno{#1}
  \patchAmsMathEnvironmentForLineno{#1*}
}
\AtBeginDocument{
\patchBothAmsMathEnvironmentsForLineno{equation}
\patchBothAmsMathEnvironmentsForLineno{align}
\patchBothAmsMathEnvironmentsForLineno{flalign}
\patchBothAmsMathEnvironmentsForLineno{alignat}
\patchBothAmsMathEnvironmentsForLineno{gather}
\patchBothAmsMathEnvironmentsForLineno{multline}
\patchBothAmsMathEnvironmentsForLineno{eqnarray}
}

\usepackage{hyperref}    
\usepackage[all]{hypcap}

\def\lhcb {\mbox{LHCb}\xspace}

\def\babar  {\mbox{BaBar}\xspace}
\def\belle  {\mbox{Belle}\xspace}

\def\MagUp {\mbox{\em Mag\kern -0.05em Up}\xspace}

\ifthenelse{\boolean{uprightparticles}}
{

 \def\Ppi         {\ensuremath{\uppi}\xspace}
 
 \def\Prho        {\ensuremath{\uprho}\xspace}

 \def\PDelta      {\ensuremath{\Delta}\xspace}
 \def\PXi      {\ensuremath{\Xi}\xspace}
 \def\PLambda      {\ensuremath{\Lambda}\xspace}
 \def\PSigma      {\ensuremath{\Sigma}\xspace}
 \def\POmega      {\ensuremath{\Omega}\xspace}
 \def\PUpsilon      {\ensuremath{\Upsilon}\xspace}

 \def\PB      {\ensuremath{\mathrm{B}}\xspace}
 
 \def\PD      {\ensuremath{\mathrm{D}}\xspace}

 \def\PK      {\ensuremath{\mathrm{K}}\xspace}

 \def\Pb      {\ensuremath{\mathrm{b}}\xspace}
 \def\Pc      {\ensuremath{\mathrm{c}}\xspace}

 \def\Pi      {\ensuremath{\mathrm{i}}\xspace}

 \def\Ps      {\ensuremath{\mathrm{s}}\xspace}

}
{

 \def\Ppi         {\ensuremath{\pi}\xspace}
 
 \def\Prho        {\ensuremath{\rho}\xspace}

 \mathchardef\PDelta="7101
 \mathchardef\PXi="7104
 \mathchardef\PLambda="7103
 \mathchardef\PSigma="7106
 \mathchardef\POmega="710A
 \mathchardef\PUpsilon="7107
 
 \def\PB      {\ensuremath{B}\xspace}
 
 \def\PD      {\ensuremath{D}\xspace}

 \def\PK      {\ensuremath{K}\xspace}

 \def\Pb      {\ensuremath{b}\xspace}
 \def\Pc      {\ensuremath{c}\xspace}

 \def\Pi      {\ensuremath{i}\xspace}

 \def\Ps      {\ensuremath{s}\xspace}

}

\makeatletter
\ifcase \@ptsize \relax
  \newcommand{\miniscule}{\@setfontsize\miniscule{4}{5}}
\or
  \newcommand{\miniscule}{\@setfontsize\miniscule{5}{6}}
\or
  \newcommand{\miniscule}{\@setfontsize\miniscule{5}{6}}
\fi
\makeatother

\DeclareRobustCommand{\optbar}[1]{\shortstack{{\miniscule (\rule[.5ex]{1.25em}{.18mm})}
  \\ [-.7ex] $#1$}}

\def\squark    {{\ensuremath{\Ps}}\xspace}

\def\cquark    {{\ensuremath{\Pc}}\xspace}

\def\bquark    {{\ensuremath{\Pb}}\xspace}

\def\pion   {{\ensuremath{\Ppi}}\xspace}

\def\pip    {{\ensuremath{\pion^+}}\xspace}
\def\pim    {{\ensuremath{\pion^-}}\xspace}
\def\pipm   {{\ensuremath{\pion^\pm}}\xspace}

\def\rhomeson {{\ensuremath{\Prho}}\xspace}
\def\rhoz     {{\ensuremath{\rhomeson^0}}\xspace}

\def\kaon    {{\ensuremath{\PK}}\xspace}
  \def\Kbar    {{\kern 0.2em\overline{\kern -0.2em \PK}{}}\xspace}

\def\KorKbar    {\kern 0.18em\optbar{\kern -0.18em K}{}\xspace}

\def\Km      {{\ensuremath{\kaon^-}}\xspace}

  \def\Dbar    {{\kern 0.2em\overline{\kern -0.2em \PD}{}}\xspace}
\def\D       {{\ensuremath{\PD}}\xspace}

\def\DorDbar    {\kern 0.18em\optbar{\kern -0.18em D}{}\xspace}

\def\Dzb     {{\ensuremath{\Dbar{}^0}}\xspace}

\def\Dstarzb {{\ensuremath{\Dbar{}^{*0}}}\xspace}

\def\Dssm    {{\ensuremath{\D^{*-}_\squark}}\xspace}

\def\B       {{\ensuremath{\PB}}\xspace}
\def\Bbar    {{\ensuremath{\kern 0.18em\overline{\kern -0.18em \PB}{}}}\xspace}

\def\BorBbar    {\kern 0.18em\optbar{\kern -0.18em B}{}\xspace}
\def\Bz      {{\ensuremath{\B^0}}\xspace}

\def\Bd      {{\ensuremath{\B^0}}\xspace}
\def\Bs      {{\ensuremath{\B^0_\squark}}\xspace}

  \def\Y#1S{\ensuremath{\PUpsilon{(#1S)}}\xspace}

\def\Lz          {{\ensuremath{\PLambda}}\xspace}
\def\Lbar        {{\ensuremath{\kern 0.1em\overline{\kern -0.1em\PLambda}}}\xspace}
\def\LorLbar    {\kern 0.18em\optbar{\kern -0.18em \PLambda}{}\xspace}

\def\Lb      {{\ensuremath{\Lz^0_\bquark}}\xspace}

\def\BF         {{\ensuremath{\cal B}}\xspace}

\def\BR         {\BF}

\def\to                 {\ensuremath{\rightarrow}\xspace}

\def\CP                {{\ensuremath{C\!P}}\xspace}

\def\AT#1     {\ensuremath{A_{\mathrm{T}}^{#1}}\xspace}

\newcommand{\tev}{\ifthenelse{\boolean{inbibliography}}{\ensuremath{~T\kern -0.05em eV}\xspace}{\ensuremath{\mathrm{\,Te\kern -0.1em V}}}\xspace}
\newcommand{\gev}{\ensuremath{\mathrm{\,Ge\kern -0.1em V}}\xspace}
\newcommand{\mev}{\ensuremath{\mathrm{\,Me\kern -0.1em V}}\xspace}
\newcommand{\kev}{\ensuremath{\mathrm{\,ke\kern -0.1em V}}\xspace}
\newcommand{\ev}{\ensuremath{\mathrm{\,e\kern -0.1em V}}\xspace}
\newcommand{\gevc}{\ensuremath{{\mathrm{\,Ge\kern -0.1em V\!/}c}}\xspace}
\newcommand{\mevc}{\ensuremath{{\mathrm{\,Me\kern -0.1em V\!/}c}}\xspace}
\newcommand{\gevcc}{\ensuremath{{\mathrm{\,Ge\kern -0.1em V\!/}c^2}}\xspace}
\newcommand{\gevgevcccc}{\ensuremath{{\mathrm{\,Ge\kern -0.1em V^2\!/}c^4}}\xspace}
\newcommand{\mevcc}{\ensuremath{{\mathrm{\,Me\kern -0.1em V\!/}c^2}}\xspace}

\def\mum  {\ensuremath{{\,\upmu\rm m}}\xspace}

\def\gsim{{~\raise.15em\hbox{$>$}\kern-.85em
          \lower.35em\hbox{$\sim$}~}\xspace}
\def\lsim{{~\raise.15em\hbox{$<$}\kern-.85em
          \lower.35em\hbox{$\sim$}~}\xspace}

\def\ptot       {\mbox{$p$}\xspace}
\def\pt         {\mbox{$p_{\rm T}$}\xspace}

\def\evtgen     {\mbox{\textsc{EvtGen}}\xspace}

\def\geant      {\mbox{\textsc{Geant4}}\xspace}

\def\photos     {\mbox{\textsc{Photos}}\xspace}

\def\pythia     {\mbox{\textsc{Pythia}}\xspace}

\def\tell1  {TELL1\xspace}
\def\ukl1   {UKL1\xspace}

\newcommand{\ie}{\mbox{\itshape i.e.}\xspace}

\def\dbpim    {\Dzb\pim}
\def\pipi    {\pip\pim}
\def\mdbpips      {m^2(\Dzb \pip)}
\def\mdbpims      {m^2(\Dzb \pim)}
\def\mpipis      {m^2(\pip\pim)}

\def\mdbpim     {m(\Dzb\pim)}
\def\mdpipm     {m(\Dzb\pipm)}
\def\btodpipi  {\Bz \to \Dzb \pip \pim}

\def\btodrhoz  {\Bz \to \Dzb \rhoz}

\def\btodrho {B \to D \rho}

\def\DstarZero   {\D^{*}_{0}(2400)^{-}}
\def\DstarTwo  {\D^{*}_{2}(2460)^{-}}

\usepackage{cite} 
\usepackage{mciteplus}

\begin{document}
\renewcommand{\thefootnote}{\fnsymbol{footnote}}
\setcounter{footnote}{1}

\begin{titlepage}
\pagenumbering{roman}

\vspace*{-1.5cm}
\centerline{\large EUROPEAN ORGANIZATION FOR NUCLEAR RESEARCH (CERN)}
\vspace*{1.5cm}
\hspace*{-0.5cm}
\begin{tabular*}{\linewidth}{lc@{\extracolsep{\fill}}r}
\ifthenelse{\boolean{pdflatex}}
{\vspace*{-2.7cm}\mbox{\!\!\!\includegraphics[width=.14\textwidth]{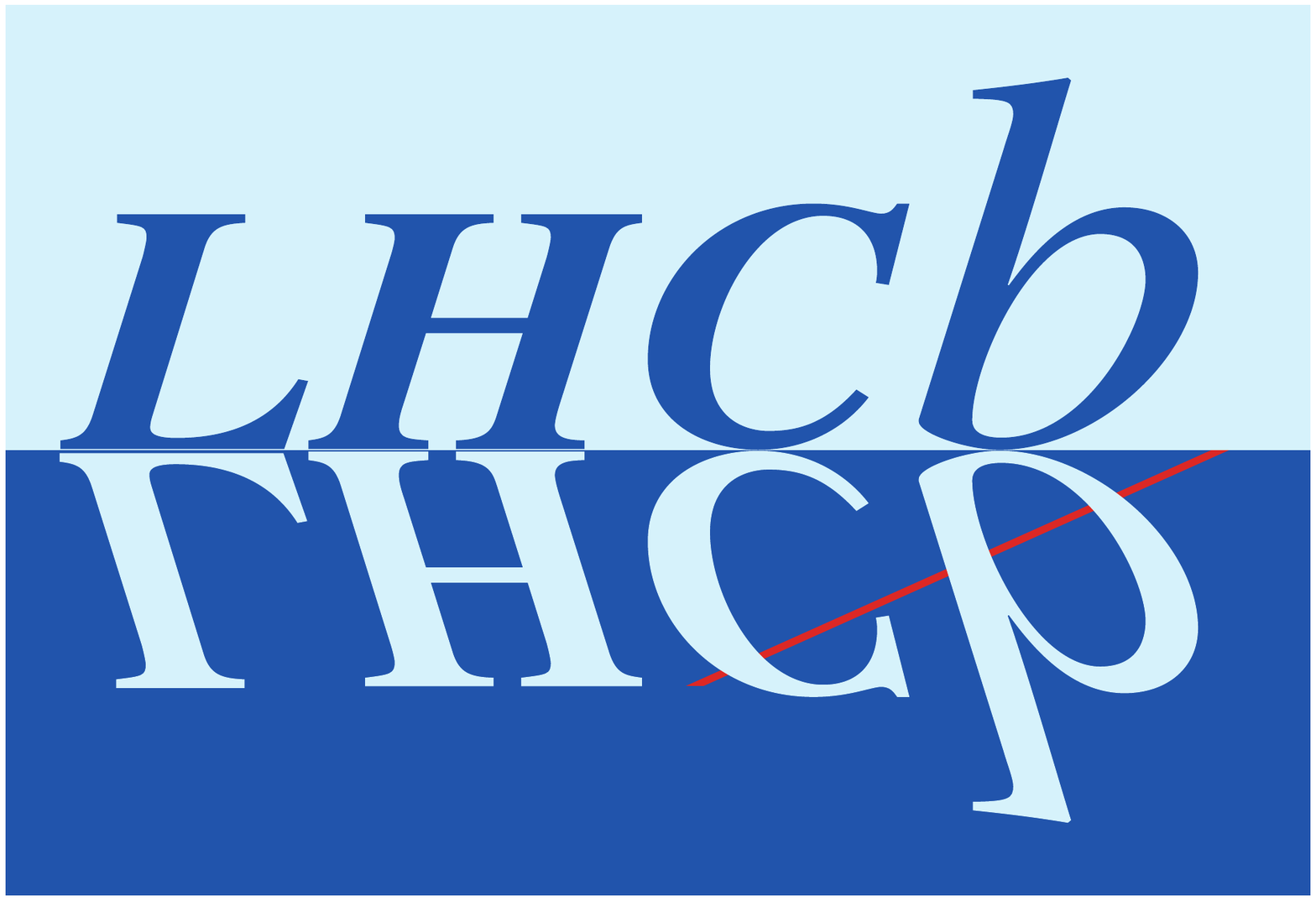}} & &}
{\vspace*{-1.2cm}\mbox{\!\!\!\includegraphics[width=.12\textwidth]{lhcb-logo.eps}} & &}
\\
 & & CERN-PH-EP-2015-110 \\  
 & & LHCb-PAPER-2014-070 \\  
 & & 7 May, 2015 \\ 
 & & \\
\end{tabular*}

\vspace*{2.0cm}

{\bf\boldmath\huge
\begin{center}
  Dalitz plot analysis of $\btodpipi$ decays
\end{center}
}

\vspace*{1.0cm}

\begin{center}
The LHCb collaboration\footnote{Authors are listed at the end of this paper.}
\end{center}

\begin{abstract}
  \noindent
The resonant substructures of $\btodpipi$ decays are studied with the Dalitz plot technique. In this study a data sample corresponding to an integrated luminosity of 3.0 fb$^{-1}$ of $pp$ collisions collected by the LHCb detector is used. The branching fraction of the $\btodpipi$ decay in the region $\mdpipm>2.1\gevcc$ is  measured to be $(8.46 \pm 0.14 \pm 0.29 \pm 0.40) \times 10^{-4}$,  where the first uncertainty is statistical, the second is systematic and the last arises from the normalisation channel $B^0 \to D^*(2010)^-\pi^+$. The $\pi^+\pi^-$ S-wave components are modelled with the Isobar and K-matrix formalisms.
Results of the Dalitz plot analyses using both models are presented. A resonant structure at $\mdbpim\approx 2.8 \gevcc$ is confirmed and its spin-parity is determined for the first time as $J^P = 3^-$. The branching fraction, mass and width of this structure are determined together with those of the $\DstarZero$ and $\DstarTwo$ resonances. The branching fractions of other $B^0 \to \Dzb h^0$ decay components with $h^0 \to \pi^+\pi^-$ are also reported. Many of these branching fraction measurements are the most precise to date. The first observation of the decays $\Bz \to \Dzb  f_0(500)$, $\Bz \to \Dzb  f_0(980)$, $\Bz \to \Dzb  \rho(1450)$, $\Bz \to D_3^*(2760)^- \pip$ and the first evidence of $\Bz \to \Dzb  f_0(2020)$ are presented.
\end{abstract}

\vspace*{1.0cm}

\begin{center}
  Published in Phys.~Rev.~D
\end{center}

\vspace{\fill}

{\footnotesize
\centerline{\copyright~CERN on behalf of the \lhcb collaboration, license \href{http://creativecommons.org/licenses/by/4.0/}{CC-BY-4.0}.}}
\vspace*{2mm}

\end{titlepage}

\newpage
\setcounter{page}{2}
\mbox{~}
\newpage

\cleardoublepage

\newcommand{\al}{\ensuremath{\kern 0.5em }}
\newcommand{\all}{\ensuremath{\kern 0.25em }}
\renewcommand{\thefootnote}{\arabic{footnote}}
\setcounter{footnote}{0}

\pagestyle{plain} 
\setcounter{page}{1}
\pagenumbering{arabic}

\section{Introduction}
\label{sec:Introduction}

The study of the  Cabibbo-Kobayashi-Maskawa (CKM) mechanism~\cite{PhysRevLett.10.531,PTP.49.652}  is a central topic in flavour physics.
Accurate measurements of the various CKM matrix parameters through different processes provide sensitivity to new physics effects,
by  testing the global consistency of the Standard Model. Among them, the  CKM angle $\beta$ is  expressed in terms of the CKM matrix elements
as $\arg(-V_{cd}V^*_{cb}/V_{td}V_{tb}^*)$. The most precise measurements have been obtained with the $B^0 \to (c\bar{c}) K^{(\ast)0}$ decays by \babar~\cite{Babarsin2beta}, \belle~\cite{Bellesin2beta} and more recently by LHCb~\cite{LHCb-PAPER-2015-004}. The decay~\footnote{The inclusion of charge conjugate states is implied throughout the paper.} $\btodpipi$ through the $b \to c\bar{u}d$ transition has sensitivity to the CKM angle $\beta$~\cite{Charles:1998vf,Timsin2beta,BabarDCPh0,BabarDGGSZh0,BelleDGGSZh0} and  to new physics effects~\cite{PLB395-241,IJMPA12-2459,PLB407-61,PRL79-61}.

\begin{figure}[!tbh]
\centering
\includegraphics[scale=0.63 ]{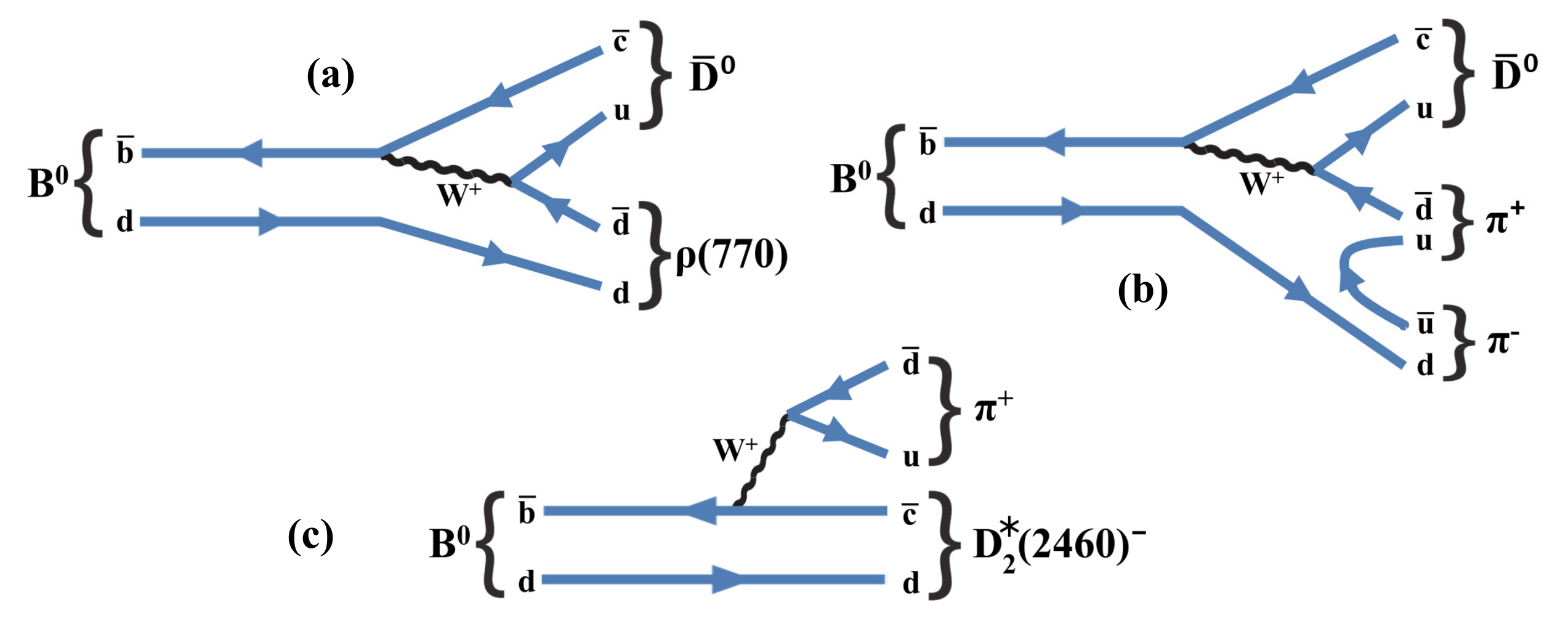}
\caption{Examples of tree diagrams via $\bar{b} \to \bar{c}u\bar{d}$ transition to produce (a) $\pipi$ resonances, (b) nonresonant three-body decay and (c) $\dbpim$ resonances. } \label{fig: Feyman}
\end{figure}

The Dalitz plot analysis~\cite{Dalitz}  of  $\btodpipi$ decays, with the $\Dzb \to K^+ \pi^-$ mode, is presented as
the first step towards an alternative method to measure the CKM angle $\beta$. Two sets of results are given, where the $\pi^+\pi^-$ S-wave components are
modelled with the Isobar~\cite{Isobar1, Isobar2, Isobar3} and K-matrix~\cite{KMatrixTheory} formalisms. Dalitz plot analyses of the decay $\btodpipi$ have already been performed by \belle~\cite{BelleDpipi1,BelleDpipi2} and \babar~\cite{BaBarDpipi}.  Similar studies for the charged \B decays  $ B^- \to D^{(*)+} \pi^- \pi^-$ have been published by the $B$-factories~\cite{Abe:2003zm,Aubert:2009wg}. The LHCb dataset offers a larger and almost pure signal sample. Feynman diagrams of the dominant tree level amplitudes contributing to the decay $\btodpipi$ are shown in Fig.~\ref{fig: Feyman}.

In addition to the interest for the CKM parameter measurements, the analysis of the Dalitz plot  of the  $\btodpipi$ decay is motivated by its rich resonant structure. The decay $\btodpipi$ contains information about excited $D$ mesons  decaying to $D\pi$, with natural spin and parity $J^P = 0^+$, $1^-$, $2^+$, ... A complementary  Dalitz plot analysis of the decay $\Bs \to \Dzb \Km \pip$ was recently published by LHCb~\cite{LHCb-PAPER-2014-035,LHCb-PAPER-2014-036}, and constrains the phenomenology of the  $\Dzb\Km$ ($D_{sJ}^-$) and $\Km\pip$ states. The spectrum of excited $D$ mesons is predicted by theory~\cite{DsTheory1,DsTheory2} and contains the known states $D^*(2010), D^*_0(2400), D^*_2(2460)$, as well as other  unknown states not yet fully explored. An extensive  discussion on theory predictions for the $c\bar{u}$, $c\bar{d}$ and $c\bar{s}$ mass spectra is provided in Refs.~\cite{LHCb-PAPER-2013-026,LHCb-PAPER-2014-036}.
More recent measurements performed in inclusive decays by  \babar~\cite{NewDstarBaBar}  and  LHCb~\cite{LHCb-PAPER-2013-026},  have led to  the observation of several new states: $D^*(2650), D^*(2760)$, and $D^*(3000)$.  However, their spin and parity are difficult to determine from inclusive studies. Orbitally excited $D$ mesons  have also been studied in semi-leptonic $B$ decays (see a review in Ref.~\cite{HFAGsin2beta}) with limited precision. These are of prime interest  both in the extraction of  the CKM parameter $|V_{cb}|$, where longstanding differences remain between exclusive and inclusive methods (see review  in Ref.~\cite{PDG}), and in recent studies of $B \to D^{(*)}\tau \bar{\nu}_\tau$~\cite{BaBarDtauNu} which have generated much theoretical discussion (see, e.g., Refs.~\cite{Dtaunu1,Dtaunu2}).

A measurement of the branching fraction of the decay $B^0\to  \Dzb\rho^0$ is also presented. 
This study helps in understanding the effects of  colour-suppression in $B$ decays, which is due to the requirement that the colour quantum numbers of the quarks produced from the virtual $W$ boson must match those of the spectator quark to form a $\rho^0$ meson~\cite{Bauer:1986bm,Neubert:2001sj,Chua:2001br,Mantry:2003uz,BaBarColSup}. Moreover, using isospin symmetry to relate the decay amplitudes of $B^0\to  \Dzb\rho^0$, $B^0\to  D^-\rho^+$ and $B^+\to  \Dzb\rho^+$, effects of final state interactions (FSI) can be studied in those decays (see a review in Refs.~\cite{Rosner:1999zm,Neubert:2001sj}). The previous measurement for the branching fraction of $B^0\to  \Dzb\rho^0$ has limited precision, $(3.2\pm 0.5)\times10^{-4}$\cite{BelleDpipi2},  and is in agreement with  theoretical predictions that range from $1.7$ to $3.4 \times10^{-4}$~\cite{Chua:2001br,Keum:2003js}.

Finally, a study of the $\pi^+\pi^-$  system is performed  on a broad phase-space range in $\btodpipi$ from $280 \mevcc$  ($\approx 2 m_{\pi}$) to $3.4 \gevcc$  ($\approx m_{B^0_d}-m_{D^0}$), which is much larger than that accessible in charmed meson decays such as  $D^0 \to K^0_ {\rm S}\pi^+\pi^-$~\cite{Aubert:2008bd,delAmoSanchez:2010xz,Peng:2014oda} or  in $B$ decays such as $B^0_{(s)} \to J/\psi\pi^+\pi^-$~\cite{LHCb-PAPER-2012-005,LHCb-PAPER-2012-045,LHCb-PAPER-2013-069,LHCb-PAPER-2014-012}. The nature of the light scalar $\pi^+\pi^-$  states below $1~\gevcc$   ($J^{PC}=0^{++}$), and in particular the $f_0(500)$ and $f_0(980)$ states, has been a longstanding debate (see, e.g., Refs.~\cite{Amsler:2013wea,Jaffe:2004ph,Klempt:2007cp}). Popular interpretations include tetraquarks, meson-meson bound states (molecules), or some other mixtures, where the iso-singlets   $f_0(500)$ and $f_0(980)$ can mix, therefore leading to a non-trivial nature (e.g. pure  $s\bar{s}$ state) of the $f_0(980)$ and complicating the determination of the CKM phase $\phi_s$ from $B^0_{s} \to J/\psi\pi^+\pi^-$ decays~\cite{Zhang:2012zk,LHCb-PAPER-2013-069,LHCb-PAPER-2014-019}.  In the tetraquark picture, the mixing angle, $\omega_{\textrm{mix}}$, between the $f_0(980)$ and $ f_0(500)$ states is predicted to be $|\omega_{\textrm{mix}}| \approx  20^{\circ}$~\cite{Maiani:2004uc,Hooft:2008we} (recomputed with the latest average of the mass of the $\kappa$ meson  $682  \pm 29 \mevcc$~\cite{PDG}). Other theory models based on QCD factorisation and its extensions~\cite{Wang:2009azc,Li:2012sw} predict that the $f_0(500)$ and $f_0(980)$ mixing angle $\varphi_{\textrm{mix}}$ for the $q\bar{q}$ model is $20^{\circ} \lesssim \varphi_{\textrm{mix}}  \lesssim 45^{\circ}$. The LHCb experiment, in the study of $B^0_{(s)} \to J/\psi\pi^+\pi^-$ decays~\cite{LHCb-PAPER-2012-045,LHCb-PAPER-2013-069,LHCb-PAPER-2014-012}, has already set stringent upper bounds on $\varphi_{\textrm{mix}}$  in $B^0$ ($B^0_{s}$)  decay: $\varphi_{\textrm{mix}}<17^{\circ}$ ($<7.7^{\circ}$) at $90\%$ CL. For the first time, the $f_0(500) - f_0(980)$ mixing  in the  $\btodpipi$ decay, both in $q\bar{q}$ and tetraquark pictures, is studied.

The analysis of the decay  $\btodpipi$ presented in this paper is based on a data sample corresponding to an integrated luminosity of $3.0  \,{\rm fb}^{-1}$ of $pp$ collision data collected with the LHCb detector. Approximately one third of the data was obtained during 2011 when the collision centre-of-mass energy was $\sqrt{s} = 7 \tev$ and the rest during 2012 with $\sqrt{s} = 8 \tev$.

The paper is organised as follows. A brief description of the LHCb detector as well as the reconstruction and simulation software is given in Sec.~\ref{sec:Detector}. The selection of signal candidates and the fit to the \Bd candidate invariant mass distribution used to separate and to measure signal and background yields are described in Sec.~\ref{sec:Selection}. An overview of the Dalitz plot analysis formalism is given in Sec.~\ref{sec:DalitzFormulism}. Details and results of the amplitude analysis fits are presented in Sec.~\ref{sec:DalitzFit}. In Sec.~\ref{sec:BFDstarpi} the measurement of the $\btodpipi$ branching fraction is documented. The evaluation of systematic uncertainties is described in Sec.~\ref{sec:Systematic}. The results are given in Sec.~\ref{sec:Results}, and a summary concludes the paper in Sec.~\ref{sec:conclusion}.

\section{The LHCb detector}
\label{sec:Detector}

The \lhcb detector~\cite{Alves:2008zz} is a single-arm forward
spectrometer covering the \mbox{pseudorapidity} range $2<\eta <5$,
designed for the study of particles containing \bquark or \cquark
quarks. The detector includes a high-precision tracking system
consisting of a silicon-strip vertex detector surrounding the $pp$
interaction region~\cite{LHCb-DP-2014-001}, a large-area silicon-strip detector located
upstream of a dipole magnet with a bending power of about
$4{\rm\,Tm}$, and three stations of silicon-strip detectors and straw
drift tubes~\cite{LHCb-DP-2013-003} placed downstream of the magnet.
The tracking system provides a measurement of momentum, \ptot,  with
a relative uncertainty that varies from 0.4\% at low momentum to 0.6\% at 100\gevc.
The minimum distance of a track to a primary vertex, the impact parameter (IP), is measured with a resolution of $(15+29/\pt)\mum$,
where \pt is the component of \ptot transverse to the beam, in \gevc.
Different types of charged hadrons are distinguished using information
from two ring-imaging Cherenkov detectors~\cite{LHCb-DP-2012-003}. Photon, electron and
hadron candidates are identified by a calorimeter system consisting of
scintillating-pad and preshower detectors, an electromagnetic
calorimeter and a hadronic calorimeter. Muons are identified by a
system composed of alternating layers of iron and multiwire
proportional chambers~\cite{LHCb-DP-2012-002}.

The online event selection is performed by a trigger which consists of a 
hardware stage, based on information from the calorimeter and muon systems, 
followed by a software stage, which applies a full event reconstruction.
At the hardware trigger stage, events are required to have a muon with high
\pt or a hadron, photon or electron with high transverse energy in the
calorimeters.
For hadrons, the transverse energy threshold is 3.5 GeV.
The software trigger requires a two-, three- or
four-track secondary vertex with a significant displacement from the primary
$pp$ interaction vertices~(PVs). At least one charged particle must have a
transverse momentum $\pt > 1.7 \gevc$ and be inconsistent with originating from
a PV. A multivariate algorithm~\cite{BBDT} is used for the identification of
secondary vertices consistent with the decay of a \bquark hadron.
The \pt of the photon from \Dssm decay is too low to contribute to the trigger
decision.

Simulated events are used to characterise the detector response to signal and certain types of background events.
In the simulation, $pp$ collisions are generated using \pythia~\cite{Sjostrand:2006za,*Sjostrand:2007gs} with a specific \lhcb configuration~\cite{LHCb-PROC-2010-056}.  Decays of hadronic particles are described by \evtgen~\cite{Lange:2001uf}, in which final-state
radiation is generated using \photos~\cite{Golonka:2005pn}. The interaction of the generated particles with the detector, and its
response, are implemented using the \geant toolkit~\cite{Allison:2006ve, *Agostinelli:2002hh} as described in Ref.~\cite{LHCb-PROC-2011-006}.

\section{Event selection}
\label{sec:Selection}

Signal \Bz candidates are formed by combining  \Dzb candidates, reconstructed in the decay channel $K^+\pi^-$, with two additional pion candidates of opposite charge. Reconstructed tracks are required to be of good quality and to be inconsistent with originating from a PV. They are also required to have sufficiently high $p$ and \pt and to be
within kinematic regions where reasonable particle identification (PID) performance is achieved, as determined by calibration samples of
 $D^{*+} \to D^0(K^-\pi^+) \pi^+$ decays.  The four final state tracks are required to be positively identified by the PID system. The \Dzb daughters are required to form a good quality vertex and to have an invariant mass within 100 \mevcc of the known \Dzb mass~\cite{PDG}. The \Dzb candidates and the two charged pion candidates are required to form a good vertex. The reconstructed \Dzb and \Bz vertices are required to be significantly displaced from the PV. To improve the \Bz candidate invariant mass resolution, a kinematic fit~\cite{Hulsbergen:2005pu} is used, constraining the \Dzb candidate to  its known mass~\cite{PDG}.

By requiring the reconstructed \Dzb vertex to be displaced downstream from the reconstructed \Bz vertex, backgrounds from both charmless $B$ decays and direct prompt charm production coming from the PV are reduced to a negligible level. Background from $D^{*}(2010)^{-}$ decays is removed by requiring $m(\Dzb\pi^{\pm})>2.1$\gevcc. Backgrounds from doubly mis-identified $\Dzb \to K^+ \pi^-$ or doubly Cabibbo-suppressed $\Dzb \to K^- \pi^+$ decays are also removed by this requirement.

To further distinguish signal from combinatorial background, a multivariate analysis based on a Fisher discriminant~\cite{Fisher:1936et} is applied. The \textit{sPlot} technique~\cite{splot} is used to statistically separate signal and background events with the \Bz candidate mass used as the discriminating variable. Weights obtained from this procedure are applied to the candidates to obtain signal and background distributions that are used to train the discriminant. The Fisher discriminant uses information about the event kinematic properties, vertex quality, IP and \pt of the tracks and flight distance from the PV. It is optimised by maximising the purity of the signal events.

Signal candidates are retained for the Dalitz plot analysis if the invariant mass of the \Bz meson lies in the range [5250, 5310] $\mevcc$ and that of the \Dzb meson in the range [1840, 1890] $\mevcc$ (called the signal region). Once all selection requirements are applied, less than 1\,\% of the events contain multiple candidates, and in those cases one candidate is chosen randomly.

Background contributions from decays with the same topology, but having one or two mis-identified particles, are estimated to be less than 1\,\% and are not considered in the Dalitz analysis. These background contributions include decays like $B^0 \to \Dzb K^+ \pi^-$, $B_s^0 \to \Dzb K^- \pi^+$ \cite{LHCb-PAPER-2013-022}, $\Lb \to D^0 p \pi^-$  \cite{LHCb-PAPER-2013-056} and $\btodpipi$ with $\Dzb \to \pi^+\pi^-$ or $\Dzb \to K^+ K^-$.

Partially reconstructed decays of the type $\btodpipi X$, where one or more particles are not reconstructed, have similar efficiencies to the signal channel decays. They are distributed in the region below the \Bz mass.
By requiring the invariant mass of \Bz candidates to be larger than 5250 \mevcc, these backgrounds are reduced to a negligible level, as determined by simulated samples of $B^0 \to \Dstarzb \pi^+ \pi^-$ and $B^0 \to \Dstarzb \rho(770)$ with $\Dstarzb$ decaying into $\Dzb \gamma$ or $\Dzb \pi^0$ under different hypotheses for the $\Dstarzb$ helicity.

\begin{figure}
  \centering
  \includegraphics[height=7cm]{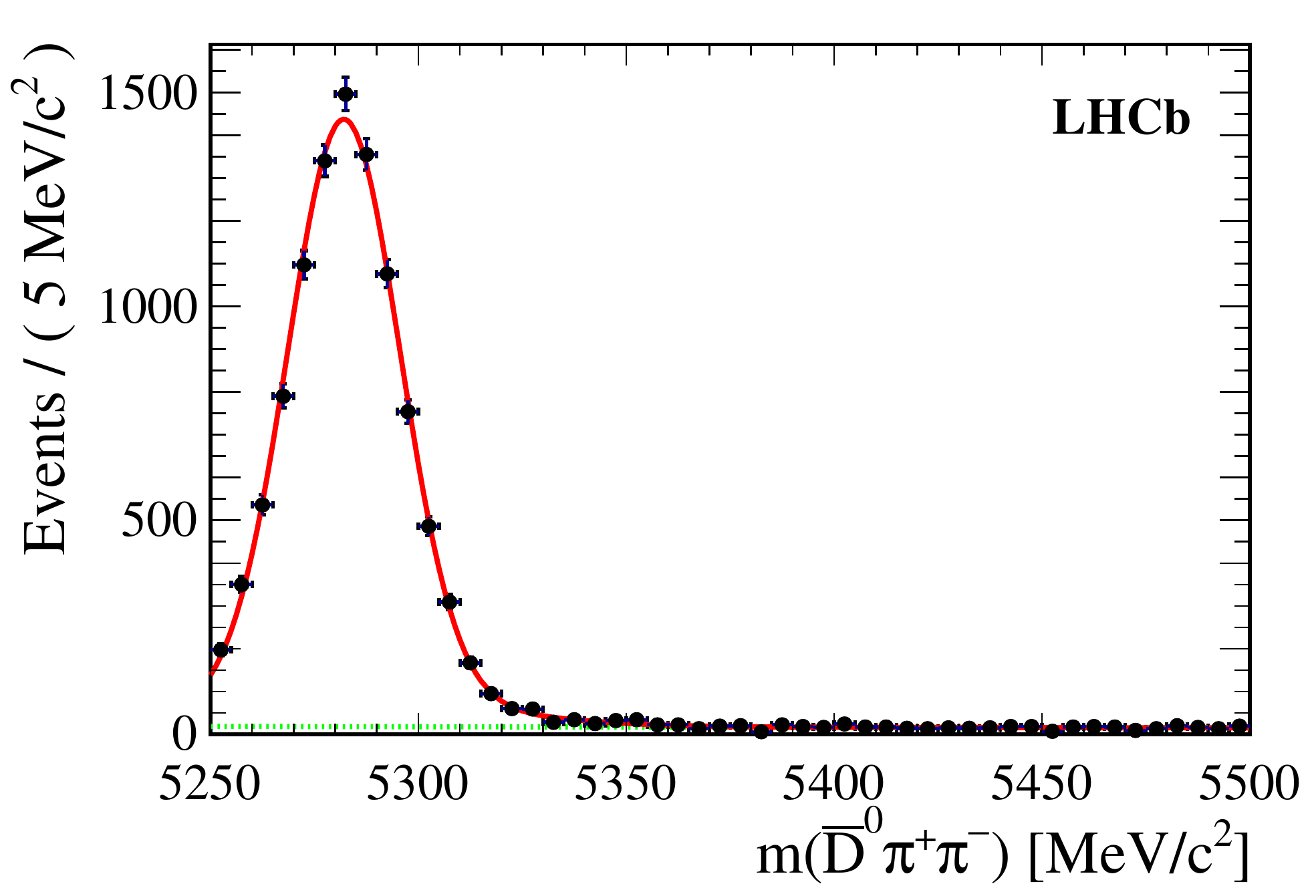}
\caption{Invariant mass distribution of $B^0 \to \Dzb \pi^+\pi^-$ candidates.
Data points are shown in black.
The fit is shown as a  solid (red) line with the background component displayed as  dashed (green) line.
}
\label{fig: MassFit}
\end{figure}

The signal and combinatorial background yields are determined using an unbinned extended maximum likelihood fit to the invariant mass distribution of $\Bz$ candidates. The invariant mass distribution is shown in Fig.~\ref{fig: MassFit}, with the fit result superimposed. The fit uses a Crystal Ball (CB) function~\cite{CBFunction} convoluted with a Gaussian function for the signal distribution and a linear function for the combinatorial background distribution in the mass range of [5250, 5500]~\mevcc.
Simulated studies validate this choice of signal shape and the tail parameters of the CB function are fixed to those determined from simulation.
Table~\ref{tab: MassFit} summarises the fit results on the free parameters, where $\mu_{B^0}$ is the mean peak position and $\sigma_\textrm{G}$ is the width of the Gaussian function.
The parameter $\sigma_{\textrm{CB}}$ is the width of the Gaussian core of the CB function. The parameters $f_{\textrm{CB}}$ and $p_1$ give the fit fraction of the CB function and the slope of the linear function that describes the background distribution. The yields of signal ($\nu_s^0$) and background ($\nu_b^0$) events given in Table~\ref{tab: MassFit} are calculated within the signal region. The purity is $(97.8\pm 0.2)\,\%$.

\begin{table}[!tbh]
\centering
\caption{Results of the fit to the invariant mass distribution of $B^0 \to \Dzb \pi^+\pi^-$ candidates.
Uncertainties are statistical only.}
\label{tab: MassFit}
\begin{tabular}{lrcl}
\hline
Parameter & \multicolumn{3}{c}{Value}  \\
\hline
$\mu_{B^0}$ & $5282.1$ & $\pm$ & $0.2$ \mevcc \\
$\sigma_\textrm{G}$ & $33.6$ & $ \pm $ & $5.4$ \mevcc \\
$\sigma_{\textrm{CB}}$ & $13.4$ & $ \pm $ & $0.3$ \mevcc \\
$f_{\textrm{CB}}$ & $0.908$ &  $\pm$ & $0.025$ \\
$p_1$ & $-0.152$ & $ \pm $ & $0.035$ (\gevcc)$^{-1}$ \\
$\nu_s^0$ & $9565$ & $ \pm$ & $116$ \\
$\nu_b^0$ & $215$ &$\pm$ & $19$ \\
\hline
\end{tabular}
\end{table}

\section{Dalitz plot analysis formalism}
\label{sec:DalitzFormulism}
The analysis of the distribution of decays across the  Dalitz plot~\cite{Dalitz} allows a determination of the amplitudes contributing to the three-body $\btodpipi$ decay.
Two of the three possible two-body invariant mass-squared combinations, which are connected by
\begin{equation}
\mdbpips + \mdbpims + \mpipis = m^2_{B^0} + m^2_{D^0} + 2 m^2_{\pi},
\end{equation}
are sufficient to describe the kinematics of the system.
The two observables $\mdbpims$ and $\mpipis$, where resonances are expected to appear, are chosen in this paper.
These observables are calculated with the masses of the \Bz and \Dzb mesons constrained to their known values~\cite{PDG}.
The invariant mass resolution has negligible effect and therefore it is not modeled in the Dalitz plot analysis.

The total decay amplitude is described by a coherent sum of amplitudes from resonant or nonresonant intermediate processes as
\begin{equation}
M(\vec{x}) =  \sum_i c_i A_i(\vec{x}). \label{fun: matrixelement}
\end{equation}
The complex coefficient $c_i$ and amplitude $A_i(\vec{x})$ describe the relative contribution and dynamics of the $i$-th intermediate state,
where $\vec{x}$ represents the $(\mdbpims,\mpipis)$ coordinates in the Dalitz plot.
The Dalitz plot analysis determines the coefficients $c_i$.
In addition, fit fractions and interference fit fractions are also calculated to give a convention-independent representation of the population of the Dalitz plot.
The fit fractions are defined as
\begin{equation}
{\cal{F}}_{i} = \frac{\int |c_i A_i(\vec{x})|^2 \, d\vec{x}}{ \int |\sum_i c_i A_i(\vec{x})|^2 \, d\vec{x}},  \label{fun: fitfrac}
\end{equation}
and the interference fit fractions between the resonances $i$ and $j$ ($i < j$) are defined as
\begin{equation}
{\cal{F}}_{ij} = \frac{\int 2 \textrm{Re}[c_i c_j^* A_i(\vec{x})A_j^*(\vec{x})] d\vec{x}}{\int |\sum_i c_i A_i(\vec{x})|^2 \, d\vec{x}},
\end{equation}
where the integration is performed over the full Dalitz plot with $m(\Dzb\pi^{\pm}) > 2.1$ \gevcc.
Due to these interferences between different contributions, the sum of the fit fractions is not necessarily equal to unity.

The amplitude $A_i(\vec{x})$ for a specific resonance $r$ with spin $L$ is written as
\begin{equation}
A_i(\vec{x}) = F_B^{(L)}(q, q_0) \times  F_r^{(L)}(p, p_0) \times T_L (\vec{x})\times R(\vec{x}).  \label{fun: DalitzForm}
\end{equation}
The functions $F_B^{(L)}(q, q_0)$ and $F_r^{(L)}(p, p_0)$ are the Blatt-Weisskopf barrier factors~\cite{BWFactor} for the production, $\Bz \to r h_3$, and the decay, $r \to h_1 h_2$, of the resonance, respectively. The parameters $p$ and $q$ are the momenta of one of the resonance daughters ($h_1$ or $h_2$) and of the bachelor particle ($h_3$), respectively, both evaluated in the rest frame of the resonance. The value $p_0$ ($q_0$) represents the value of $p$ ($q$) when the invariant mass of the resonance is equal to its pole mass.
The spin-dependent $F_B$ and $F_r$ functions are defined as
\begin{eqnarray}
 L = 0 &:& F^{(0)}(z, z_0) = 1, \nonumber \\
 L = 1 &:& F^{(1)}(z, z_0) = \sqrt{\frac{1+z_0}{1+z}},  \nonumber \\
 L = 2 &:& F^{(2)}(z, z_0) =\sqrt{\frac{(z_0-3)^2 + 9z_0}{(z-3)^2 + 9z}},  \\
 L = 3 &:& F^{(3)}(z, z_0) =\sqrt{\frac{z_0(z_0 -15)^2 + 9(2z_0-5)}{z(z-15)^2 + 9(2z-5)}}, \nonumber \\
 L = 4 &:& F^{(4)}(z, z_0) =\sqrt{\frac{(z_0^2 -45z_0 +105)^2 + 25z_0(2z_0-21)^2}{(z^2-45z+105)^2 + 25z(2z-21)^2}}, \nonumber
\end{eqnarray}
where $z_{(0)}$ is equal to $(r_{\textrm{BW}}\times q_{(0)})^2$ or $(r_{\textrm{BW}}\times p_{(0)})^2$.
The value for the radius of the resonance, $r_{\textrm{BW}}$, is taken to be 1.6 $\gev^{-1}\times \hbar c$ ($= 0.3$ fm)~\cite{LHCb-PAPER-2014-014}.

The function $T_L(\vec{x})$ represents the  angular distribution for  the decay of a spin $L$ resonance. It is defined as
\begin{eqnarray}
 L = 0 &:& T_0 = 1, \nonumber \\
 L = 1 &:& T_1 = \sqrt{1+y^2}\cos{\theta} \times qp,  \nonumber \\
 L = 2 &:& T_2 = (y^2 + 3/2)(\cos^2{\theta} -1/3) \times q^2p^2, \\
 L = 3 &:& T_3 = \sqrt{1+y^2}(1+2y^2/5)(\cos^3{\theta}-3\cos({\theta})/5) \times q^3p^3, \nonumber \\
 L = 4 &:& T_4 = (8y^4/35 + 40y^2/35 + 1)(\cos^4{\theta}-30\cos^2({\theta})/35+3/35) \times q^4 p^4. \nonumber
\end{eqnarray}
The helicity angle, $\theta$, of the resonance is defined as the angle between the direction of the momenta $p$ and $q$.
The $y$ dependence accounts for relativistic transformations between the \Bz and the resonance rest frames~\cite{Covariant1, Covariant2}, where
\begin{equation}
1+y^2=\frac{m_{B^0}^2 + m^2(h_1h_2) -m_{h_3}^2}{2m(h_1h_2)m_{B^0}}.
\end{equation}

Finally, $R(\vec{x})$ is the resonant lineshape and is described by the relativistic Breit-Wigner (RBW) function unless specified otherwise,
\begin{equation}
\textrm{RBW}(s) = \frac{1}{m_r^2 - s - im_r \Gamma^{(L)}(s)},
\end{equation}
where $s = m^2(h_1h_2)$ and $m_r$ is the pole mass of the resonance; $\Gamma^{(L)}(s)$,  the mass-dependent width, is defined as
\begin{eqnarray}
\Gamma^{(L)}(s)& = &\Gamma_0\left(\frac{p}{p_0}\right)^{2L+1} \left(\frac{m_r}{\sqrt{s}}\right)F_r^{(L)}(p, p_0)^2, 
\end{eqnarray}
where $\Gamma_0$ is the partial width of the resonance, i.e., the width at the peak mass $s = m_r$.

The lineshapes of $\rho(770)$, $\rho(1450)$ and $\rho(1700)$ are described by the Gounaris-Sakurai (GS) function~\cite{GSFunction},
\begin{equation}
\textrm{GS}(s) = \frac{m_{r}^2(1+ \Gamma_{0}g/m_{r})}{m_{r}^2 - s + f(s) -i m_{r}\Gamma_{\rho}(s)},
\end{equation}
where
\begin{eqnarray}
f(s) &=& \Gamma_{0} \frac{m_{r}^2}{p_0^3} \left[ (h(s) - h(m_{r}^2))p^2 + (m_{r}^2 - s) p_0^2 \frac{dh}{ds}\bigg|_{s = m_{r}^2} \right], \nonumber \\
h(s) &=& \frac{2}{\pi} \frac{p}{\sqrt{s}} \textrm{log} \left( \frac{\sqrt{s} + 2p}{2m_{\pi}} \right), \nonumber \\
g &=& \frac{3}{\pi} \frac{m_{\pi}^2}{p_{0}^2} \textrm{log} \left(\frac{m_{r} + 2p_{0}}{2m_{\pi}} \right) + \frac{m_{r}}{2\pi p_{0}} - \frac{m_{\pi}^2m_{r}}{\pi p_{0}^3}, \nonumber \\
{\rm and \ }  \Gamma_{\rho}(s) &=& \Gamma_{0} \left[ \frac{p}{p_{0}}  \right]^3 \left[ \frac{m_{r}^2}{s} \right]^{1/2}. 
\end{eqnarray}
The $\rho-\omega$ interference is taken into account by
\begin{equation}
R_{\rho-\omega}(s) = \textrm{GS}_{\rho(770)}(s) \times (1 + a e^{i\theta}  \textrm{RBW}_{\omega(782)}(s)),
\label{eq: RhoOmega}
\end{equation}
where $\Gamma_0$ is used, instead of the mass-dependent width $\Gamma^{(L)}(s)$, for $\omega(782)$~\cite{rhoomega}.

The $D^*(2010)^-$ contribution is vetoed as described in Sec.~\ref{sec:Selection}.
Possible remaining contributions from the $D^*(2010)^-$ RBW tail or general $\Dzb\pi^-$ P-waves are modelled as
\begin{equation}
R_{D^*(2010)}(\mdbpips) = e^{-(\beta_1 + i\beta_2)\mdbpips},\label{eq: D2010}
\end{equation}
where $\beta_1$ and $\beta_2$ are free parameters.

The $\pi^+\pi^-$ S-wave contribution is modelled  using two alternative approaches, the Isobar model~\cite{Isobar1, Isobar2, Isobar3} or the K-matrix model~\cite{KMatrixTheory}.  Contributions from the $f_0(500)$, $f_0(980)$, $f_0(2020)$ resonances and a nonresonant component are parametrised separately in the Isobar model and globally by one amplitude in the K-matrix model.

In the Isobar model, the $f_0(2020)$ resonance is modelled by a RBW function and the modelling of the $f_0(500)$, $f_0(980)$ resonances and the nonresonant contribution are described as follows.
The Bugg resonant lineshape~\cite{Bugg} is employed for the $f_0(500)$ contribution,
\begin{equation}\label{eq: Bugg}
R_{f_0(500)}(s) = 1/\left[m_r^2 - s - g_1^2\frac{s-s_A}{m_r^2 - s_A}z(s) -i m_r\Gamma_{\rm tot}(s)\right],
\end{equation}
where
\begin{eqnarray}
 m_r\Gamma_1(s) &=& g_1^2 \frac{s-s_A}{m_r^2 - s_A}\rho_1(s),  \nonumber \\
g_1^2(s) & =& m_r(b_1 + b_2s)\exp(-(s-m_r^2)/A), \nonumber \\
z(s) &=& j_1(s) - j_1(m_r^2), \nonumber \\
j_1(s) & =& \frac{1}{\pi} \left[2+ \rho_1 \textrm{log} \left( \frac{1-\rho_1}{1+\rho_1}\right) \right], \nonumber\\
m_r\Gamma_2(s) &=& 0.6g_1^2(s)(s/m_r^2)\exp(-\alpha|s-4m_K^2|)\rho_2(s), \nonumber \\
m_r\Gamma_3(s) &=& 0.2g_1^2(s)(s/m_r^2)\exp(-\alpha|s-4m_{\eta}^2|)\rho_3(s), \nonumber \\
m_r\Gamma_4(s) &=& m_rg_{4\pi}\rho_{4\pi}(s)/\rho_{4\pi}(m_r^2), \nonumber \\
 \rho_{4\pi}(s)  &=& 1/\left[1+\exp(7.082 - 2.845s)\right], \nonumber\\
{\rm and \ }\Gamma_{\rm tot}(s) & = &\sum_{i=1}^{4}\Gamma_i(s). 
\end{eqnarray}
The parameters are fixed to $m_r$ = $0.953$\gevcc, $s_A = 0.41 \ m_{\pi}^2$, $b_1 = 1.302$\gevgevcccc, $b_2 = 0.340$, $A = 2.426$\gevgevcccc and $g_{4\pi} = 0.011$\gevcc~\cite{Bugg}.
The phase-space factors of the decay channels $\pi\pi$, $KK$ and $\eta\eta$ correspond to $\rho_{1,2,3}(s)$, respectively and are defined as
\begin{equation}~\label{eq: phf}
\rho_{1,2,3}(s) = \sqrt{1 - 4\frac{m^2_{1,2,3}}{s}}, \qquad 1,2,  \textrm{ and }  3 = \pi, K \textrm{ and } \eta.
\end{equation}

The Flatt\'e formula~\cite{Flatte:1976xu} is used to describe the $f_0(980)$ lineshape,
  \begin{equation} \label{eq: flatt}
R_{f_0(980)}(s) = \frac{1}{m_r^2 - s - im_r(\rho_{\pi\pi}(s) g_1 + \rho_{KK}(s) g_2)},
\end{equation}
where
\begin{eqnarray}
\rho_{\pi\pi}(s) &=& \frac{2}{3}\sqrt{1-\frac{4m_{\pi^{\pm}}^2}{s}} + \frac{1}{3}\sqrt{1-\frac{4m_{\pi^0}^2}{s}}, \nonumber\\
{\rm and \ }\rho_{KK}(s) &=& \frac{1}{2}\sqrt{1-\frac{4m_{K^{\pm}}^2}{s}} + \frac{1}{2}\sqrt{1-\frac{4m_{K^0}^2}{s}}. 
\end{eqnarray}
The parameters $g_{1(2)}$ and $m_r$~\cite{LHCb-PAPER-2012-005} are $m_r = 939.9 \pm 6.3$\mevcc, $g_1 = 199 \pm 30$\mev and $g_2/g_1 = 3.0 \pm 0.3$.

The nonresonant contribution is described by
\begin{equation}
R_{NR}(\mpipis, \mdbpips) = e^{i\alpha \mpipis}. \label{eq: nonres}
\end{equation}
Its modulus equals unity, and a slowly varying phase over $m^2(\pi^+\pi^-)$ accounts for rescattering effects of the $\pi^+\pi^-$ final state and $\alpha$ is a free parameter of the model.

The  K-matrix formalism~\cite{KMatrixTheory}  describes the production, rescattering and decay of the $\pi^+\pi^-$ S-wave in a coherent way. The scattering matrix $S$, from an initial state to a final state, is
\begin{equation}
S = I +2i \left( {\rho^{\dag}} \right)^{1/2}T{\rho}^{1/2},
\end{equation}
where $I$ is the identity matrix, $\rho$ is a diagonal phase-space matrix and $T$ is the transition matrix.
The unitarity requirement $SS^{\dag} = I$ gives
\begin{equation}
(T^{-1} + i\rho)^{\dag} = T^{-1} + i\rho.
\end{equation}
The K-matrix is a Lorentz-invariant Hermitian matrix, defined as ${\rm K}^{-1} = T^{-1} + i\rho$.
The amplitude for a decay process,
\begin{equation}
A_i = (I - i{\rm K}\rho)_{ij}^{-1} P_j,
\end{equation}
is computed by combining the K-matrix obtained from scattering experiment with a production vector to describe process-dependent contributions.
The K-matrix is modelled as a five-pole structure,
\begin{equation}
{\rm K}_{ij}(s) = \left(  \sum_\alpha\frac{g_i^{\alpha} g_j^{\alpha} }{m_{\alpha}^2  - s}  + f_{ij}^{\rm scatt} \frac{ 1 -  s_0^{\textrm{scatt}}}{s - s_0^{\textrm{scatt}}}  \right) \frac{1 -s_{A0}}{s - s_{A0}}\left(s - \frac{s_A m_{\pi}^2}{2}\right),
\end{equation}
where the indexes $i, j = 1, 2, 3, 4, 5$ correspond to five decay channels: $\pi\pi$, $K\Kbar$, $\eta\eta$, $\eta\eta'$ and multi-meson (mainly 4$\pi$ states) respectively.
The coupling constant of the bare state $\alpha$ to the decay channel $i$, $g_i^{\alpha}$, is obtained from a global fit of scattering data and is listed in Table~\ref{tab:KMatrixPara}. The mass $m_{\alpha}$ is the bare pole mass and is in general different from the resonant mass of the RBW function. The parameters $f_{ij}^{\textrm{scatt}}$ and $s_0^{\textrm{scatt}}$ are used to describe smooth scattering processes. The last factor of the K-matrix, $ \frac{1 -s_{A0}}{s - s_{A0}}\left(s - \frac{s_A m_{\pi}^2}{2}\right)$, regulates the singularities near the $\pi^+\pi^-$ threshold, the so-called ``Adler zero"~\cite{Adler:1964um,Adler:1965ga}. The Hermitian property of the K-matrix imposes the relation $f_{ij}^{\textrm{scatt}} = f_{ji}^{\textrm{scatt}}$, and since only $\pi^+\pi^-$ decays are considered, if $i \ne 1$ and $ j \ne 1$, $f_{ij}^{\textrm{scatt}} $ is set to 0.
The production vector is modelled with
\begin{equation}
P_j = \left[ f_{1j}^{\textrm{prod}} \frac{1 - s_0^{\textrm{prod}}}{s - s_0^{\textrm{prod}}} + \sum_{\alpha} \frac{\beta_{\alpha}g_j^{\alpha}}{m_{\alpha}^2 -s} \right],\label{eq:kmatrix}
\end{equation}
where $f_{1j}^{\textrm{prod}}$ and $\beta_{\alpha}$ are free parameters. The singularities in the K-matrix and the production vector cancel when calculating the amplitude matrix element.

\begin{table}[!tbh]
\centering
\caption{The K-matrix parameters used in this paper are taken from a global analysis of $\pi^+\pi^-$ scattering data~\cite{BaBarDpipi}. Masses and coupling constants are in units of \gevcc.}
\label{tab:KMatrixPara}
\begin{tabular}{rrrrrr}
\hline
$m_{\alpha}$ & $g_{\pi^+\pi^-}^{\alpha}$ & $g_{K\Kbar}^{\alpha}$ & $g_{4\pi}^{\alpha}$ & $g_{\eta\eta}^{\alpha}$ & $g_{\eta\eta'}^{\alpha}$ \\
0.65100 & 0.22889 & $-0.55377$ & 0.00000 & $-0.39899$ & $-0.34639$ \\
1.20360 & 0.94128 & 0.55095 & 0.00000 & 0.39065 & 0.31503 \\
1.55817 & 0.36856 & 0.23888 & 0.55639 & 0.18340 & 0.18681 \\
1.21000 & 0.33650 & 0.40907 & 0.85679 & 0.19906 & $-0.00984$ \\
1.82206 & 0.18171 & $-0.17558$ & $-0.79658$ & $-0.00355$ & 0.22358 \\
\hline
& $f_{11}^{\textrm{scatt}}$  &$f_{12}^{\textrm{scatt}}$ & $f_{13}^{\textrm{scatt}}$ & $f_{14}^{\textrm{scatt}}$  & $f_{15}^{\textrm{scatt}}$  \\
& 0.23399 & 0.15044 & $-0.20545$ & 0.32825 & 0.35412\\
& $s_{0}^{\textrm{scatt}}$  & $s_{0}^{\textrm{prod}}$ & $s_{A0}$ & $s_{A}$ & \\
& $-3.92637$ & $-3.0$ & $-0.15$ & 1 & \\
\hline
\end{tabular}
\end{table}

\section{Dalitz plot fit}
\label{sec:DalitzFit}

An unbinned extended maximum likelihood fit is performed to the Dalitz plot distribution.
The likelihood function is defined by
\begin{equation}\label{eq: Constrain}
\mathcal{L} = \mathcal{L}' \times \frac{1}{\sqrt{2\pi}\sigma_s} \exp\left(-\frac{(\nu_s - \nu^0_s)^2}{2\sigma^2_s}\right) \times \frac{1}{\sqrt{2\pi}\sigma_{b}} \exp\left(-\frac{(\nu_{b} - \nu^0_{b})^2}{2\sigma^2_{b}}\right),
\end{equation}
where
\begin{equation}
\mathcal{L}' = \frac{e^{-(\nu_s + \nu_{b})} (\nu_s + \nu_{b})^N}{N!} \prod_{i=1}^N \left[ \frac{\nu_s}{\nu_s + \nu_{b}} f_s(\vec{x}_i ; \theta_s) + \frac{\nu_{b}}{\nu_s + \nu_{b}} f_{b}(\vec{x}_i; \theta_b) \right].
\label{eq: likelihood1}
\end{equation}
The background probability density function (PDF) is given by $f_{b}(\vec{x}; \theta_b)$ and is described in Sec.~\ref{sec:DalitzFitBgdModel}. The signal PDF, $f_s(\vec{x}_i ; \theta_s)$, is described by 
\begin{equation}
\frac{M(\vec{x}_i ; \theta_s)  \varepsilon(\vec{x}_i)}{\int M(\vec{x} ; \theta_s) \varepsilon(\vec{x}) d\vec{x}},
\end{equation}
where the decay amplitude, $M(\vec{x} ; \theta_s)$, is described in Sec.~\ref{sec:DalitzFormulism} and the efficiency variation over the Dalitz plot, $\varepsilon(\vec{x})$, is described in Sec.~\ref{sec:DalitzFitEffModel}. The fit parameters, $\theta_{s}$ and $\theta_{b}$, include complex coefficients and resonant parameters like masses and widths. The value $N$ is the total number of reconstructed candidates in the signal region. The number of signal and background events, $\nu_s$ and $\nu_{b}$, are floated and constrained by the yields, $\nu_s^0$ and $\nu_b^0$, determined by the $\Dzb\pi^+\pi^-$ mass fit and shown with their statistical uncertainties in Table~\ref{tab: MassFit}.

\subsection{Background modelling}
\label{sec:DalitzFitBgdModel}
\begin{figure}[!tbh]
  \centering
  \includegraphics[height=6cm]{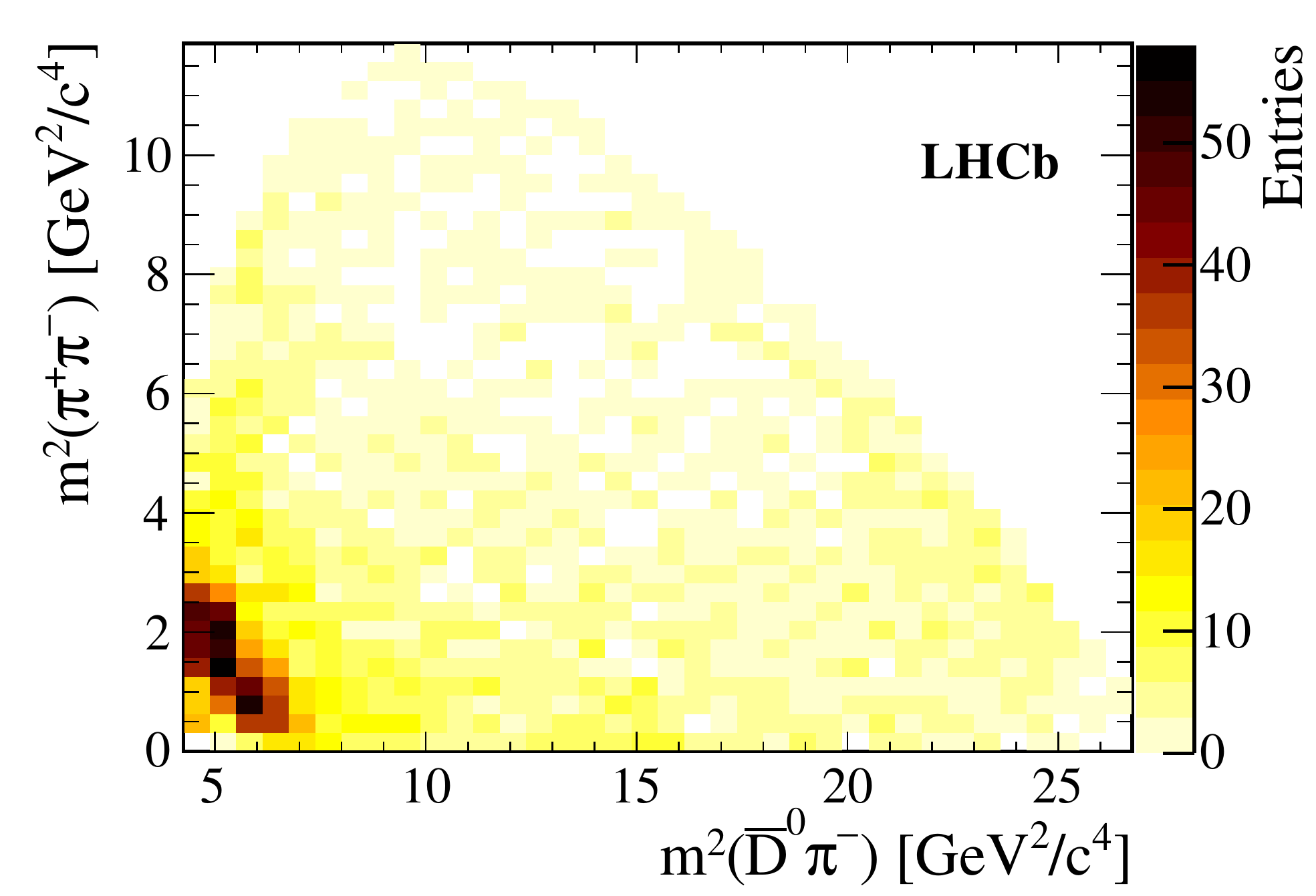}
\caption{Density profile of the combinatorial background events in the Dalitz plane obtained from the upper $m(\Dzb\pi^+\pi^-)$ sideband with a looser selection applied on the Fisher discriminant.}
\label{fig: Background}
\end{figure}

The only significant source of candidates in the signal region, other than $\btodpipi$ decays, is from combinatorial background. It is modelled using candidates in the upper $m(\Dzb\pi^+\pi^-)$ sideband ([5350, 5450] $\mevcc$) with a looser requirement on the Fisher discriminant, and is shown in Fig.~\ref{fig: Background}. The looser requirement gives a similar distribution in the Dalitz plane but with lower statistical fluctuations. The Dalitz plot distribution of the combinatorial background events lying in the upper-mass sideband is considered to provide a reliable description of that in the signal region, as no dependence on $m(\Dzb\pi^+\pi^-)$ is found by studying the Dalitz distribution in a different upper-mass sideband region. The combinatorial background is modelled with an interpolated non-parametric PDF~\cite{RooFit_1,RooFit_2}  using an adaptive kernel-estimation algorithm~\cite{Kernel}.

\subsection{Efficiency modelling}
\label{sec:DalitzFitEffModel}
The efficiency function $\varepsilon(\vec{x})$ accounts for effects of reconstruction, triggering and selection of the $\btodpipi$ signal events, and varies across the Dalitz plane.
Two  simulated samples are generated to describe its variation with several data-driven corrections. One is uniformly distributed over the phase space of the Dalitz plot and the other is uniformly distributed over the square Dalitz plot, which models efficiencies more precisely at the kinematic boundaries. The square Dalitz plot is parametrised by two variables $m'$ and $\theta'$ that each varies between 0 and 1 and are defined as
\begin{equation}
m' = \frac{1}{\pi} \arccos\left(2\frac{m(\pi^+\pi^-) - m(\pi^+\pi^-)_{\textrm{min}}}{m(\pi^+\pi^-)_{\textrm{max}} - m(\pi^+\pi^-)_{\textrm{min}}} - 1 \right) \qquad \textrm{and} \qquad \theta' = \frac{1}{\pi}\theta(\pi^+\pi^-),
\end{equation}
where $m(\pi^+\pi^-)_{\textrm{max}} = m_{B^0} - m_{D^0}$, $m(\pi^+\pi^-)_{\textrm{min}} = 2m_{\pi}$ and $\theta(\pi^+\pi^-)$ is the helicity angle of the $\pi^+\pi^-$ system.

\begin{figure}[!tbh]
  \centering
  \includegraphics[height=6cm]{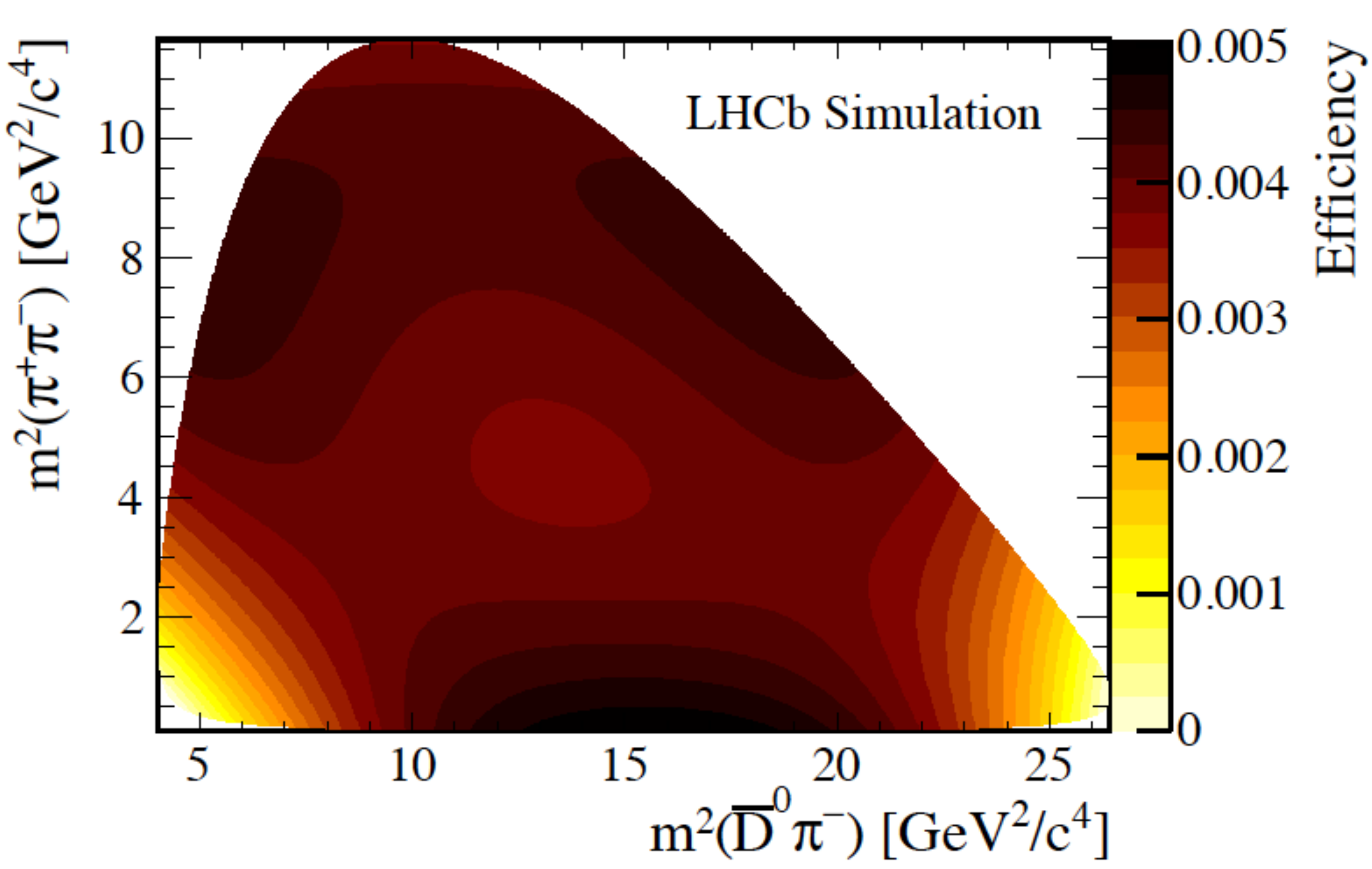}
\caption{Efficiency function for the Dalitz variables obtained in a fit to the LHCb simulated samples.}
\label{fig: Efficiency}
\end{figure}

The two samples are fitted simultaneously with common fit parameters. A 4th-order polynomial function is used to describe the efficiency variation over the Dalitz plot.
As the efficiency in the simulation is approximately symmetric over $\mdbpips$ and $\mdbpims$, the polynomial function is defined as
\begin{eqnarray}
\varepsilon(x,y) &\propto & 1.0 + a_0 (x+y) + a_1 (x+y)^2 + a_2 (xy) + a_3 (x+y)^3   \nonumber \\
	&& + a_4 (x+y)xy + a_5 (x+y)^4 + a_6 (x+y)^2xy + a_7 x^2y^2, 
	\end{eqnarray}
where
	\begin{eqnarray}
 x = \frac{\mdbpips - m^2_0}{(m_{B^0}-m_{\pi})^2 - m^2_0}  \quad \textrm{   and   }\quad  y = \frac{\mdbpims - m^2_0}{(m_{B^0}-m_{\pi})^2 - m^2_0}, 
\end{eqnarray}
with $m^2_0$ defined as $[(m_{D^0} + m_{\pi})^2 + (m_{B^0} - m_{\pi})^2]/2$.
The fitted efficiency distribution over the Dalitz plane is shown in Fig.~\ref{fig: Efficiency}.

The efficiency is corrected using dedicated control samples with data-driven methods.
The corrections applied to the simulated samples include known differences between simulation and data that originate from the trigger, PID and tracking. 
\subsection{Results of the Dalitz plot fit}
\label{sec:resonances}
\begin{figure}[!tbh]
\centerline{\includegraphics[width=8.5cm]{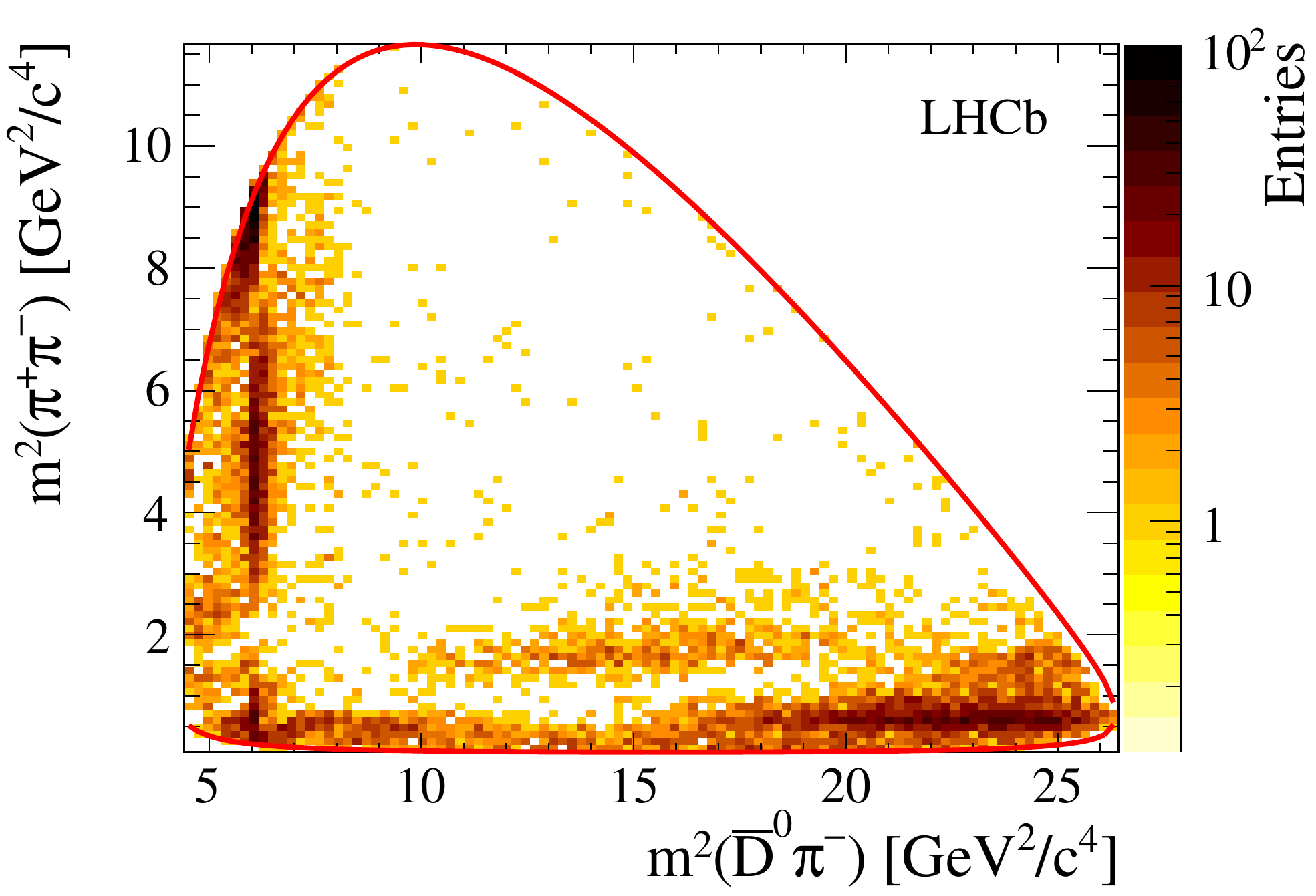}}
\caption{Dalitz plot distribution of candidates in the signal region, including background contributions.
The red line shows the Dalitz plot kinematic boundary.}
\label{fig: Binning}
\end{figure}

\begin{figure}
\begin{minipage}{0.5 \linewidth}
\centerline{\includegraphics[width=1.0\linewidth]{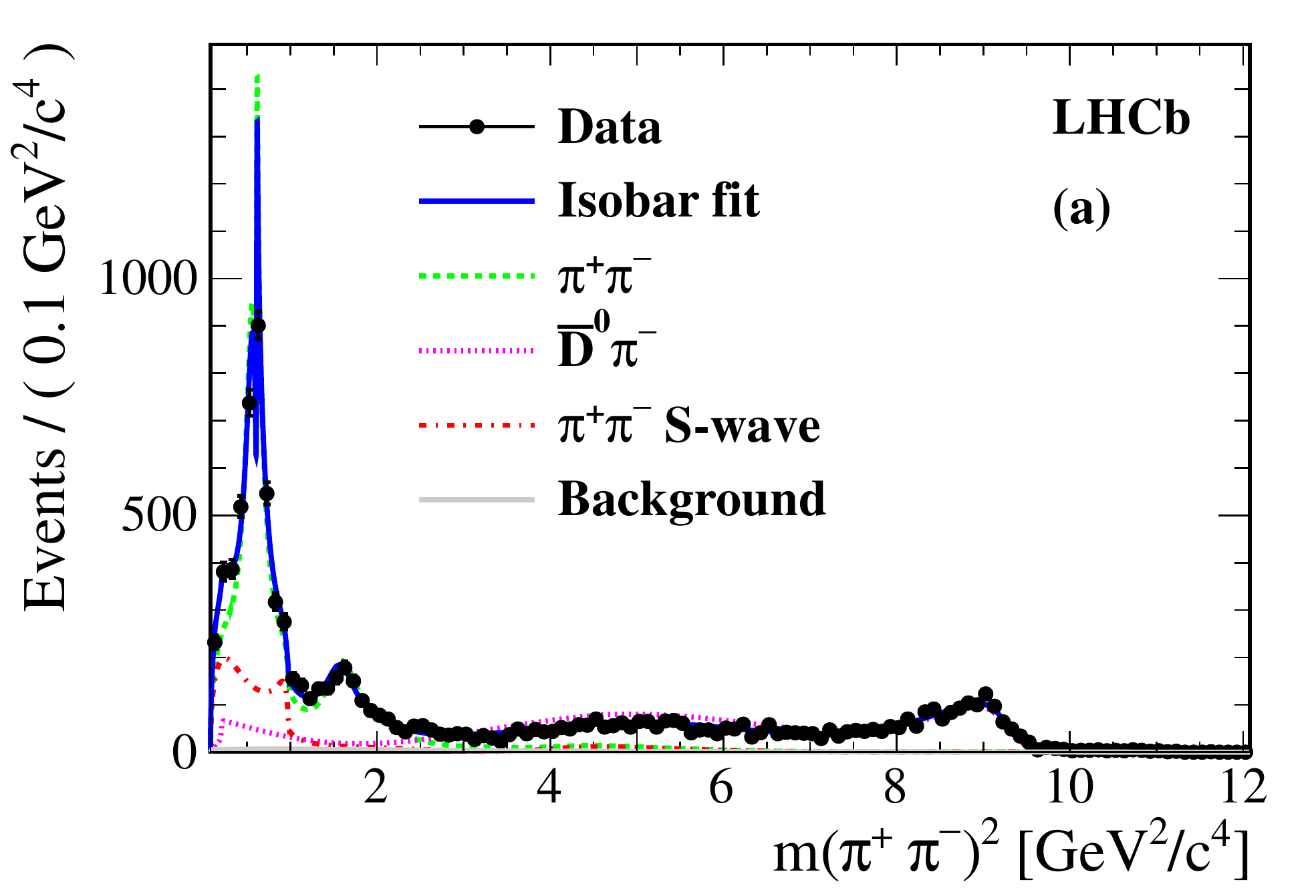}}
\end{minipage}
\begin{minipage}{0.5\linewidth}
\centerline{\includegraphics[width=1.0\linewidth]{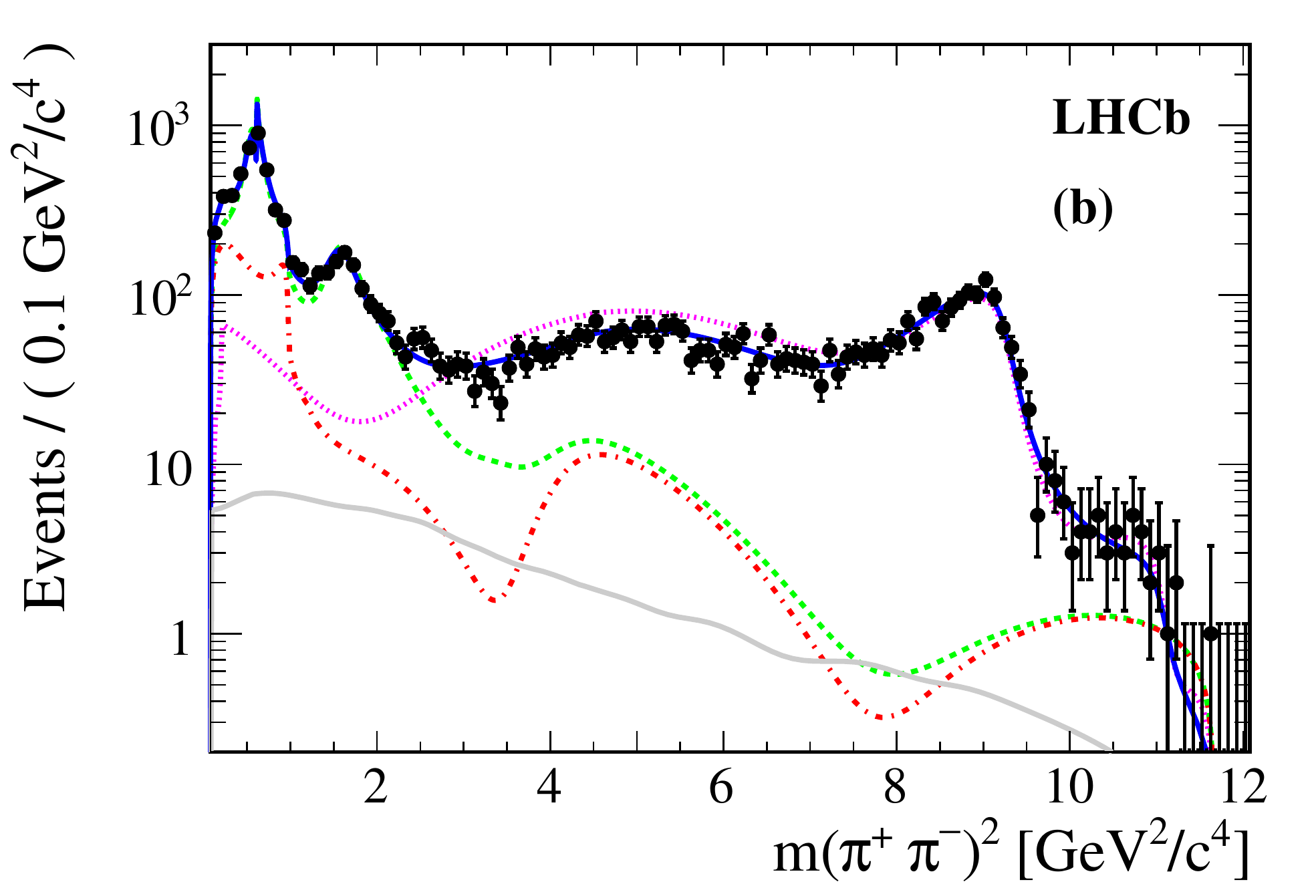}}
\end{minipage}
\begin{minipage}{0.5 \linewidth}
\centerline{\includegraphics[width=1.0\linewidth]{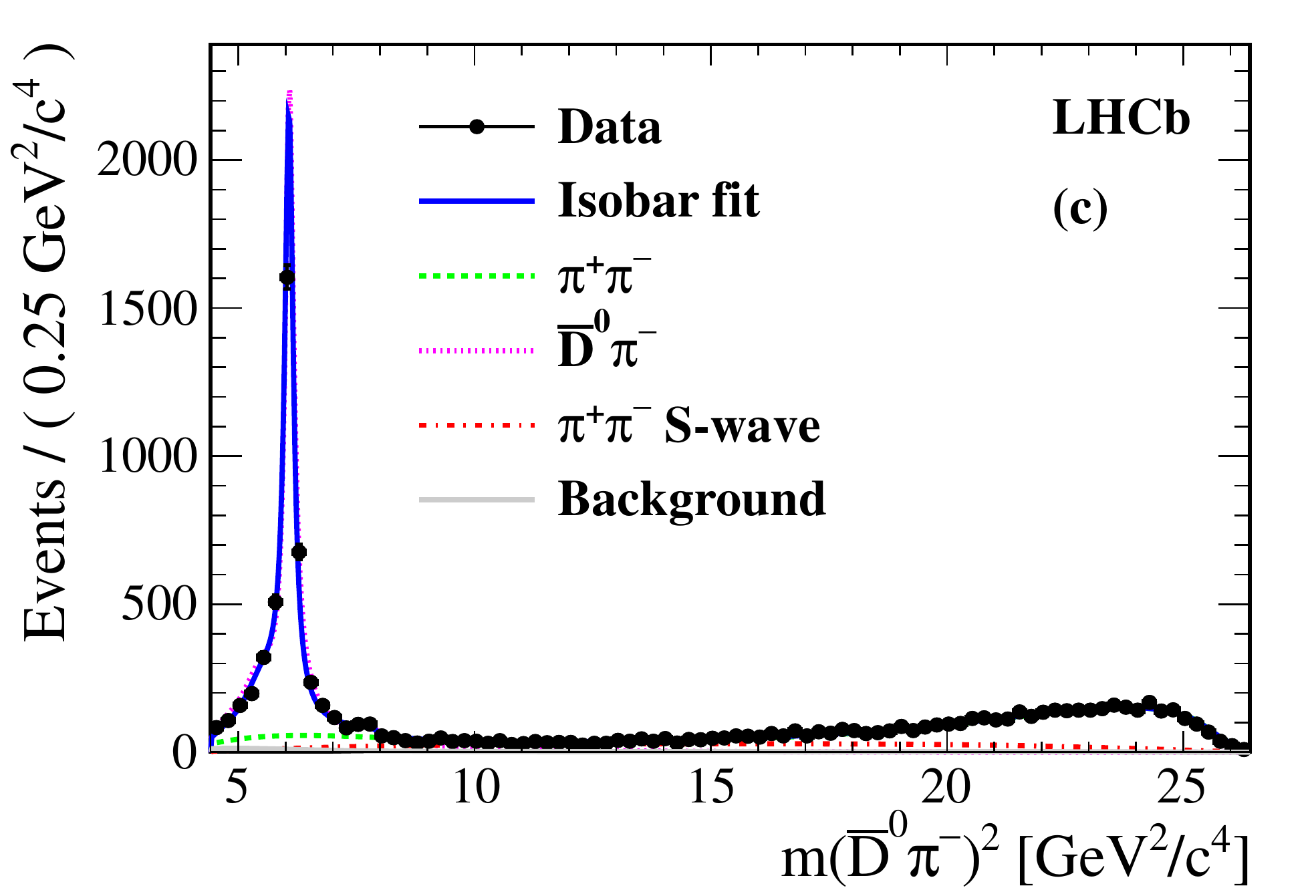}}
\end{minipage}
\begin{minipage}{0.5\linewidth}
\centerline{\includegraphics[width=1.0\linewidth]{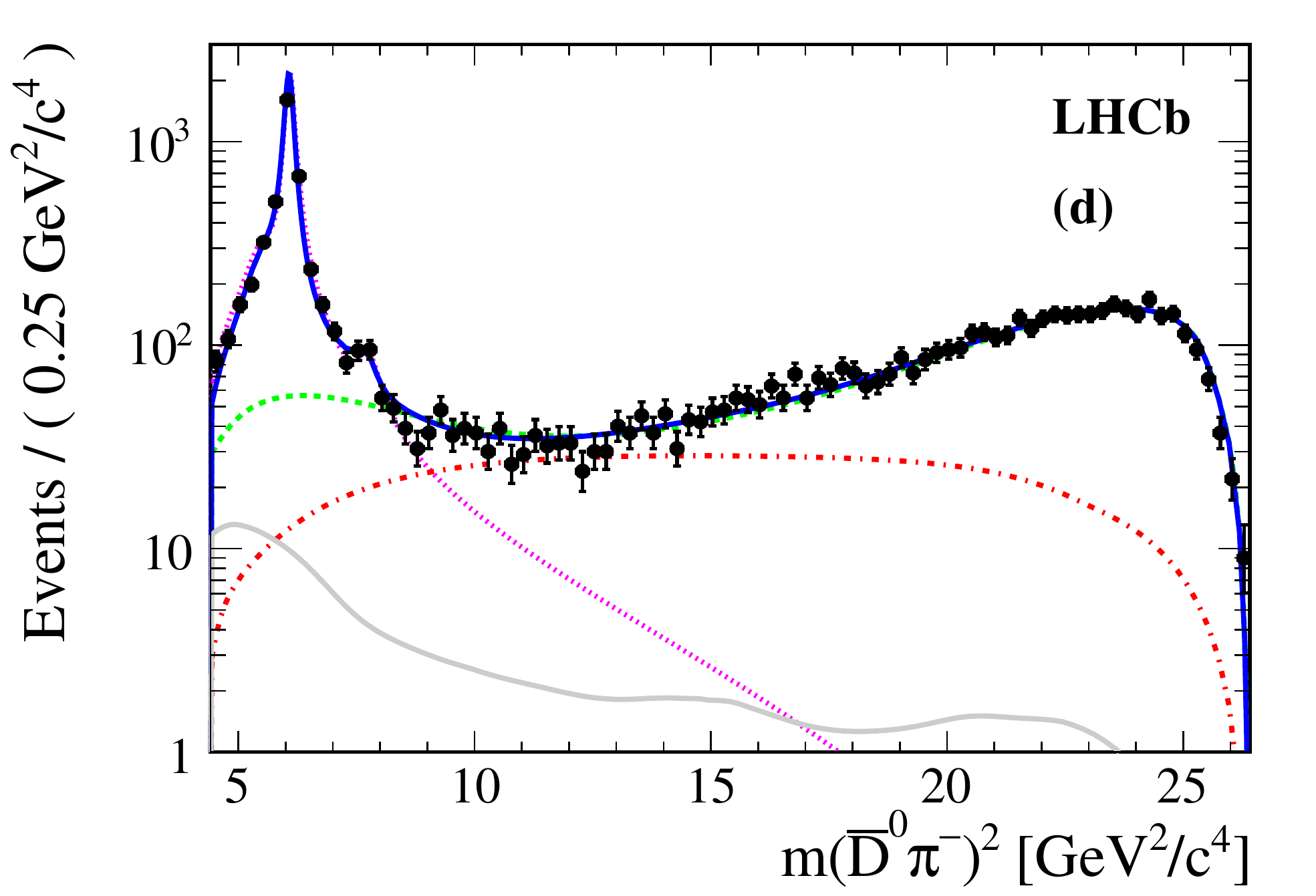}}
\end{minipage}
\caption{Projections of the data and Isobar fit onto (a) $m^2(\pi^+\pi^-)$ and (c) $m^2(\Dzb\pi^-)$ with a linear scale.
Same projections shown in (b) and (d) with a logarithmic scale. Components are described in the legend.
The lines denoted $\Dzb\pi^-$ and $\pi^+\pi^-$ include the coherent sums of all $\Dzb\pi^-$ resonances, $\pi^+\pi^-$ resonances, and $\pi^+\pi^-$ S-wave resonances.
The various contributions do not add linearly due to interference effects.
}
\label{fig: fitIso}
\end{figure}

\begin{figure}
\begin{minipage}{0.5 \linewidth}
\centerline{\includegraphics[width=1.0\linewidth]{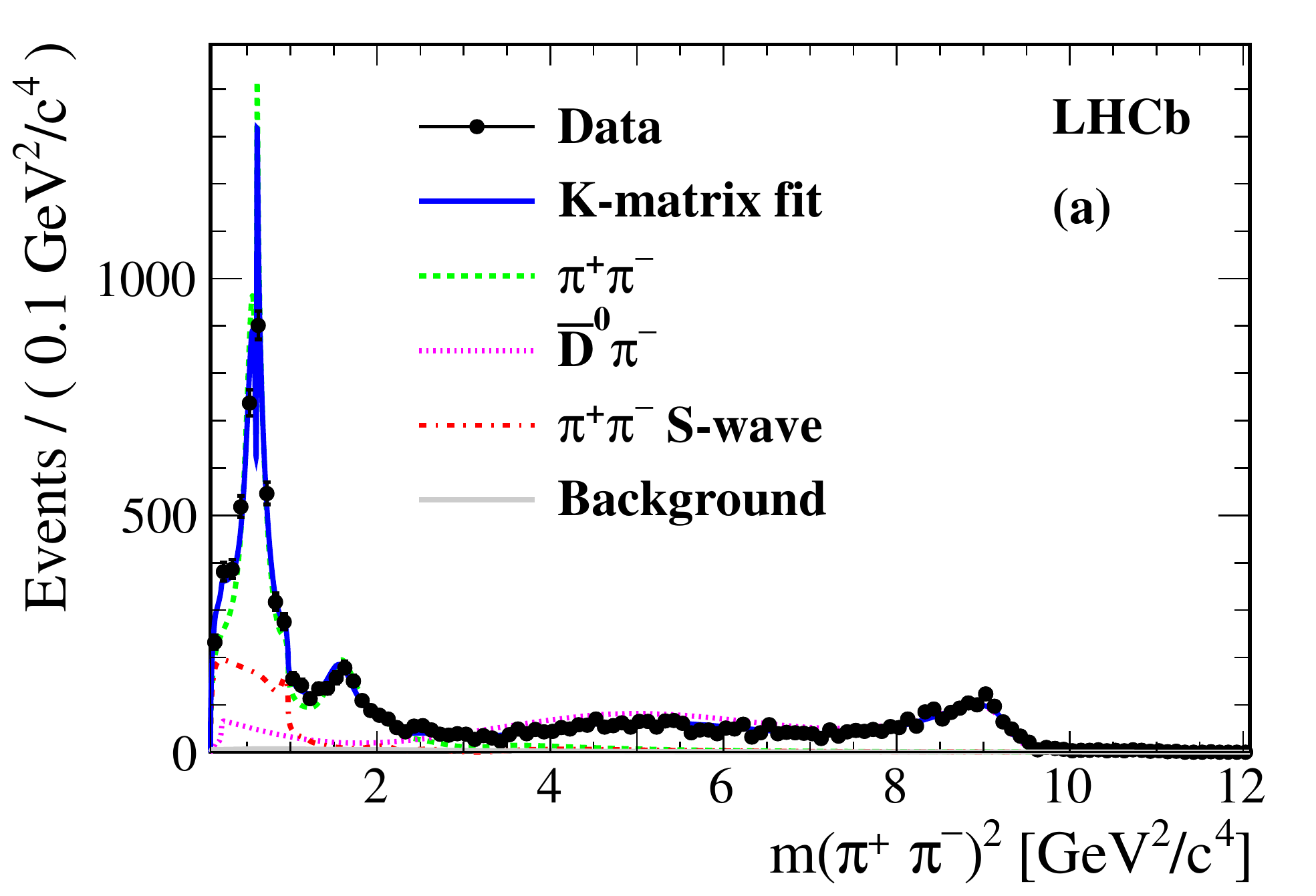}}
\end{minipage}
\begin{minipage}{0.5\linewidth}
\centerline{\includegraphics[width=1.0\linewidth]{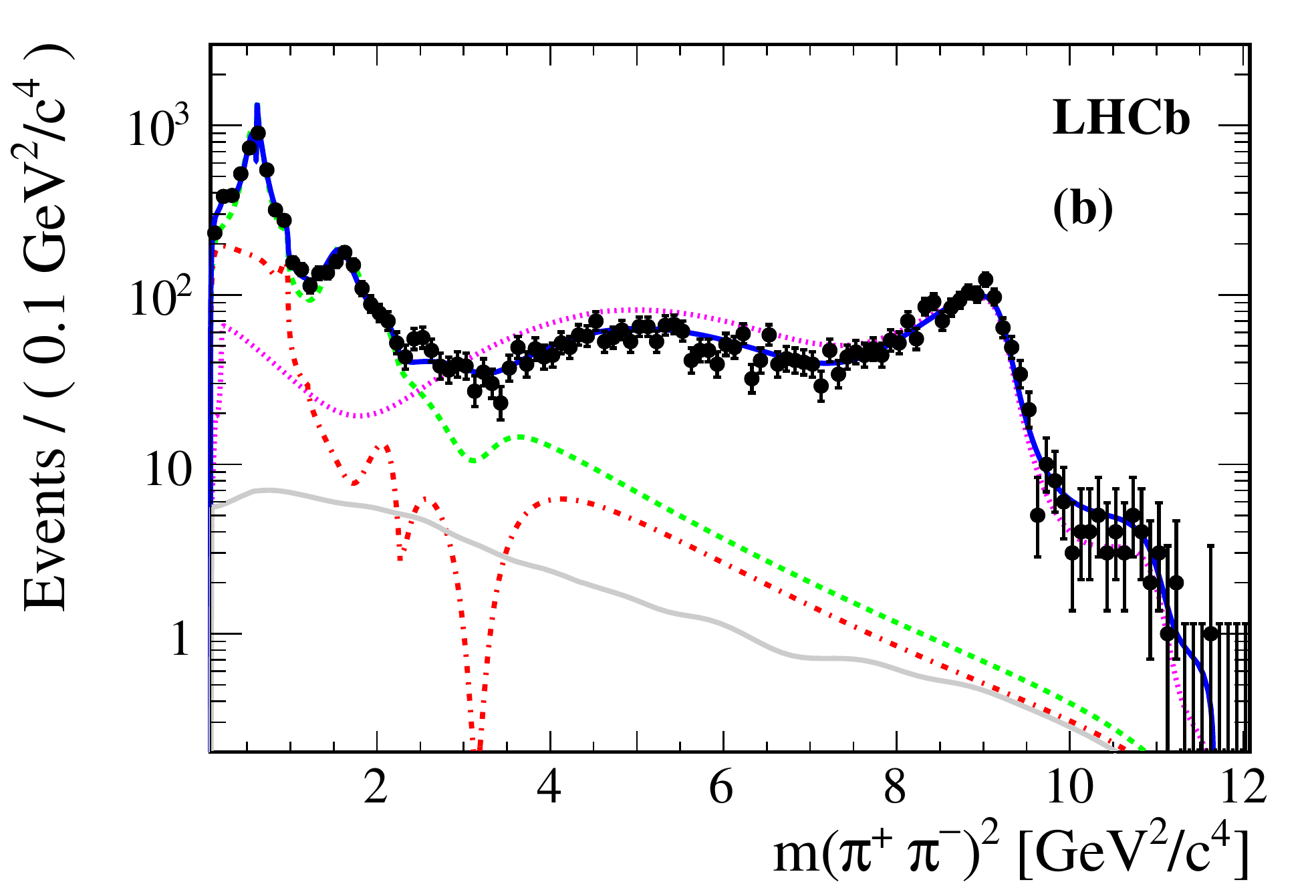}}
\end{minipage}
\begin{minipage}{0.5 \linewidth}
\centerline{\includegraphics[width=1.0\linewidth]{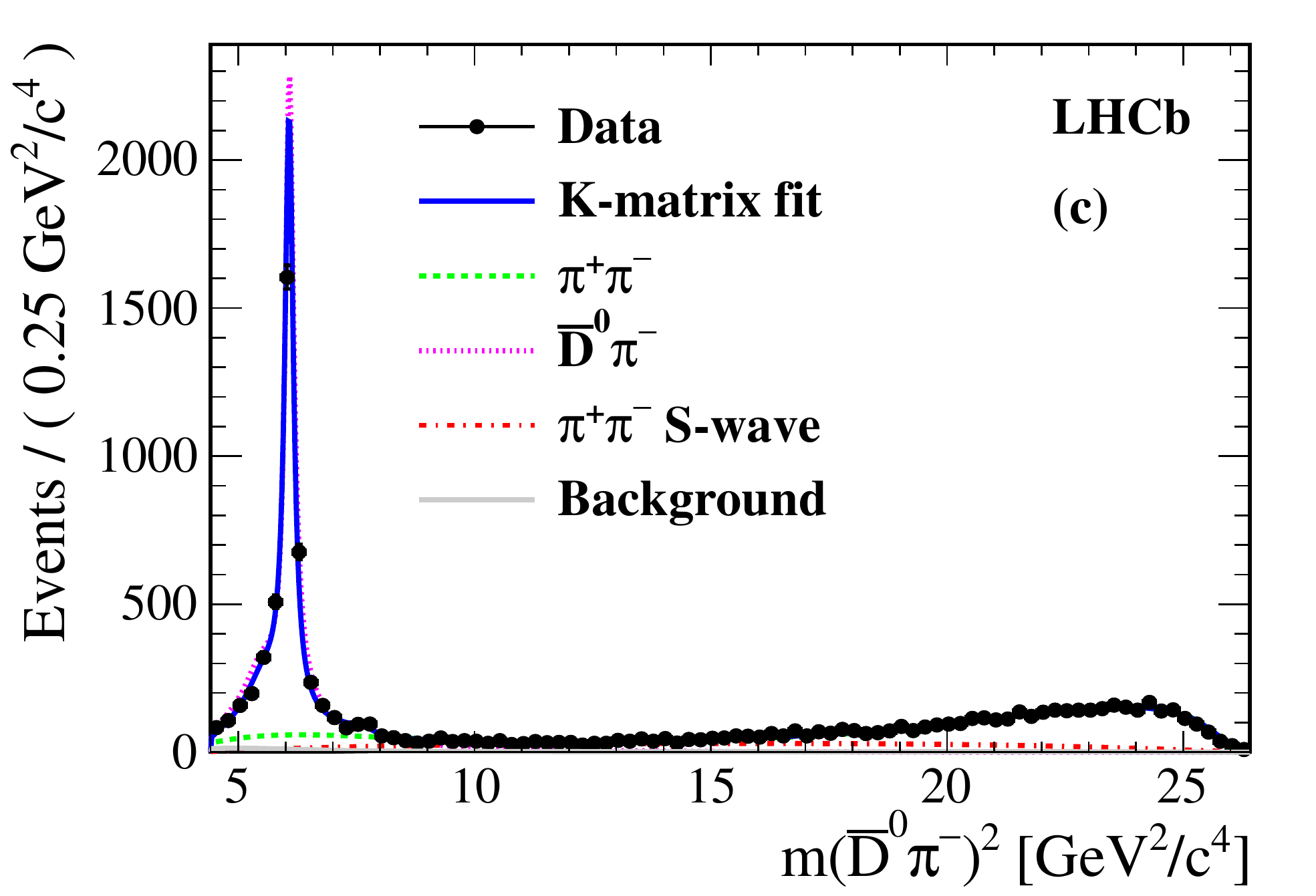}}
\end{minipage}
\begin{minipage}{0.5\linewidth}
\centerline{\includegraphics[width=1.0\linewidth]{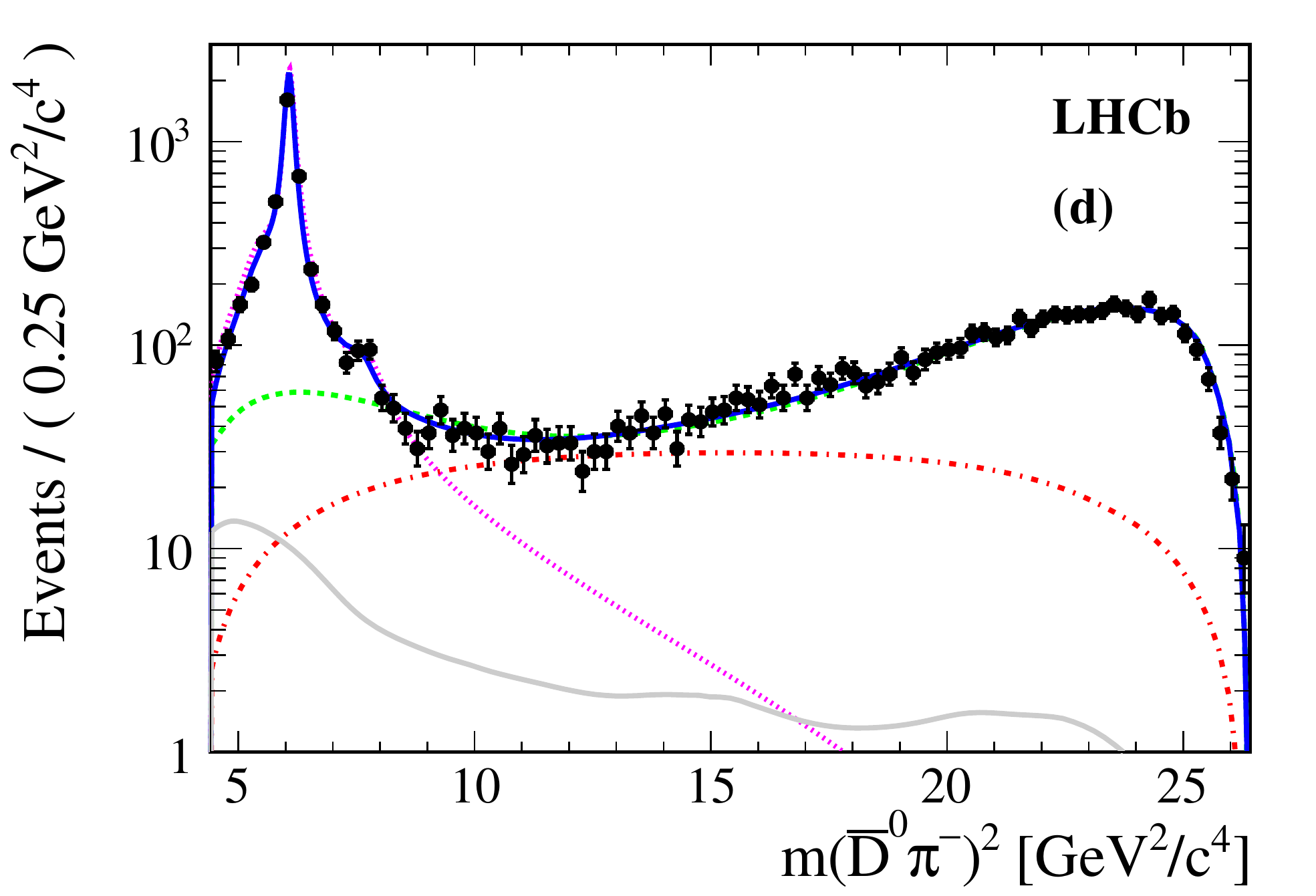}}
\end{minipage}
\caption{Projections of the data and K-matrix fit onto (a) $m^2(\pi^+\pi^-)$ and (c) $m^2(\Dzb\pi^-)$ with a linear scale.
Same projections shown in (b) and (d) with a logarithmic scale. Components are described in the legend.
The lines denoted $\Dzb\pi^-$ and $\pi^+\pi^-$ include the coherent sums of all $\Dzb\pi^-$ resonances, $\pi^+\pi^-$ resonances, and $\pi^+\pi^-$ S-wave resonances.
The various contributions do not add linearly due to interference effects.
}
\label{fig: fitkma}
\end{figure}

The Dalitz plot distribution from data in the signal region is shown in Fig.~\ref{fig: Binning}. The analysis is performed using the Isobar model and the K-matrix model. The nominal fit model in each case is defined by considering many possible resonances and removing those that do not significantly contribute to the Dalitz plot analysis. The resulting resonant contributions are given in Table~\ref{tab: Resonances} while the projections of the fit results are shown in Fig.~\ref{fig: fitIso} (Fig.~\ref{fig: fitkma}) for the Isobar (K-matrix) model.

\begin{table}[!tbh]
\centering
\caption{Resonant contributions to the nominal fit models and their properties. Parameters and uncertainties of $\rho(770)$, $\omega(782)$, $\rho(1450)$ and $\rho(1700)$ come from Ref.~\cite{Babarrhoomega}, and  those of $f_2(1270)$ and $f_0(2020)$ come from Ref.~\cite{PDG}. Parameters of $f_0(500)$, $f_0(980)$ and K-matrix formalism are described in Sec.~\ref{sec:DalitzFormulism}. }
\label{tab: Resonances}
\begin{tabular}{lcccc}
\hline
Resonance & Spin  & Model & $m_r$ (\mevcc) &  $\Gamma_0$ (\mev)  \\
\hline
$\Dzb \pi^-$ P-wave & 1  & Eq.~\ref{eq: D2010} & \multicolumn{2}{c}{Floated} \\
$D_0^*(2400)^-$ & 0  & RBW & \multicolumn{2}{c}{Floated} \\
$D_2^*(2460)^-$ & 2  & RBW & \multicolumn{2}{c}{Floated} \\
$D_J^*(2760)^-$ & 3  & RBW & \multicolumn{2}{c}{Floated} \\
\hline
$\rho(770)$ & 1  & GS & $775.02 \pm 0.35$ & $149.59 \pm 0.67$\\
$\omega(782)$ & 1 &  Eq.~\ref{eq: RhoOmega} & $781.91 \pm 0.24$ & $\al\al8.13 \pm 0.45$\\
$\rho(1450)$ & 1 & GS & $\al\all1493 \pm \al\all15$ & $\al\al\all427\pm \al\all31$ \\
$\rho(1700)$ & 1 & GS & $\al\all1861 \pm \al\all17$ & $\al\al\all316 \pm \al\all26$ \\
$f_2(1270)$ & 2 & RBW & $1275.1 \pm \al1.2$ & $\al185.1^{\all+\al\al2.9}_{\all-\al\al2.4}$ \\
\hline
$\pi\pi$ S-wave & 0 & K-matrix & \multicolumn{2}{c}{See Sec.~\ref{sec:DalitzFormulism}} \\
\hline
$f_0(500)$ & 0 & Eq.~\ref{eq: Bugg} & \multicolumn{2}{c}{See Sec.~\ref{sec:DalitzFormulism}} \\
$f_0(980)$ & 0 & Eq.~\ref{eq: flatt} & \multicolumn{2}{c}{See Sec.~\ref{sec:DalitzFormulism}} \\
$f_0(2020)$ & 0 & RBW & $\al\all1992 \pm \al\all16$ & $\al\al\all442 \pm \al\all60$\\
Nonresonant & 0 & Eq.~\ref{eq: nonres} & \multicolumn{2}{c}{See Sec.~\ref{sec:DalitzFormulism}} \\
\hline
\end{tabular}
\end{table}

\begin{figure}[!tbh]
\begin{minipage}{0.5 \linewidth}
\centerline{\includegraphics[width=1.0\linewidth]{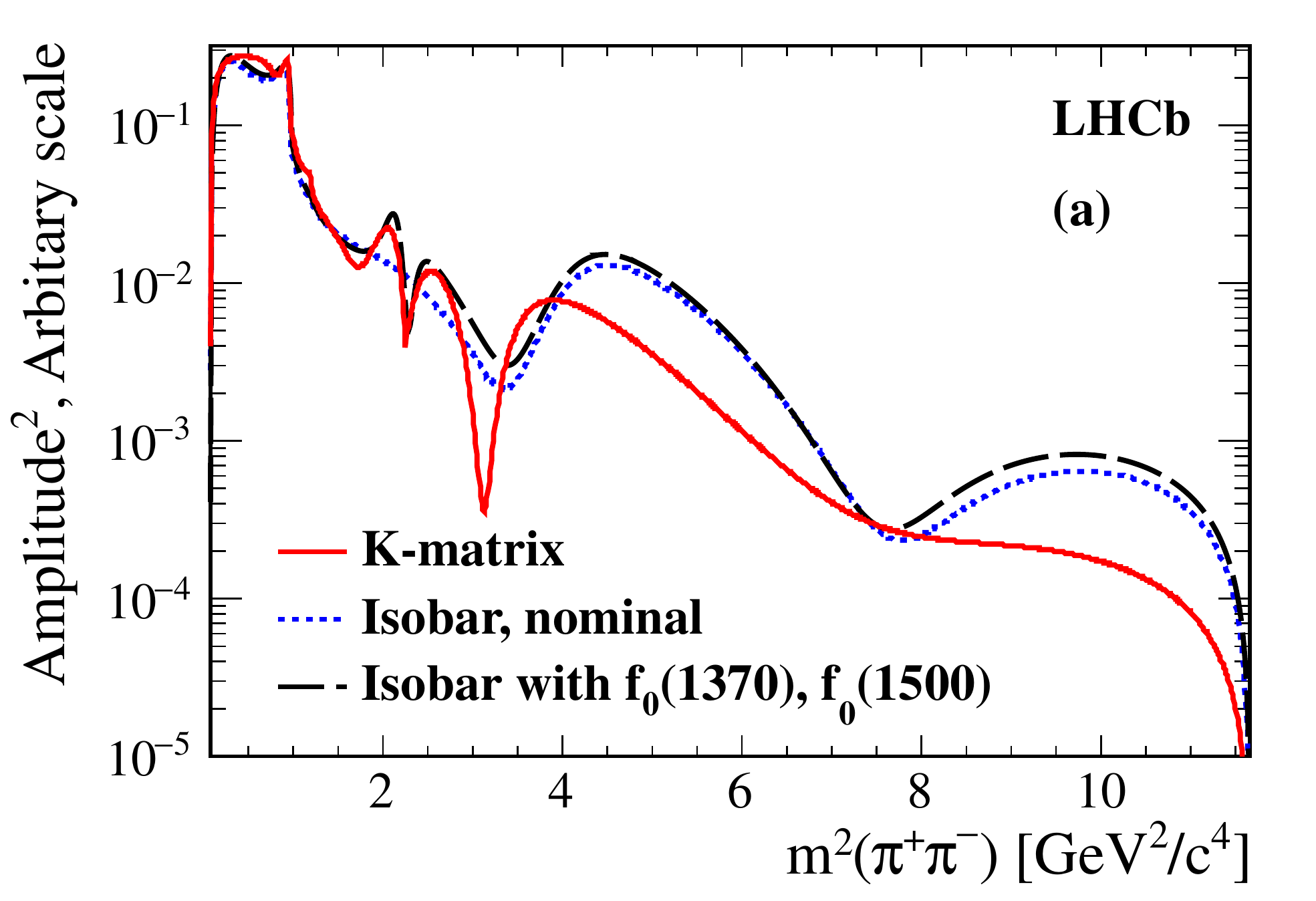}}
\end{minipage}
\begin{minipage}{0.5\linewidth}
\centerline{\includegraphics[width=1.0\linewidth]{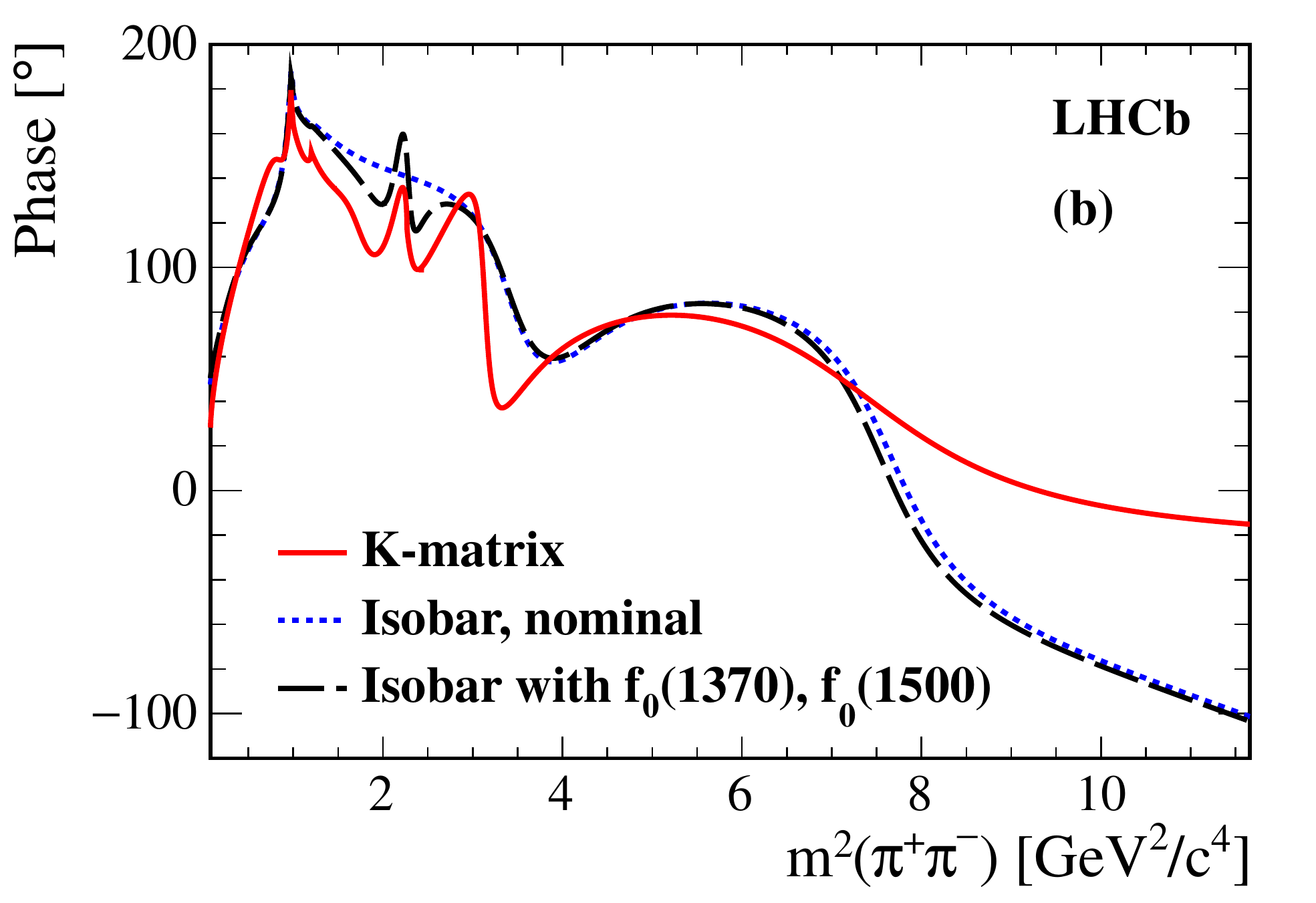}}
\end{minipage}
\caption{
Comparison of the $\pi^+\pi^-$ S-wave obtained from the Isobar model and the K-matrix model, for (a) amplitudes and (b) phases. 
The K-matrix model is shown by the red solid line, two scenarios for the Isobar model with (black long dashed line) and without (blue dashed line) $f_0(1370)$ and $f_0(1500)$  are shown.}
\label{fig: IsoVsKMa}
\end{figure}

The comparisons of the S-wave results for the Isobar model and the K-matrix model are shown in Fig.~\ref{fig: IsoVsKMa}. The results from the two models agree reasonably well for the amplitudes and phases. In the $\pi^+\pi^-$ mass-squared region of $[1.5, 4.0] \ {\rm GeV}^2/c^4$, small structures are seen in the K-matrix model, indicating possible contributions from $f_0(1370)$ and $f_0(1500)$ states. These contributions are not significant in the Isobar model and are thus not included in the nominal fit: adding them results in marginal changes and shows similar qualitative behaviour to the K-matrix model as displayed on Fig.~\ref{fig: IsoVsKMa}. The measured S-waves from both models qualitatively agree with predictions given in Ref.~\cite{rescattering1}.

\begin{figure}[!tbh]
\begin{minipage}{0.5 \linewidth}
\centerline{\includegraphics[width=1.0\linewidth]{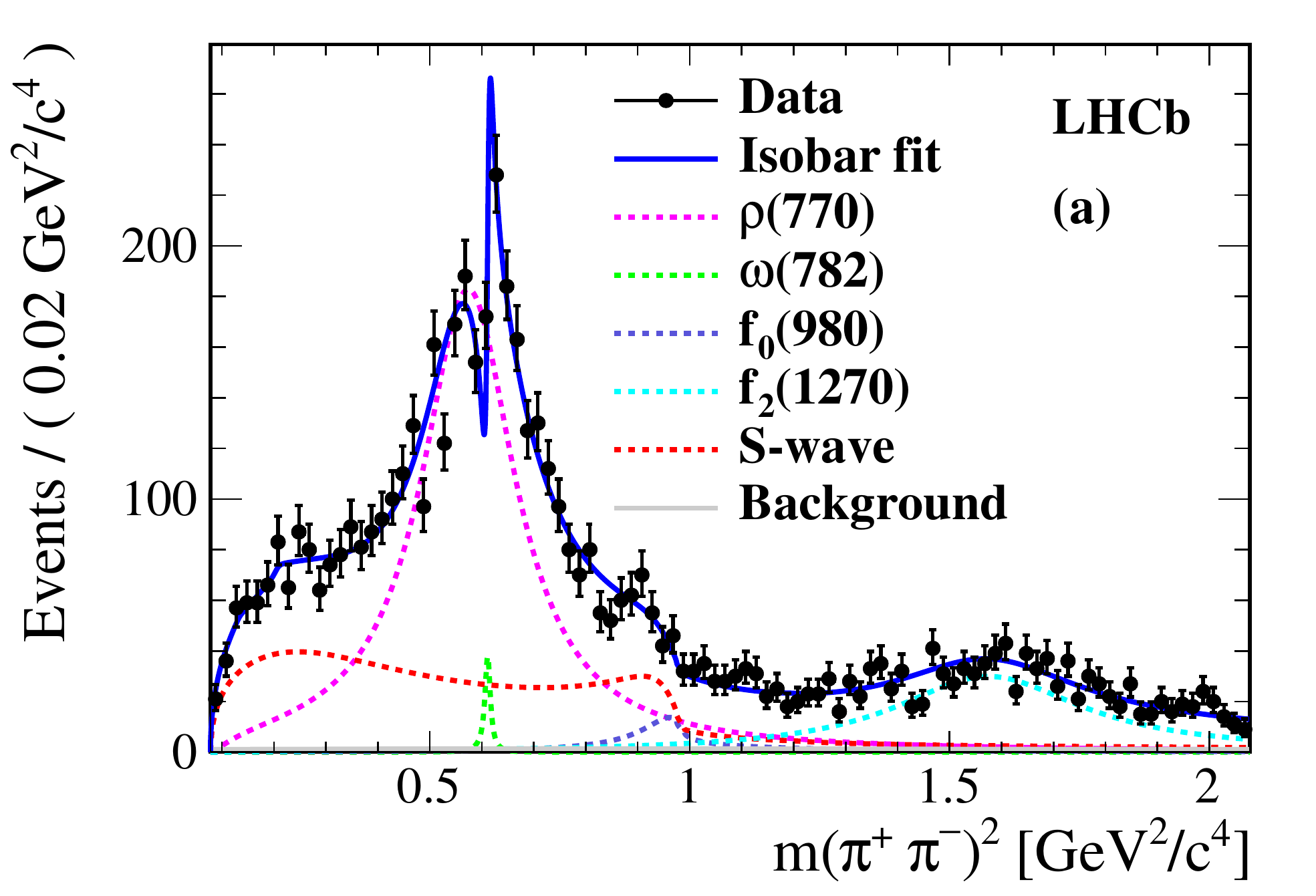}}
\end{minipage}
\begin{minipage}{0.5\linewidth}
\centerline{\includegraphics[width=1.0\linewidth]{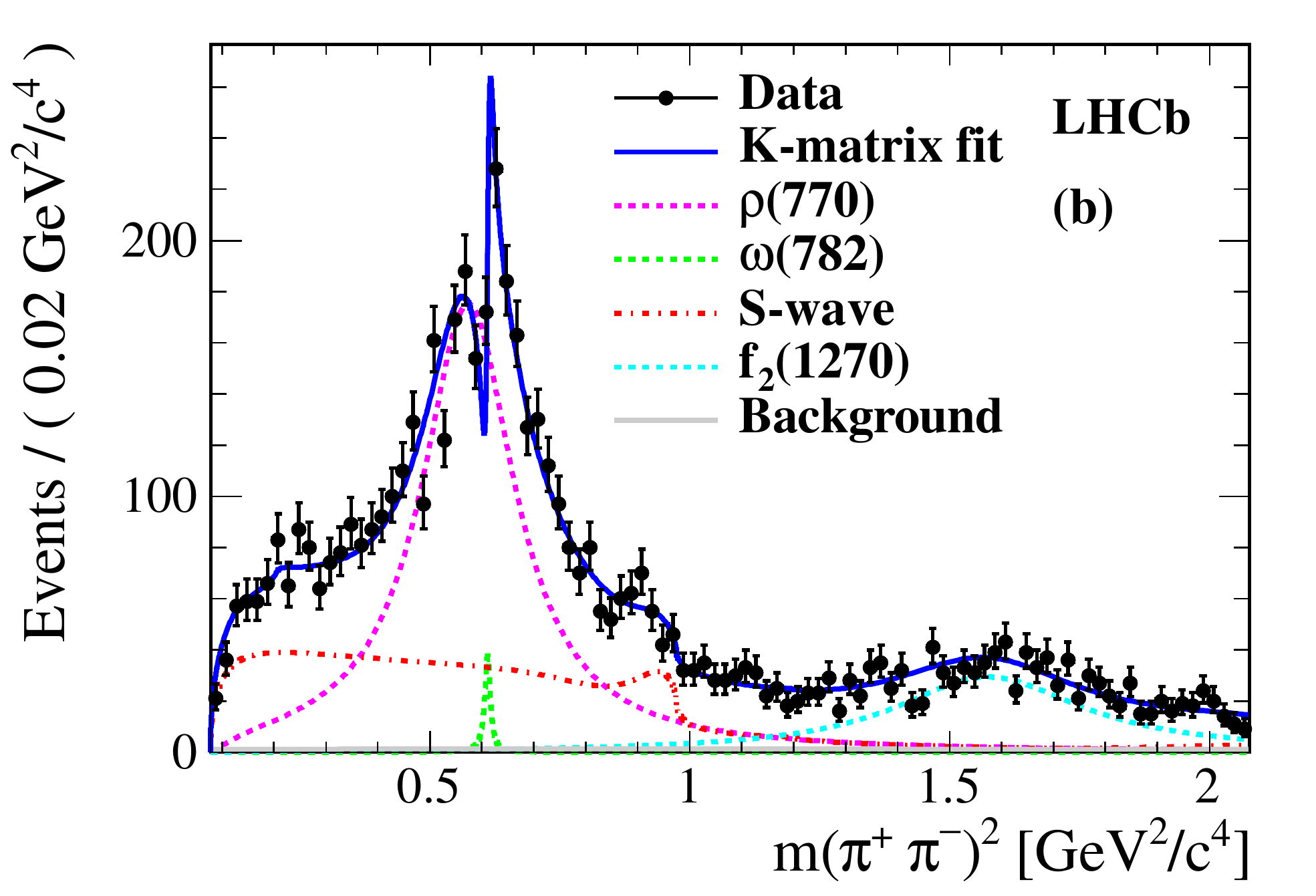}}
\end{minipage}
\caption{Distributions of $m^2(\pi^+\pi^-)$ in the $\rho(770)$ mass region. The different fit components are described in the legend.
Results from (a) the Isobar model and (b) the K-matrix model are shown.}
\label{fig: fit770}
\end{figure}

To see more clearly the resonant contributions in the region of the $\rho(770)$ resonance, the data are plotted in the $\pi^+\pi^-$ invariant mass-squared region $[0.0, 2.1] \  {\rm GeV}^2/c^{4}$ in Fig.~\ref{fig: fit770}. In the region around 0.6 GeV$^2/c^{4}$, interference between the $\rho(770)$ and $\omega(782)$ resonances is evident. In the $\pi^+\pi^-$ S-wave distributions of both the Isobar model and the K-matrix model, a peaking structure is seen in the region $[0.9, 1.0] \ {\rm GeV}^2/c^4$,  which corresponds to the $f_0(980)$ resonance. The structure in the region $[1.3, 1.8]  \ {\rm GeV}^2/c^4$ corresponds to the spin-2 $f_2(1270)$ resonance.

\begin{figure}[!tbh]
\begin{minipage}{0.5 \linewidth}
\centerline{\includegraphics[width=1.0\linewidth]{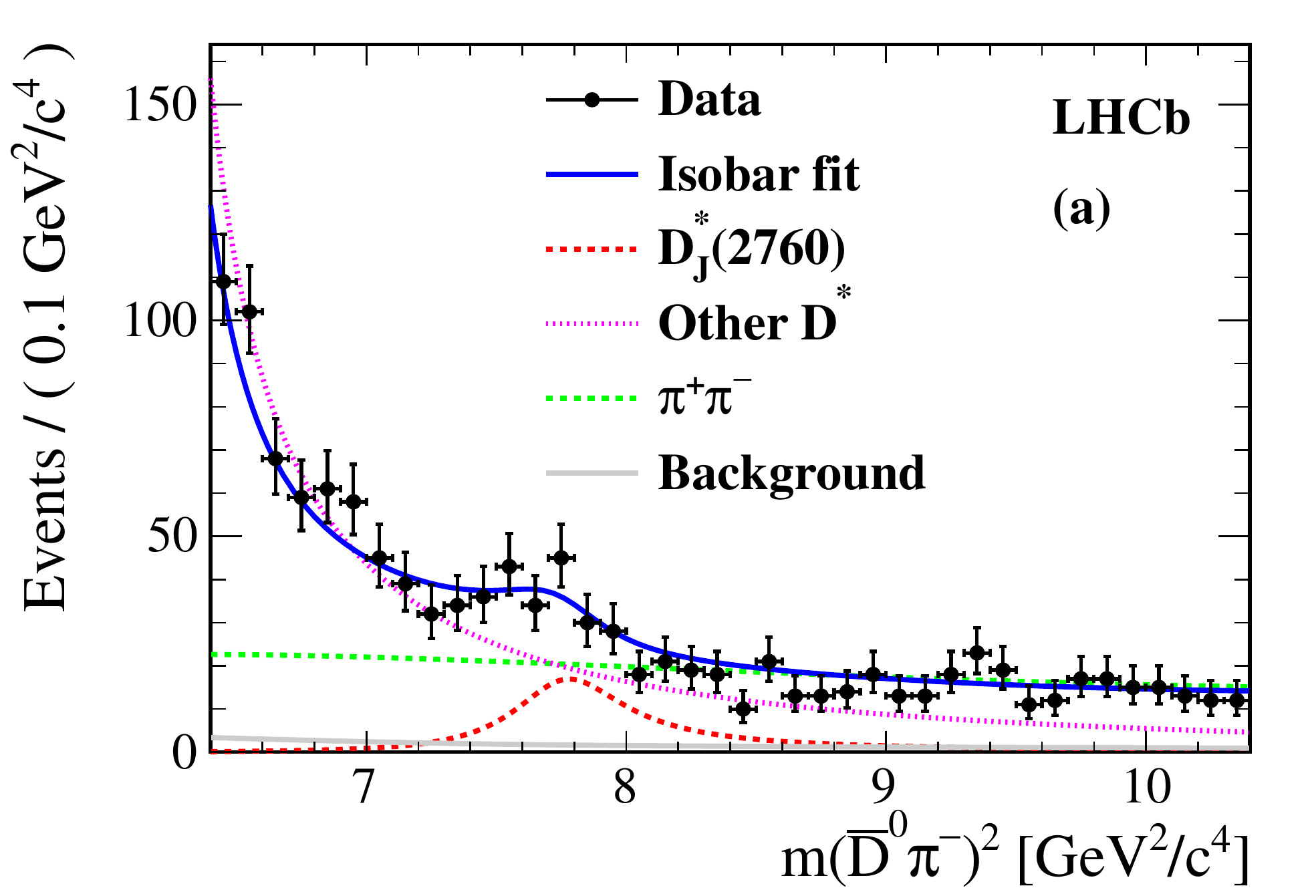}}
\end{minipage}
\begin{minipage}{0.5\linewidth}
\centerline{\includegraphics[width=1.0\linewidth]{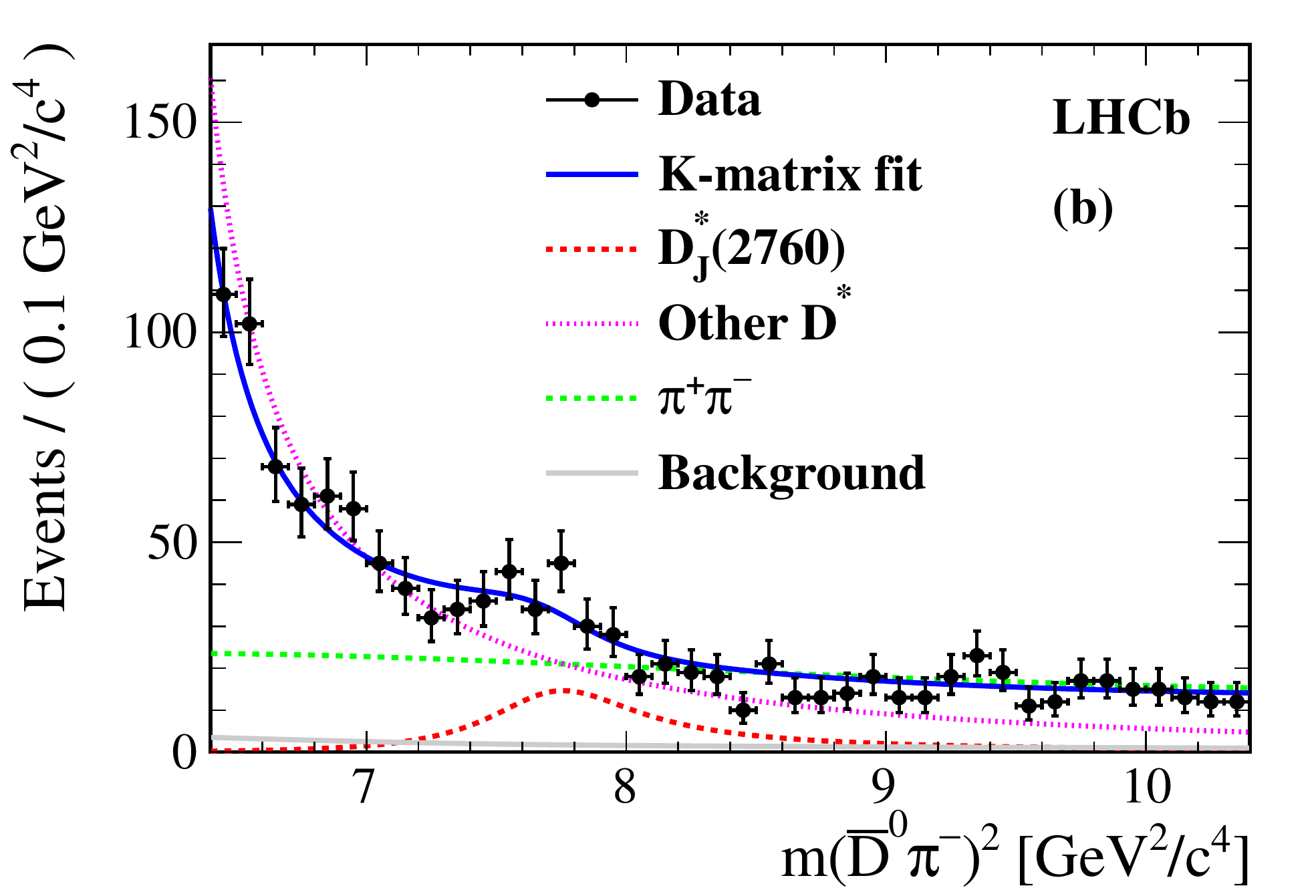}}
\end{minipage}
\caption{
Distributions of $\mdbpims$ in the $D_J^*(2760)^-$ mass region. The different fit components are described in the legend.
Both results from (a) the Isobar model and (b) the K-matrix model are shown.
}
\label{fig: fit2760}
\end{figure}

Distributions in the invariant mass-squared region $[6.4, 10.4] \ {\rm GeV}^2/c^4$ of $m^2(\Dzb\pi^-)$ are shown in Fig.~\ref{fig: fit2760}.
There is a significant contribution from the $D_J^*(2760)^-$ resonance observed in Ref.~\cite{LHCb-PAPER-2013-026} and a spin-3 assignment gives the best description.
A detailed discussion on the determination of the spin of $D_J^*(2760)$ is provided  in Sec.~\ref{sec:spin}.

The fit quality is  evaluated by determining a $\chi^2$ value by comparing the data and the fit model 
in $N_{\textrm{bins}} = 256$ bins that are defined adaptively to ensure approximately equal population with a minimum bin content of 37 entries.
A value of 287 (296) is found for the Isobar (K-Matrix) model based on statistical uncertainties only.
The effective number of degrees of freedom (nDoF) of the $\chi^2$ is bounded by $N_{\textrm{bins}} -1$ and $N_{\textrm{bins}} - N_{\textrm{pars}} - 1$, where $N_{\textrm{pars}}$ is the number of parameters determined by the data.
Pseudo experiments give an effective number of 234 (235) nDoF.

Further checks of the consistency between the fitted models and the data are performed with the unnormalised Legendre polynomial
weighted moments as a function of $\mdbpims$ and $m^2(\pi^+\pi^-)$. The corresponding distributions are shown in Appendix~\ref{sec:legendre}.

\section{Measurement of the {\boldmath $\btodpipi$} branching fraction}
\label{sec:BFDstarpi}

Measuring the branching fractions of the different resonant contributions requires  knowledge of the $\btodpipi$ branching fraction. This branching fraction is normalised relative to the $B^0 \to D^*(2010)^-\pi^+$ decay that has the same final state, so systematic uncertainties are reduced. Identical selections are applied to select $B^0 \to D^*(2010)^-\pi^+$ and $\btodpipi$ candidates, the only difference being that $m(\Dzb\pi^-)<2.1$ \gevcc is  used to select $D^*(2010)^-$ candidates. The kinematic constraints remove backgrounds from doubly mis-identified $\Dzb \to K^+ \pi^-$ or doubly Cabibbo-suppressed $\Dzb \to K^- \pi^+$ decays and no requirement is applied on $m(\Dzb\pi^+)$.

\begin{figure}[!tbh]
\begin{minipage}{0.5 \linewidth}
\centerline{\includegraphics[width=1.0\linewidth]{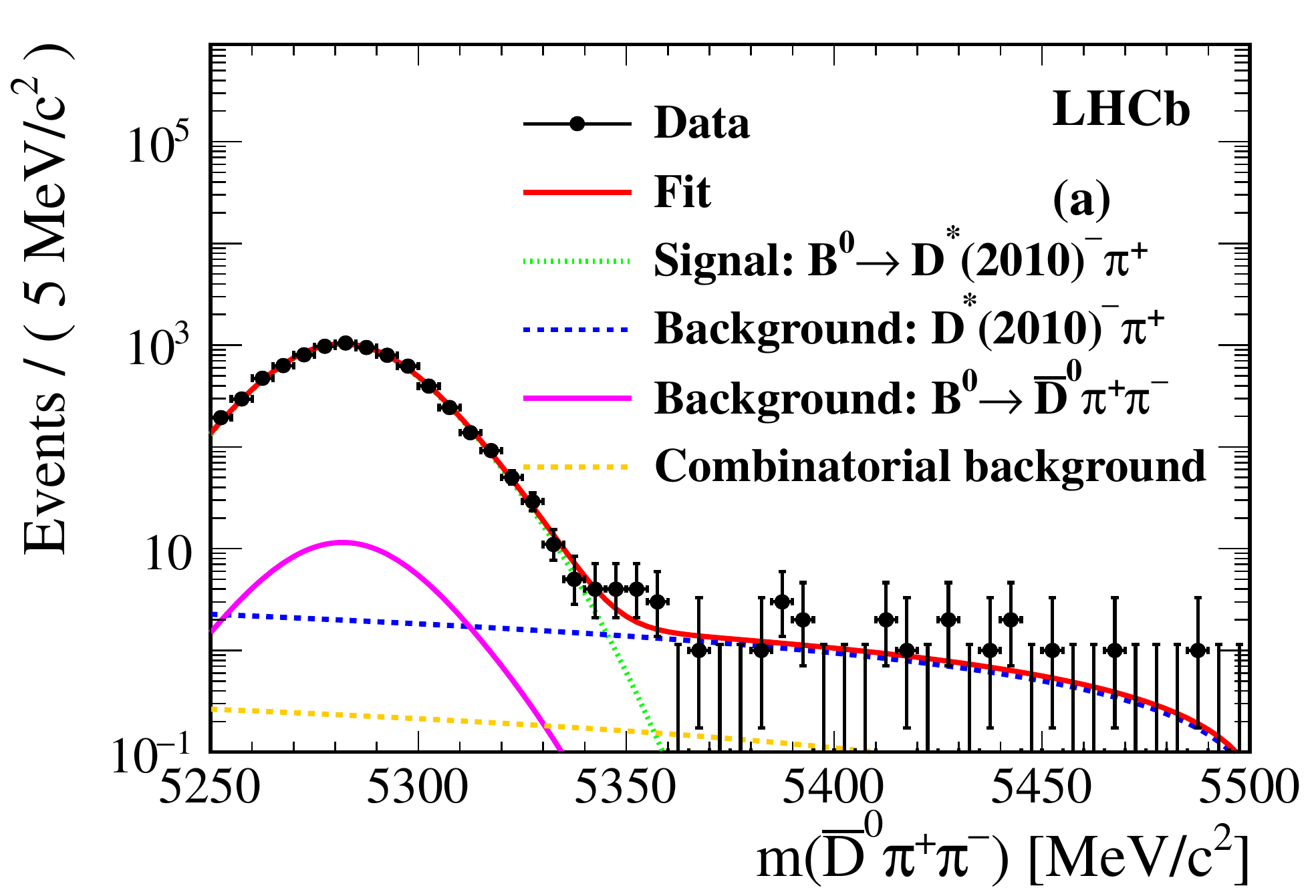}}
\end{minipage}
\begin{minipage}{0.5\linewidth}
\centerline{\includegraphics[width=1.0\linewidth]{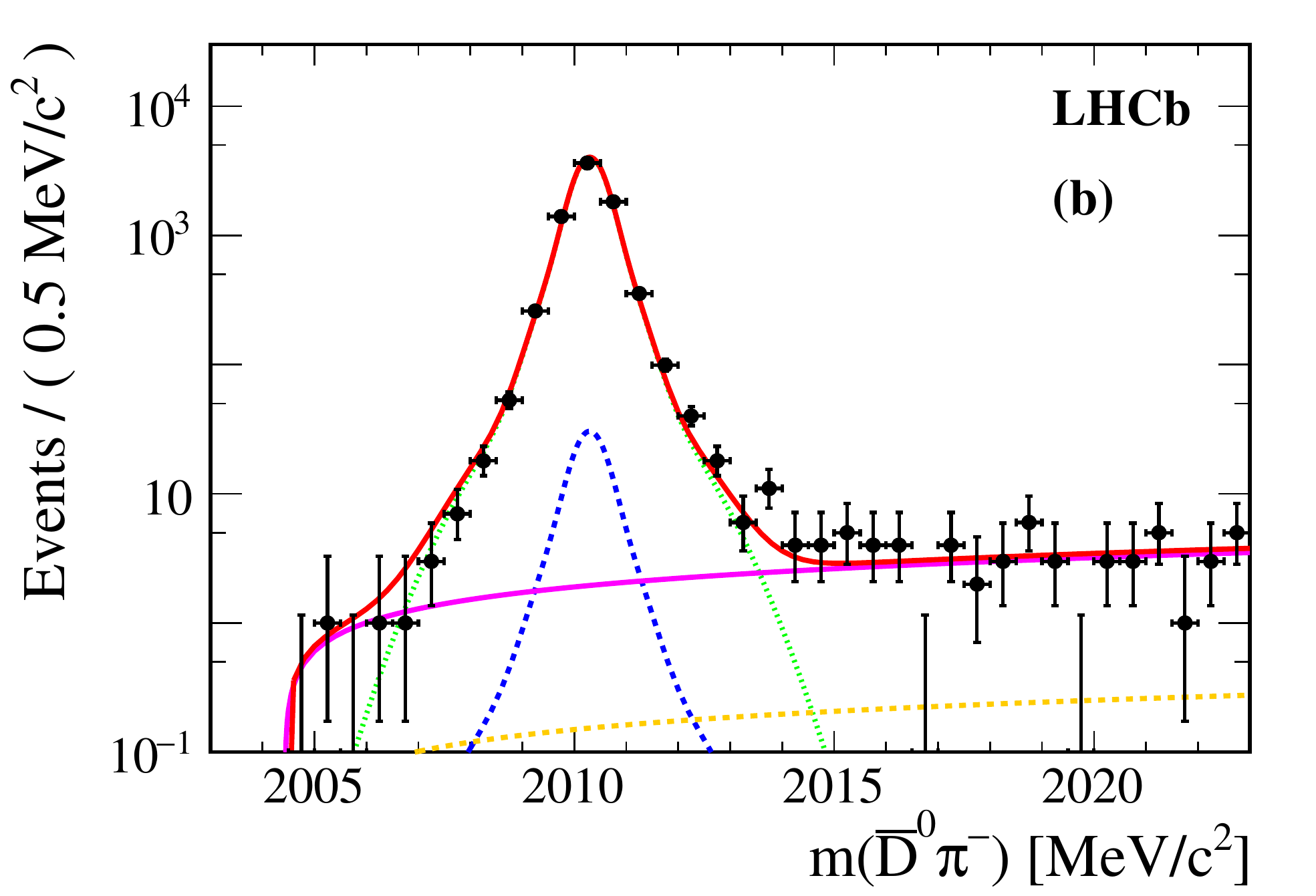}}
\end{minipage}
\caption{
Invariant mass distributions of (a) $m(\Dzb\pi^+\pi^-)$ and (b) $m(\Dzb\pi^-)$ for $B^0 \to D^*(2010)^-\pi^+$ candidates. The data is shown as black
points with the fit superimposed as red solid lines. 
}
\label{fig:MassDstar}
\end{figure}

The invariant mass distributions of $m(\Dzb\pi^-)$ and $m(\Dzb\pi^+\pi^-)$ for the $B^0 \to D^*(2010)^-\pi^+$ candidates are shown in Fig.~\ref{fig:MassDstar} and are fitted simultaneously to determine the signal and background contributions.  The $D^*(2010)^-$ signal distribution is modelled by three Gaussian functions to account for resolution effects while its background is modelled by a phase-space factor.
The modelling of the signal and background shapes in the $m(\Dzb\pi^+\pi^-)$ distribution are described in Sec.~\ref{sec:Selection}. The $B^0 \to D^*(2010)^- \pi^+$ yield in the signal region is $7327 \pm 85$.

The efficiencies for selecting $B^0 \to D^*(2010)^- \pi^+$ and $B^0 \to \Dzb \pi^+\pi^-$ decays are obtained from simulated samples. To take into account
the resonant distributions in the Dalitz plot, the $B^0 \to \Dzb \pi^+\pi^-$ simulated sample is weighted using the model described in the previous sections.
The average efficiencies are $(1.72 \pm 0.05) \times 10^{-4}$ and $(4.96 \pm 0.05) \times 10^{-4}$ for the $B^0 \to D^*(2010)^- \pi^+$ and $B^0 \to \Dzb \pi^+\pi^-$ decays.

Using the branching fractions of $\BR(B^0 \to D^*(2010)^- \pi^+) = (2.76 \pm 0.13) \times 10^{-3}$  and $\BR( D^*(2010)^- \to \Dzb \pi^-) =(67.7 \pm 0.5) \%$~\cite{PDG}, the derived branching fraction of $B^0 \to \Dzb \pi^+\pi^-$ in the kinematic region $m(\Dzb \pi^{\pm}) >2.1$ \gevcc is  $(8.46 \pm 0.14 \pm 0.40) \times 10^{-4}$, where the first uncertainty is statistical and the second uncertainty comes from the branching fraction of the normalisation channel.

\section{Systematic uncertainties}
\label{sec:Systematic}

\subsection{Common systematic uncertainties and checks}
\label{sec:CommonSystematic}

Two categories of systematic uncertainties are considered, each of which is quoted separately.  They originate from  the imperfect knowledge of the experimental conditions  and from the assumptions made in the Dalitz plot fit model. The Dalitz model-dependent uncertainties also account for the precision on the external parameters. The various sources are assumed to be independent and summed in quadrature to give the total.

Experimental systematic uncertainties arise from the efficiency and background modelling and from the veto on the $D^{*}(2010)^-$ resonance. Those corresponding to the signal efficiency are due to imperfect estimations of PID, trigger, tracking reconstruction effects, and to the finite size of the simulated samples. Each of these effects is evaluated by  the differences between the results using efficiencies computed from the simulation and from the data-driven methods. The systematic uncertainties corresponding to the modelling of the small residual background are estimated by using different sub-samples of backgrounds. The systematic uncertainty due to the veto on the $D^*(2010)^-$ resonance is assigned by changing the selection requirement from $m^2(\Dzb\pi^{\pm}) > 2.10 \gevcc$ to $2.05 \gevcc$.

The systematic uncertainties related to the Dalitz models considered (see Sec.~\ref{sec:DalitzFormulism}) include effects from other possible resonant contributions that are not included in the nominal fit, from the modelling of resonant lineshapes and from  imperfect knowledge of the parameters of the modelling, \ie, the masses and widths of the $\pi^+\pi^-$ resonances considered, and the resonant radius.

The non-significant resonances added to the model for systematic studies are the $f_0(1300)$, $f_0(1500)$, $f'_2(1525)$, and $D^*(2650)^-$ ($f'_2(1525)$ and $D^*(2650)^-$) mesons for the Isobar (K-matrix) model~\cite{PDG,LHCb-PAPER-2013-026,LHCb-PAPER-2013-069,LHCb-PAPER-2014-012}. The spin of the $D^*(2650)^-$ resonance  is set to 1. The differences between each alternative model and the nominal model are conservatively assigned as systematic uncertainties.

The radius of the resonances $(r_{\textrm{BW}})$ is set to a unique value of $1.6 \gev^{-1}\times \hbar c$ in the nominal fit. In the systematic studies, it is floated as a free parameter and its best fit value is $1.84 \pm 0.05$ GeV$^{-1}\times \hbar c$ ($1.92 \pm 0.31$ GeV$^{-1}\times \hbar c$) for the Isobar (K-matrix) model.  The value 1.85 GeV$^{-1}\times \hbar c$ is chosen to estimate the systematic uncertainties due to the imperfect knowledge of this parameter.

The masses and widths of the $\pi^+\pi^-$ resonances considered are treated as free parameters with Gaussian constraints according to the inputs listed in Table~\ref{tab: Resonances}. The differences between the results from those fits and those of the nominal fits are assigned as systematic uncertainties.

For the Isobar model, additional systematic uncertainties due to the modelling of the $f_0(500)$ and $f_0(980)$ resonances are considered. The Bugg model~\cite{Bugg} for the  $f_0(500)$ resonance and the Flatt\'e model~\cite{Flatte:1976xu} for the $f_0(980)$ resonance, used in the nominal fit, are replaced by more conventional RBW functions. The masses and widths, left as free parameters, give $553 \pm 15 \mevcc$ and $562 \pm 39 \mev$,  for the $f_0(500)$ meson and $981 \pm 13\mevcc$ and $191 \pm 39\mev$, for the $f_0(980)$ meson. The resulting differences to the nominal fit are assigned as systematic uncertainties.

The kinematic variables are calculated with the masses of the $\Dzb$ and $B^0$ mesons constrained to their known values~\cite{PDG}. These kinematic constraints affect the extraction of the masses and widths of the $\dbpim$ resonances. The current world average value for the $B^0$ meson mass is $5279.58 \pm 0.17 \mevcc$ and for the $\Dzb$ meson is $1864.84 \pm 0.07\mevcc$~\cite{PDG}. A conservative and direct estimation of the systematic uncertainties on the masses and widths of the $\dbpim$ resonances is provided by the sum in quadrature of the $B^0$ and $\Dzb$ mass uncertainties. The effects of mass constraints are found to be negligible for the fit fractions, moduli and phases of the complex coefficients.

The systematic uncertainties are summarised for the Isobar (K-matrix) model Dalitz  analysis in Appendix~\ref{sec:Systematicuncertainties}. Systematic uncertainties related to the measurements performed with the Isobar formalism are listed in Tables~\ref{tab:IsoMW} to \ref{tab:IsoFraction}, while those for the  K-matrix formalism  are given in Tables~\ref{tab:KMaMW} to \ref{tab:KMaFraction}.  In most of cases, the dominant systematic uncertainties are due to the $D^*(2010)^-$ veto and the model uncertainties related to other resonances not considered in the nominal fit. In the Isobar model, the modelling of the   $f_0(500)$ and $f_0(980)$ resonances  also have non-negligible systematic effects.

Several cross-checks have been performed to study the stability of the results. The analysis was repeated for different Fisher discriminant selection criteria, different trigger requirements and different sub-samples,  corresponding to the two data-taking periods and to the two half-parts of the $\Dzb\pi^+\pi^-$ invariant mass signal region, above and below the $B^0$ mass~\cite{PDG}. Results from those checks demonstrate good consistency with respect to the nominal fit results. No bias is seen, therefore no correction is applied, nor is any related uncertainty assigned.

\subsection{Systematic uncertainties on the {\boldmath $\B^0 \to \Dzb \pi^+\pi^-$} branching fraction }
 \label{sec:DpipiBFSyst}

The systematic uncertainties related to the measurement of the $\btodpipi$ branching fraction are listed  in Table~\ref{tab:sysons}.
The systematic uncertainties on the PID, trigger, reconstruction and statistics of the simulated samples are calculated in a similar way to those of the Dalitz plot analysis.
Other systematic uncertainties are discussed below.
\begin{table}
\centering
\caption{
Systematic uncertainties on  $\BR(\btodpipi)$.
\label{tab:sysons}}
\begin{tabular}{l.} 
\hline
Source & \multicolumn{1}{c}{Uncertainty $(\times 10^{-4})$} \\
\hline
PID & 0.02 \\
Trigger & 0.13\\
Reconstruction & <0.01 \\
Size of simulated sample & 0.26 \\
$B^0$, $D^{*}(2010)^{-}$ mass model & <0.01\\
Dalitz structure & 0.04\\
\hline
Total & 0.29 \\
\hline
\end{tabular}
\end{table}

The systematic uncertainty on the modelling of the $\Dzb \pi^-$ and $D^*(2010)^- \pi^+$ invariant mass distributions is estimated by counting the number of signal events in the $B^0$ signal region assuming a flat background contribution. The $D^*(2010)^-$ mass region is restricted to the range [2007, 2013] \mevcc for this estimate. The calculated branching fraction is nearly identical to that from the mass fit and thus has a negligible contribution to the systematic uncertainty. The signal purity of $B^0 \to D^{*}(2010)^{-} \pi^+$ is more than 99\%.

To account for the effect of resonant structures on the signal efficiency, the data sample is divided using an adaptive binning scheme. 
The average efficiency is calculated in a model independent way as
\begin{equation}
\varepsilon _{\textrm{ave}} =  \frac{\sum_i N_i}{\sum_i N_i/\varepsilon_i},
\end{equation}
where $N_i$ is the number of events in bin $i$ and $\varepsilon_i$ is the average efficiency in bin $i$ calculated from the efficiency model. The difference between this model-independent method and the nominal is assigned as a systematic uncertainty.

\section{Results}
\label{sec:Results}

\subsection{Significance of resonances}
\label{sec:signif}

The Isobar and K-matrix models employed to describe the Dalitz plot of the $\btodpipi$ decay include all of the resonances listed in Table~\ref{tab: Resonances}.  The statistical significances of well-established $\pi^+\pi^-$ resonances are calculated directly with their masses and widths fixed to the world averages. They are computed as the relative change of the minimum of the negative logarithm of the likelihood (NLL) function with and without a given resonance.
 Besides the $\pi^+\pi^-$ resonances listed in Table~\ref{tab: Resonances}, the significances of the $f_0(1370)$, $f_0(1500)$ and $f'_2(1525)$ are also given. The results, expressed as multiples of Gaussian standard deviations ($\sigma$), are summarised in Table~\ref{tab: fit_contribution}. All of the other $\pi^+\pi^-$ resonances not listed in this Table have large statistical significances, well above fivestandard deviations.
\begin{table}
\centering
\footnotesize
\caption{Statistical significance ($\sigma$) of $\pi^+\pi^-$ resonances in the Dalitz plot analysis. For the statistically significant resonances, the effect of adding dominant systematic uncertainties is shown (see text). }
\label{tab: fit_contribution}
\begin{tabular}{lcccccccc}
\hline
Resonances  & $\omega(782)$ & $f_0(980)$ & $f_0(1370)$ & $\rho(1450)$ & $f_0(1500)$ & $f'_2(1525)$ & $\rho(1700)$ & $f_0(2020)$ \\\hline
Isobar & 8.0 & 10.7 & 1.1 &  8.7 & 1.1 & 3.6 & 4.5 & 10.2 \\
K-matrix & 8.1 & n/a & n/a & 8.6 & n/a & 2.6 & 2.2 & n/a \\
With syst. & 7.7 & 7.0 & n/a & 8.7 & n/a & n/a & n/a & 4.3 \\
\hline
\end{tabular}
\end{table}

To test the significance of the $D_J^*(2760)^-$ state, where $J=3$ (see Sec.~\ref{sec:spin}), an ensemble of pseudo experiments is generated with the same number of events as in the data sample, using parameters obtained from the fit with the $D_J^*(2760)^-$ resonance excluded. The difference of the minima of the NLL when fitting with and without $D_J^*(2760)^-$ is used as a test statistic. It corresponds to  11.4\,$\sigma$  (11.5\,$\sigma$) for the for the Isobar (K-matrix) model and confirms the observation of $D_J^*(2760)^-$ reported in Ref.~\cite{LHCb-PAPER-2013-026}.
The two other orbitally excited $D$ resonances, $D_J^*(2650)^-$ and $D_J^*(3000)^-$, whose observations are presented in the same paper, are added into the nominal fit model with different spin hypotheses and tiny improvements are found. They also do not describe the data in the absence of the  $D_J^*(2760)^-$. Those resonances are thus not confirmed by this analysis. Finally, an extra $\Dzb\pi^-$ resonance, with different spin hypotheses ($J=0, 1, 2, 3, 4$) and with its mass and width allowed to vary, is added to the nominal fit model and no significant contribution is found.

The significance of each of the significant $\omega(782)$,  $f_0(980)$, $\rho(1450)$, $f_0(2020)$ and  $D_J^*(2760)^-$ states is checked while including the dominant systematic uncertainties (see Sec.~\ref{sec:CommonSystematic}), namely, the modelling of the $f_0(500)$ and $f_0(980)$ resonances, the addition of other resonant contributions and the modification of the $D^*(2010)^-$ veto criteria. In all configurations, the significances of the $\omega(782)$, $f_0(980)$, $\rho(1450)$ and $D_J^*(2760)^-$ resonances are greater than $7.7$\,$\sigma$,  $7.0$\,$\sigma$,  $8.7$\,$\sigma$, and $10.8$\,$\sigma$, respectively. The significance of the $f_0(2020)$ drops to $4.3$\,$\sigma$ when using a RBW lineshape for the $f_0(500)$ resonance. The abundant $f_0(500)$ contribution is highly significant under all of the applied changes.

\subsection{Spin of resonances}
\label{sec:spin}
\begin{figure}
\begin{minipage}{0.5 \linewidth}
\centerline{\includegraphics[width=1.0\linewidth]{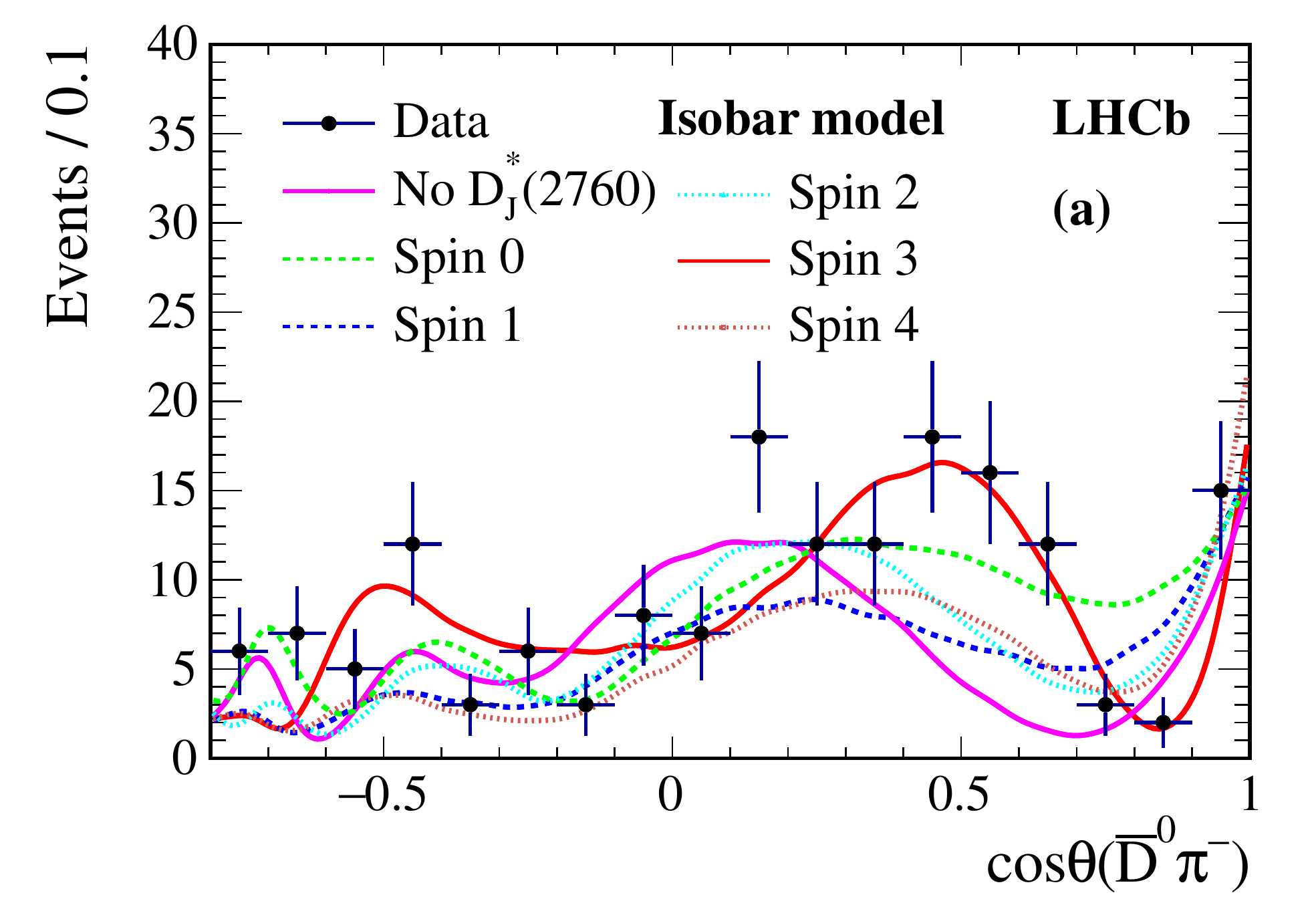}}
\end{minipage}
\begin{minipage}{0.5\linewidth}
\centerline{\includegraphics[width=1.0\linewidth]{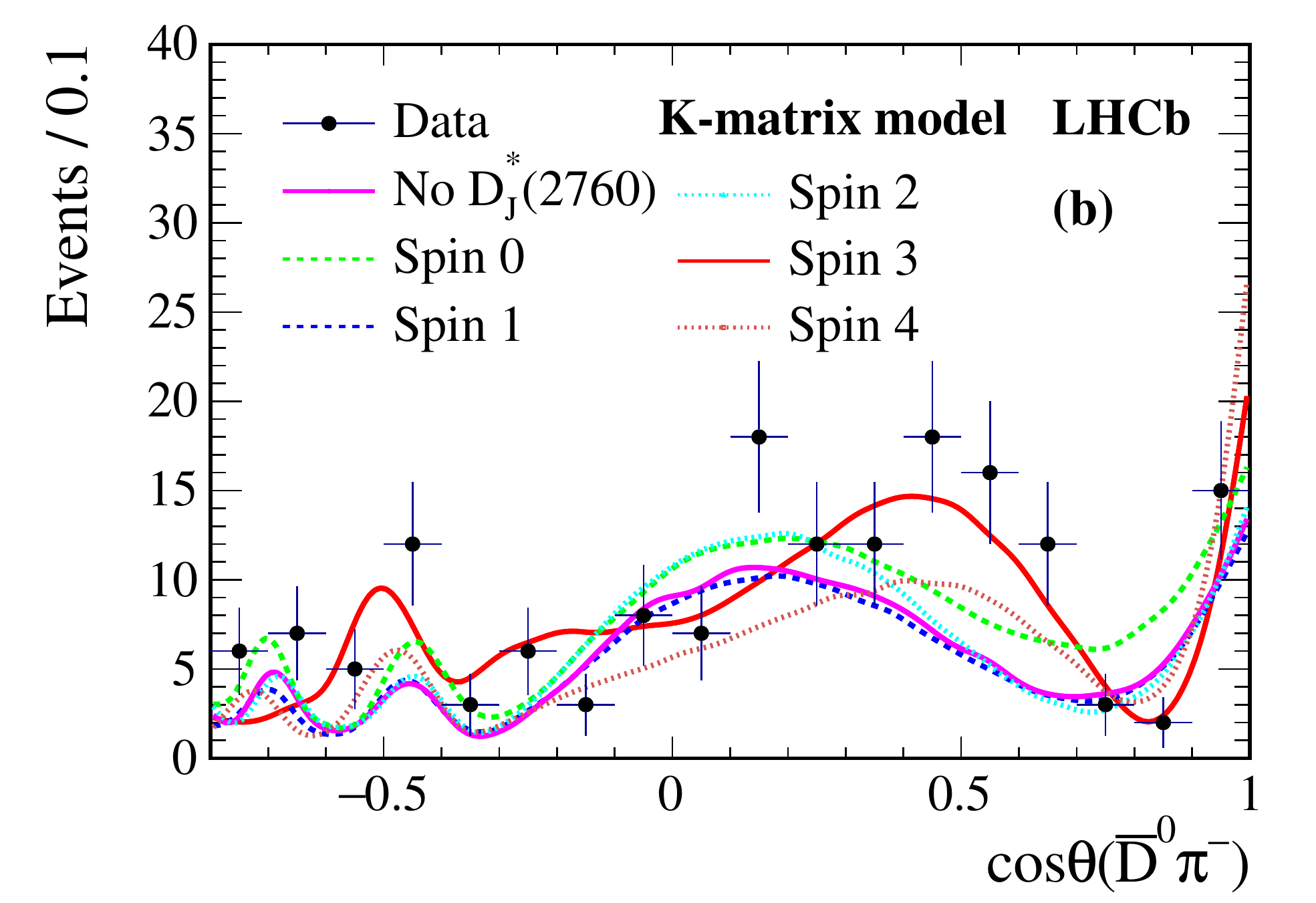}}
\end{minipage}
\caption{ Cosine of the helicity angle distributions in the $m^2(\Dzb\pi^-)$ range [7.4, 8.2] GeV$^2/c^4$ for (a) the Isobar model and (b) the K-matrix model.
The data are shown as black points. The helicity angle distributions of the Dalitz plot fit results, without the $D^*_J(2760)^-$ and with the different spin
hypotheses of $D^*_J(2760)^-$, are superimposed.
}
\label{fig: spintest}
\end{figure}

As described in Sec.~\ref{sec:resonances}, a spin-3 $D_J^*(2760)^-$ contribution gives the best description of the data.  To obtain the significance of the spin-3 hypothesis with respect to other spin hypotheses ($J=0, \ 1, \ 2, \ 4$), test statistics are built. Their computations are based on the shift of the minimum of the NLL with respect to the nominal fit model when using different spin hypotheses. The mass and width of the $D_J^*(2760)^-$ resonance are floated in all the cases. Pseudo experiments are generated using the fit parameters obtained using the other spin hypotheses. Significances are calculated according to the distributions obtained from the pseudo experiments of the test statistic  and its values from data. These studies indicate that data are inconsistent with other spin hypotheses by more than 10\,$\sigma$. Following the discovery of the $D^*_{sJ}(2860)^-$ meson, which is interpreted as the superposition of two particles with spin 1 and spin 3~\cite{LHCb-PAPER-2014-035,LHCb-PAPER-2014-036}, a similar configuration for the $D_J^*(2760)^-$ has been tested and is found to give no significant improvement in the description of the data. To illustrate the preference of the spin-3 hypothesis, the cosine of the helicity angle distributions in the mass-squared region of [7.4, 8.2] GeV$^2/c^4$ for $m^2(\Dzb\pi^-)$ are shown in Fig.~\ref{fig: spintest} under the various scenarios. Based on our result, $D_J^*(2760)^-$ is interpreted as the $D_3^*(2760)^-$ meson.
 Recently, LHCb  observed a neutral spin-1 $D^*(2760)^0$ state~\cite{LHCb-PAPER-2015-007}. The current analysis does not preclude a charged spin-1 $D^*$ state at around the same mass, but it is not sensitive to it with the current data sample size.

Studies have  also been performed to validate the spin-0 hypothesis of the $D_0^*(2400)^-$ resonance, as the spin of this state has never previously been confirmed in experiment~\cite{PDG}. When moving to other spin hypotheses, the minimum of the NLL increases by more than 250 units in all cases, which confirms the expectation of
spin 0 unambiguously.

\subsection{Results of the Dalitz plot analysis}

The shape parameters of the $\pi^+\pi^-$ resonances are fixed from previous measurements except for the nonresonant contribution in the Isobar model. The fitted value of the parameter $\alpha$ defined in Eq.~(\ref{eq: nonres}) is  $-0.363 \pm 0.027$, which corresponds to a 10\,$\sigma$ statistical significance compared to the case where there is no varying phase.
An expansion of the model by including a varying phase in the $\Dzb\pi^-$ axis is also investigated but no significantly varying phase in that system is seen.
The results indicate a weak, but non-negligible, rescattering effect in the $\pi^+\pi^-$ states, while the rescattering in the $\Dzb\pi^-$ states is not significant.
The masses, widths and other shape parameters of the  $\Dzb\pi^-$ contributions are allowed to vary in the analysis. The values of the shape parameters of the $\Dzb\pi^-$ P-wave component, defined in Eq.~(\ref{eq: D2010}), are $\beta_1 = 0.95 \pm 0.05$ ($0.90 \pm 0.04$) and $\beta_2 = 0.51 \pm 0.06$ ($0.43 \pm 0.05$) for the Isobar (K-matrix) model.

The measurements of  the masses and widths of the three resonances $D_0^*(2400)^-$, $D_2^*(2460)^-$ and $D_3^*(2760)^-$ are listed in Table~\ref{tab: massAndWidthsFinal}. The present precision on the mass and width of the $D_0^*(2400)^-$ resonance is improved with respect to Refs.~\cite{PDG,LHCb-PAPER-2013-026}. The result for the width of the $D_2^*(2460)^-$ meson is consistent with  previous measurements, whereas the result for the mass is above the world average which is dominated by the measurement using inclusive production by LHCb~\cite{LHCb-PAPER-2013-026}. 
In the previous LHCb inclusive analysis, the broad $D_0^*(2400)^-$ component was excluded from the fit model due to a high correlation with the background lineshape parameters, while here it is
included. The present result supersedes the former measurement. 
The Dalitz plot analysis used in this paper ensures that the background under the $D_2^*(2460)^-$ peak and the effect on the efficiency are under control, resulting in much lower systematic uncertainties compared to the inclusive approach.

\begin{table}
\centering
\small
\caption{Measured masses ($m$ in \mevcc) and widths ($\Gamma$ in \mev) of the $D_0^*(2400)^-$, $D_2^*(2460)^-$ and $D_3^*(2760)^-$ resonances,
where the first uncertainty is statistical, the second and the third are experimental and model-dependent systematic uncertainties, respectively.}
\label{tab: massAndWidthsFinal}
\begin{tabular}{l lc c}
\hline
 & &  Isobar & K-matrix \\
\hline
$D_0^*(2400)$ & $m$ &  $\al\all2349  \pm \al\all6 \pm \al\all1 \pm \al\all4$ & $\al\all2354 \pm \al\all7 \pm \all11 \pm \al\all2$   \\
& $\Gamma$ &  $\al\al\all217 \pm \all13 \pm \al\all5 \pm \all12$    &  $\al\al\all230 \pm \all15 \pm \all18 \pm \all11$   \\
$D_2^*(2460)$ & $ m$ & $2468.6 \pm 0.6 \pm 0.0 \pm 0.3$ & $2468.1 \pm 0.6 \pm 0.4 \pm 0.3$   \\
& $\Gamma$ &  $\al\al47.3 \pm 1.5 \pm 0.3 \pm 0.6$  &  $\al\al46.0 \pm 1.4 \pm 1.7 \pm 0.4$  \\
$D_3^*(2760)$ & $m$ &   $\al\all2798 \pm \al\all7 \pm \al\all1 \pm \al\all7$  &  $\al\all2802 \pm \all11 \pm \all10 \pm \al\all3$ \\
& $\Gamma$   & $\al\al\all105 \pm \all18 \pm \al\all6 \pm \all23$ & $\al\al\all154 \pm \all27 \pm \all13 \pm \al\all9$  \\
\hline
\end{tabular}
\end{table}

The moduli and the phases of the complex coefficients of the resonant contributions, defined in Eq.~(\ref{fun: matrixelement}), are displayed in Tables~\ref{tab: finalAmp} and \ref{tab: finalPha}.  Compatible results are obtained using both the Isobar and K-matrix models. The results for the fit fractions are given in Table~\ref{tab: finalfraction}, while results for the interference fit fractions are given in Appendix~\ref{sec:interferencefitfractions}. 
Pseudo experiments are used to validate the fitting procedure and no biases are found in the determination of parameter values.

\begin{table}
\centering
\small
\caption{The moduli of the complex coefficients of the resonant contributions for the Isobar model and the K-matrix model.
The first uncertainty is statistical, the second and the third are experimental and model-dependent systematic uncertainties, respectively.}
\label{tab: finalAmp}
\begin{tabular}{lcc}
\hline
Resonance & Isobar ($|c_i|$) & K-matrix ($|c_i|$) \\\hline
Nonresonance & $\al3.43 \pm \al0.22 \pm \al0.04 \pm \al0.51$ & n/a \\
$f_0(500)$ & $\al18.7 \pm \al0.70 \pm \al0.29 \pm \al0.80$ & n/a \\
$f_0(980)$ & $\al2.62 \pm \al0.25 \pm \al0.09 \pm \al0.46$ & n/a \\
$f_0(2020)$ & $\al4.41 \pm \al0.51 \pm \al0.21 \pm \al1.78$ & n/a\\
$\rho(770)$ & 1.0 (fixed) & 1.0 (fixed) \\
$\omega(782)$ & $\al0.30 \pm \al0.04 \pm \al0.00 \pm \al0.01$ &  $\al0.31 \pm \al0.04 \pm \al0.01 \pm \al0.01$\\
$\rho(1450)$ & $\al0.23 \pm \al0.03 \pm \al0.01 \pm \al0.02$ & $\al0.28 \pm \al0.03 \pm \al0.08 \pm \al0.01$ \\
$\rho(1700)$ & $0.078 \pm 0.016 \pm 0.006 \pm 0.008$ & $0.136 \pm 0.020 \pm 0.077 \pm 0.011$\\
$f_2(1270)$ & $0.072 \pm 0.002 \pm 0.000 \pm 0.005$ & $0.073 \pm 0.002 \pm 0.006 \pm 0.003$\\
$\Dzb\pi^-$ P-wave & $\al18.8 \pm \al\al0.7 \pm \al\al0.3 \pm \al\al1.9 $ &$\al19.6 \pm \al\al0.7 \pm \al\al0.7 \pm \al\al0.6$\\
$D_0^*(2400)^-$ & $\al12.1 \pm \al\al0.8 \pm \al\al0.3 \pm \al\al0.6$ & $\al13.1 \pm \al\al1.0 \pm \al\al0.8 \pm \al\al0.5$\\
$D_2^*(2460)^-$ & $\al1.31 \pm \al0.04 \pm \al0.02 \pm \al0.02$ & $\al1.31 \pm \al0.04 \pm \al0.04 \pm \al0.00$ \\
$D_3^*(2760)^-$ & $0.053^{\all+\all\al0.011}_{\all-\all\al0.006} \pm 0.003 \pm 0.008$ &$0.075^{\all+\all\al0.016}_{\all-\all\al0.008} \pm 0.005 \pm 0.003$ \\
 [+0.9ex]
\hline
\end{tabular}
\end{table}

\begin{table}
\centering
\small
\caption{The phase of the complex coefficients of the resonant contributions for the Isobar model and the K-matrix model.
The first uncertainty is statistical, the second and the third are experimental and model-dependent systematic uncertainties, respectively.
}
\label{tab: finalPha}
\begin{tabular}{lcc}
\hline
Resonance & Isobar (arg($c_i$)$^{\circ}$) & K-matrix (arg($c_i$)$^{\circ}$)\\
\hline
Nonresonance & $\al77.1 \pm \al4.5 \pm \al2.3 \pm \al5.4$ & n/a\\
$f_0(500)$ & $\al38.4 \pm \al2.7 \pm \al1.3 \pm \al3.7$ & n/a \\
$f_0(980)$ & $138.9 \pm \al4.6 \pm \al1.5 \pm10.9$ & n/a\\
$f_0(2020)$ & $258.5 \pm \al5.0 \pm \al1.1 \pm 26.8$ & n/a\\
$\rho(770)$ & 0.0 (fixed)  & 0.0 (fixed) \\
$\omega(782)$ & $176.8 \pm \al7.8 \pm \al0.6 \pm \al0.5$ &  $174.8 \pm \al8.0 \pm \al1.5 \pm \al0.5$\\
$\rho(1450)$ & $149.0 \pm \al7.5 \pm \al4.8 \pm \al4.5$ & $132.9 \pm \al7.8 \pm \al8.5 \pm \al5.5 $ \\
$\rho(1700)$ & $103.5 \pm 13.1 \pm  \al4.5 \pm \al2.4$ & $\al77.6 \pm \al9.9 \pm 23.1 \pm \al4.5$\\
$f_2(1270)$ & $158.1 \pm \al3.0 \pm \al1.6 \pm \al3.8 $ & $147.8 \pm \al2.5 \pm \al8.5 \pm \al2.6 $\\
$\Dzb\pi^-$ P-wave & $266.7 \pm \al3.7 \pm \al0.3 \pm \al7.1$ &$261.0 \pm \al4.0 \pm \al3.3 \pm \al6.7$\\
$D_0^*(2400)^-$ & $\al83.6 \pm \al4.4 \pm \al2.8 \pm \al4.6 $ & $\al78.4 \pm \al4.1 \pm 11.5 \pm \al1.7$\\
$D_2^*(2460)^-$ & $262.9 \pm \al2.9 \pm \al0.8 \pm \al3.0$ & $257.4 \pm \al3.4 \pm \al0.7 \pm \al1.9$ \\
$D_3^*(2760)^-$ & $\al91.1 \pm \al6.7 \pm \al1.4 \pm \al5.1$ &$\al92.7 \pm  \al7.3 \pm 15.2 \pm \al2.3$ \\
\hline
\end{tabular}
\end{table}

\begin{table}[!tbh]
\centering
\small
\caption{The fit fractions of the resonant contributions for the Isobar and K-matrix models with $m(\Dzb\pi^{\pm})> 2.1$ \gevcc.
The first uncertainty is statistical, the second and the third are experimental and model-dependent systematic uncertainties, respectively.
}
\label{tab: finalfraction}
\begin{tabular}{lcc}
\hline
Resonance & Isobar (${\cal{F}}_i$ \%) & K-matrix (${\cal{F}}_i$ \%)\\
\hline
Nonresonance & $\al2.82 \pm 0.34 \pm 0.07 \pm 0.80$ & n/a\\
$f_0(500)$ & $\al13.2 \pm 0.89 \pm 0.31 \pm 2.45$ & n/a \\
$f_0(980)$ & $\al1.56 \pm 0.29 \pm 0.11 \pm 0.54$ & n/a\\
$f_0(2020)$ & $\al1.58 \pm 0.36 \pm 0.15 \pm 1.00$ & n/a\\
S-wave & $16.39 \pm 0.58 \pm 0.43 \pm 1.46$ & $16.51 \pm 0.70 \pm 1.68 \pm 1.10$\\
$\rho(770)$ & $37.54 \pm 1.00 \pm 0.61 \pm 0.98$ & $36.15 \pm 1.00 \pm 2.13 \pm 0.79$\\
$\omega(782)$ & $\al0.49 \pm 0.13 \pm 0.01 \pm 0.03$ &  $\al0.50 \pm 0.13 \pm 0.01 \pm 0.02$\\
$\rho(1450)$ & $\al1.54 \pm 0.32 \pm 0.08 \pm 0.22$ & $\al2.16 \pm 0.42 \pm 0.82 \pm 0.21$ \\
$\rho(1700)$ & $\al0.38^{\all+\al\all0.25}_{\all-\al\all0.12} \pm 0.07 \pm 0.06$ & $\al0.83 \pm 0.21 \pm 0.61 \pm 0.12$\\
$f_2(1270)$ & $10.28 \pm 0.49 \pm 0.31 \pm 1.10$ & $\al9.88 \pm 0.58 \pm 0.83 \pm 0.58$\\
$\Dzb\pi^-$ P-wave & $\al9.21 \pm 0.56 \pm 0.24 \pm 1.73$ &$\al9.22 \pm 0.58 \pm 0.67 \pm 0.75$\\
$D_0^*(2400)^-$ & $\al9.00 \pm 0.60 \pm 0.20 \pm 0.35$ & $\al9.27 \pm 0.60 \pm 0.86 \pm 0.52$\\
$D_2^*(2460)^-$ & $28.83 \pm 0.69 \pm 0.74 \pm 0.50$ & $28.13 \pm 0.72 \pm 1.06 \pm 0.54$ \\
$D_3^*(2760)^-$ & $\al1.22 \pm 0.19 \pm 0.07 \pm 0.09$ &$\al1.58 \pm  0.22 \pm 0.18 \pm 0.07$ \\
\hline
\end{tabular}
\end{table}

\subsection{Branching fractions}

The measured branching fraction of the $\btodpipi$ decay in the phase-space region $m(\Dzb\pi^{\pm})>2.1$ GeV$/c^2$ is
\begin{equation}
\BR(B^0 \to \Dzb \pi^+\pi^-) = (8.46 \pm 0.14 \pm 0.29 \pm 0.40) \times 10^{-4},
\end{equation}
taking into account the systematic uncertainties reported in Table~\ref{tab:sysons}.
The first uncertainty is statistical, the second systematic, and the third the uncertainty from the branching fraction of the
$B^0 \to D^*(2010)^-\pi^+$ normalisation decay channel.
The result agrees with the previous \belle measurement $(8.4 \pm 0.4 \pm 0.8) \times 10^{-4}$~\cite{BelleDpipi2} and  the \babar measurement
$(8.81 \pm 0.18 \pm 0.76 \pm 0.78 \pm 0.11$)$\times 10^{-4}$~\cite{BaBarDpipi}, obtained in a slightly larger phase-space region.
A multiplicative factor of 94.5\% (96.2\%) is required to scale the \belle (\babar) results to the same phase-space region as in this analysis.

The branching fraction of each quasi-two-body decay, $B^0 \to r_i h_3$, with $r_i  \to h_1 h_2$, is given by
\begin{equation}
\BR(B^0 \to r_i h_3) \times \BR(r_i \to h_1 h_2) = \BR(B^0 \to \Dzb\pi^+\pi^-) \times \frac{{{\cal{F}}_i}} {\varepsilon^{\rm corr}_i},
\end{equation}
where the resonant states $(h_1h_2) = (\Dzb\pi^-), (\pi^+\pi^-)$.
The fit fractions ${\cal{F}}_i$, defined in Eq.~(\ref{fun: fitfrac}),  are obtained from the Dalitz plot analysis and are listed in Table~\ref{tab: finalfraction}. The correction factors, $\varepsilon^{\rm corr}_i$, account for the cut-off due to the $D^*(2010)^-$ veto. They are obtained by generating pseudo experiment samples for each resonance over the Dalitz plot and applying the same requirement ($m(\Dzb\pi^{\pm})>2.1$~\gevcc).  They are summarised in Table~\ref{tab: correctionfactor}. The correction factors are the same for the Isobar model and the K-matrix model. The effects due to the uncertainties of the masses and widths of the resonances are included in the uncertainties given in the table.

\begin{table}
\centering
\caption{Correction factors due to the $D^*(2010)^-$ veto.}
\label{tab: correctionfactor}
\begin{tabular}{lc}
\hline
Resonance &  $\varepsilon^{\rm corr}_i$ \% \\
\hline
$f_0(500)$ &  $99.52 \pm 0.10$  \\
$f_0(980)$ &  $98.74 \pm0.09$ \\
$f_0(2020)$ & $99.29 \pm0.05$   \\
S-wave & $98.55 \pm 0.04$ \\
$\rho(770)$ &  $98.95 \pm  0.03$  \\
$\omega(782)$ &  $99.39 \pm0.02$  \\
$\rho(1450)$ & $95.66 \pm 0.06$ \\
$\rho(1700)$ & $96.73 \pm 0.06$ \\
$f_2(1270)$ &  $91.91 \pm0.09$ \\
$D^*_0(2400)^-$ &   $98.60 \pm 0.10$ \\
$D^*_2(2460)^-$ &  $100.$ \\
$D^*_3(2760)^-$ &  $100.$  \\
\hline
\end{tabular}
\end{table}

Using the overall $\btodpipi$ decay  branching fraction, the fit fractions (${\cal{F}}_i$) and the correction factors ($\varepsilon^{\rm corr}_i$), the branching fractions of quasi two-body decays are calculated in Table~\ref{tab: finalBr}. The first observation of the decays $\Bz \to \Dzb  f_0(500)$, $\Bz \to \Dzb  f_0(980)$,  $\Bz \to \Dzb  \rho(1450)$, as well as $\Bz \to D_3^*(2760)^- \pip$, and the first evidence of $\Bz \to \Dzb  f_0(2020)$ are reported.
The present world averages~\cite{PDG} of the branching fractions $\BR(\Bz \to \Dzb   \rho(770))\times\BR(\rho(770) \to \pi^+\pi^-)$, $\BR(\Bz \to \Dzb  f_2(1270))\times\BR(f_2(1270) \to \pi^+\pi^-)$, $\BR(\Bz \to D_0^*(2400)^- \pip)\times \BR(D_0^*(2400)^- \to \Dzb\pim)$, and $\BR(\Bz \to D_2^*(2460)^- \pip)\times\BR(D_2^*(2460)^- \to \Dzb\pim)$ are improved considerably. When accounting for the branching fractions of the $\omega(782)$ and $ f_2(1270)$ to $\pi^+\pi^-$, one obtains the following results for the Isobar model
\begin{equation}
\BR(\Bz \to \Dzb  \omega(782)) =  (2.75 \pm 0.72 \pm 0.13 \pm 0.20 \pm 0.13^{+0.20}_{-0.23}) \times 10^{-4}
\end{equation}
and
\begin{equation}
\BR(\Bz \to \Dzb  f_2(1270))  =  (16.8 \pm 1.1 \pm 0.7 \pm 1.8\pm 0.7^{+0.5}_{-0.2}) \times 10^{-5}.
\end{equation}
For the K-matrix model, one obtains
\begin{equation}
\BR(\Bz \to \Dzb  \omega(782)) = (2.81 \pm 0.72 \pm 0.13 \pm 0.13 \pm 0.13^{+0.20}_{-0.24}) \times 10^{-4}
\end{equation}
and
\begin{equation}
\BR(\Bz \to \Dzb  f_2(1270))  =  (16.1 \pm 1.1 \pm 1.4 \pm 0.9 \pm 0.7^{+0.5}_{-0.2}) \times 10^{-5}.
\end{equation}
In both models, the fifth uncertainty is due to knowledge of  the $\pi^+\pi^-$ decay rates~\cite{PDG}.
The results are consistent with the measurement of the decay $\Bz \to \Dzb  \omega(782)$, using the dominant $\omega(782) \to \pi^+\pi^-\pi^0$ decay~\cite{PDG, BaBarColSup}.

\begin{table}
\centering
\small
\caption{Measured branching fractions of $\BR(\Bz \to r h_3) \times \BR(r \to h_1 h_2)$ for the Isobar and K-matrix models. The first uncertainty is statistical, the second the experimental systematic, the third the model-dependent systematic, and the fourth the uncertainty from the normalisation $B^0 \to D^*(2010)^- \pi^+$ channel.
}
\label{tab: finalBr}
\begin{tabular}{lcc}
\hline
 Resonance  & Isobar ($\times 10^{-5}$) & K-matrix ($\times 10^{-5}$)\\
\hline
$f_0(500)$ &$ 11.2 \pm \al0.8 \pm \al0.5 \pm \al 2.1 \pm \al0.5 $ & n/a \\
$f_0(980)$ &$ 1.34 \pm 0.25 \pm 0.10 \pm 0.46 \pm 0.06 $ & n/a \\
$f_0(2020)$ &$ 1.35 \pm 0.31 \pm 0.14 \pm 0.85 \pm 0.06 $ & n/a\\
 S-wave &$ 14.1 \pm \al0.5 \pm \al0.6 \pm \al1.3 \pm \al0.7 $ & $14.2 \pm \al0.6 \pm \al1.5 \pm \al0.9 \pm \al0.7 $ \\
$\rho(770)$ &$ 32.1 \pm \al1.0 \pm \al1.2 \pm \al0.9 \pm \al1.5 $ & $31.0 \pm \al1.0 \pm \al2.1 \pm \al0.7 \pm \al1.5 $  \\
$\omega(782)$ &$ 0.42 \pm 0.11 \pm 0.02 \pm 0.03 \pm 0.02 $ & $0.43 \pm 0.11 \pm 0.02 \pm 0.02 \pm 0.02$ \\
$\rho(1450)$ &$ 1.36 \pm 0.28 \pm 0.08 \pm 0.19 \pm 0.06 $ & $1.91 \pm 0.37 \pm 0.73 \pm 0.19 \pm 0.09 $ \\
$\rho(1700) $ &$ 0.33 \pm 0.11 \pm 0.06 \pm 0.05 \pm 0.02 $ & $0.73 \pm 0.18 \pm 0.53 \pm 0.10 \pm 0.03 $ \\
$f_2(1270)$ &$ \al9.5 \pm \al0.5 \pm \al0.4 \pm \al1.0 \pm \al0.4 $ & $\al9.1 \pm \al0.6 \pm \al0.8 \pm \al0.5 \pm \al0.4 $  \\
$D_0^*(2400)^-$ &$ \al7.7 \pm \al0.5 \pm \al0.3 \pm \al0.3 \pm \al0.4 $ & $\al8.0 \pm \al0.5 \pm \al0.8 \pm \al0.4 \pm \al0.4 $ \\
$D_2^*(2460)^-$ &$ 24.4 \pm \al0.7 \pm \al1.0 \pm \al0.4 \pm \al1.2 $ & $23.8 \pm \al0.7 \pm \al1.2 \pm \al0.5 \pm \al1.1 $\\
$D_3^*(2760)^- $ &$ 1.03 \pm 0.16 \pm 0.07 \pm 0.08 \pm 0.05 $ & $1.34 \pm 0.19 \pm 0.16 \pm 0.06 \pm 0.06 $ \\
\hline
\end{tabular}
\end{table}

\subsection{Structure of the {\boldmath $f_0(980)$ and  $f_0(500)$} resonances}

In the Isobar model, significant contributions from both $B^0 \to \Dzb f_0(500)$ and $B^0 \to \Dzb f_0(980)$ decays are observed.  
The related branching fraction measurements can be used to obtain information on the substructure of the $f_0(980)$ and $f_0(500)$ resonances within the factorisation approximation. As discussed in Sec.~\ref{sec:Introduction}, two models for the quark structure of those states are considered: $q\bar{q}$ or $[qq'][\bar{q}\bar{q'}]$ (tetraquarks). In both models, mixing angles between different quark states are determined using our measurements. In the $q\bar{q}$ model, the mixing between $s\bar{s}$ and $u\bar{u}$ or $d\bar{d}$ can be written as
\begin{eqnarray}
|f_0(980)\rangle &=& \cos\varphi_{\textrm{mix}} |s\bar{s}\rangle + \sin\varphi_{\textrm{mix}} |n\bar{n}\rangle ,\\
|f_0(500)\rangle &=& -\sin\varphi_{\textrm{mix}} |s\bar{s}\rangle + \cos\varphi_{\textrm{mix}} |n\bar{n}\rangle,
\end{eqnarray}
where $|n\bar{n}\rangle \equiv (|u\bar{u}\rangle + |d\bar{d}\rangle)/\sqrt{2}$ and $\varphi_{\textrm{mix}}$ is the mixing angle. In the $[qq'][\bar{q}\bar{q'}]$ model, the mixing angle, $\omega_{\textrm{mix}}$, is introduced and the mixing  becomes
\begin{eqnarray}
|f_0(980)\rangle &=&  \cos\omega_{\textrm{mix}} |n\bar{n}s\bar{s}\rangle  + \sin\omega_{\textrm{mix}} |u\bar{u}d\bar{d}\rangle,\\
|f_0(500)\rangle &=& -\sin\omega_{\textrm{mix}} |n\bar{n}s\bar{s}\rangle + \cos\omega_{\textrm{mix}} |u\bar{u}d\bar{d}\rangle.
\end{eqnarray}
In both cases, the following variable is defined
\begin{equation}
r^f = \frac{\BR(B^0 \to \Dzb f_0(980))}{\BR(B^0 \to \Dzb f_0(500))} \times \frac{\Phi(500)}{\Phi(980)},
\end{equation}
where $\Phi(500)$ and $\Phi(980)$ are the integrals of the phase-space factors computed over the resonant lineshapes and the phase-space factors are proportional to the momentum computed in the $B^0$ rest frame. The value of their ratio is $\Phi(500)/\Phi(980)=1.02 \pm 0.05.$

The value of the branching fraction  $\BR(f_0(500) \to \pi^+\pi^-) = 2/3$ is obtained from the isospin Clebsch-Gordan coefficients and assumes that there are  only contributions from $\pi\pi$ final states.
The ratio $\BR(f_0(980) \to K^+K^-)/\BR(f_0(980) \to \pi^+ \pi^-) = 0.35^{+0.15}_{-0.14}$, obtained from an average of the measurements by the \babar~\cite{BaBar980} and BES~\cite{BES980} collaborations, is used to estimate
the branching fraction $\BR(f_0(980) \to \pi^+\pi^-)$.
Assuming that the $\pi\pi$ and $KK$ decays are dominant in the  $f_0(980)$ decays, $\BR(f_0(980) \to \pi^+\pi^-) = 0.46 \pm 0.06$ is obtained.
This gives
\begin{equation*}
r^f = 0.177^{+0.066}_{-0.062},
\end{equation*}
taking into account the systematic uncertainties, as listed in Table~\ref{tab: sysmix}.

\begin{table}
\centering
\caption{Systematic uncertainties on $r^f$. The sum in quadrature of the uncertainties is also reported.}
\label{tab: sysmix}
\begin{tabular}{l.}
\hline
Source  & \multicolumn{1}{r}{$r^f$} \\
\hline
PID & 0.001\\
Trigger & 0.001\\
Reconstruction  & <0.001\\
Simulation statistic & 0.001\\
Background model  & 0.001\\
$D^*(2010)^-$ veto & 0.012\\
Other res. & 0.007\\
RBW parameters  & 0.008\\
$\pi\pi$ res. mass, width & 0.011\\
$f_0(500)$ model & 0.033\\
$f_0(980)$ model & 0.028\\
\hline
Total & 0.048\\
\hline
\end{tabular}
\end{table}

\begin{figure}[!htb]
\begin{minipage}{0.5 \linewidth}
\centerline{\includegraphics[width=1.0\linewidth]{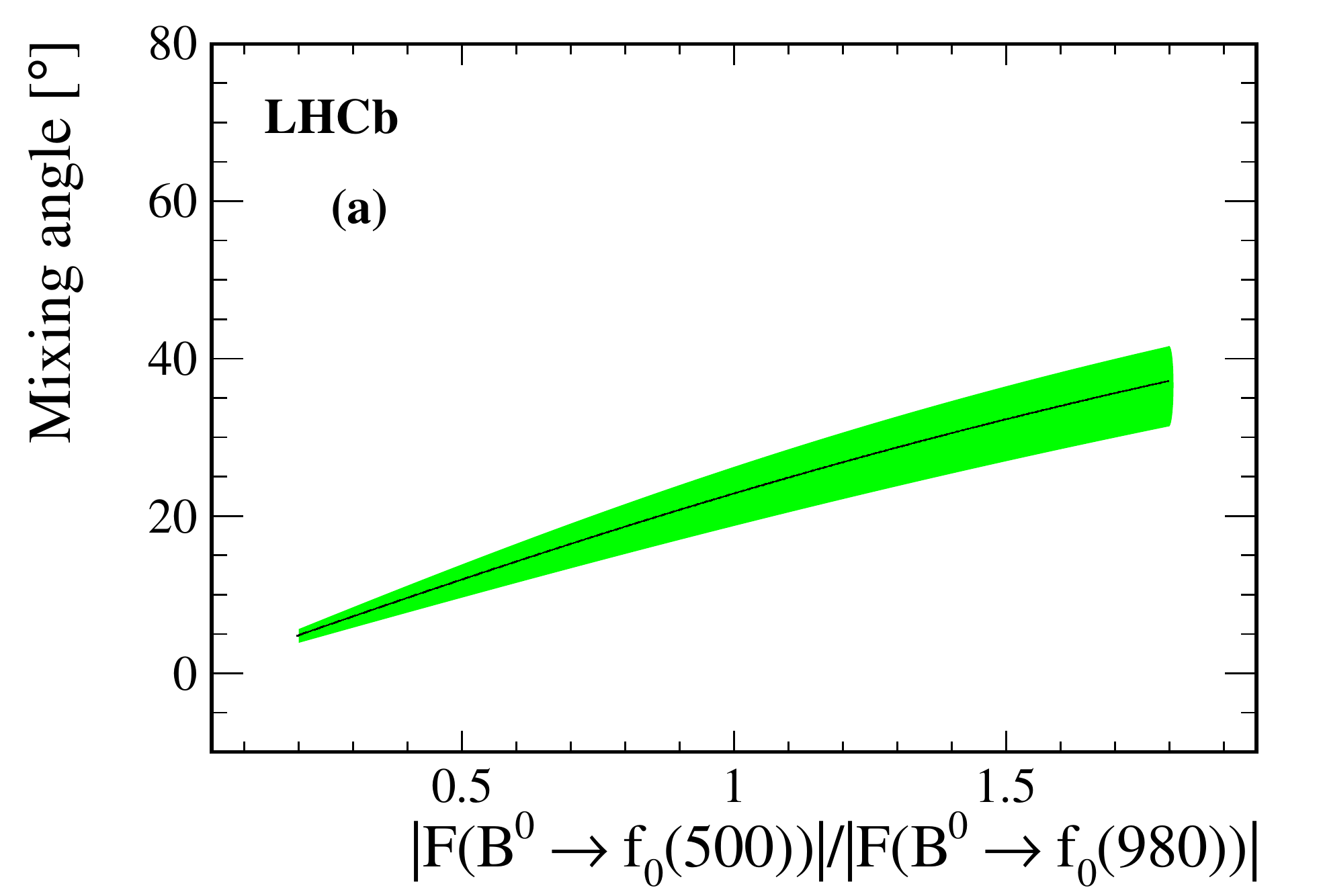}}
\end{minipage}
\begin{minipage}{0.5\linewidth}
\centerline{\includegraphics[width=1.0\linewidth]{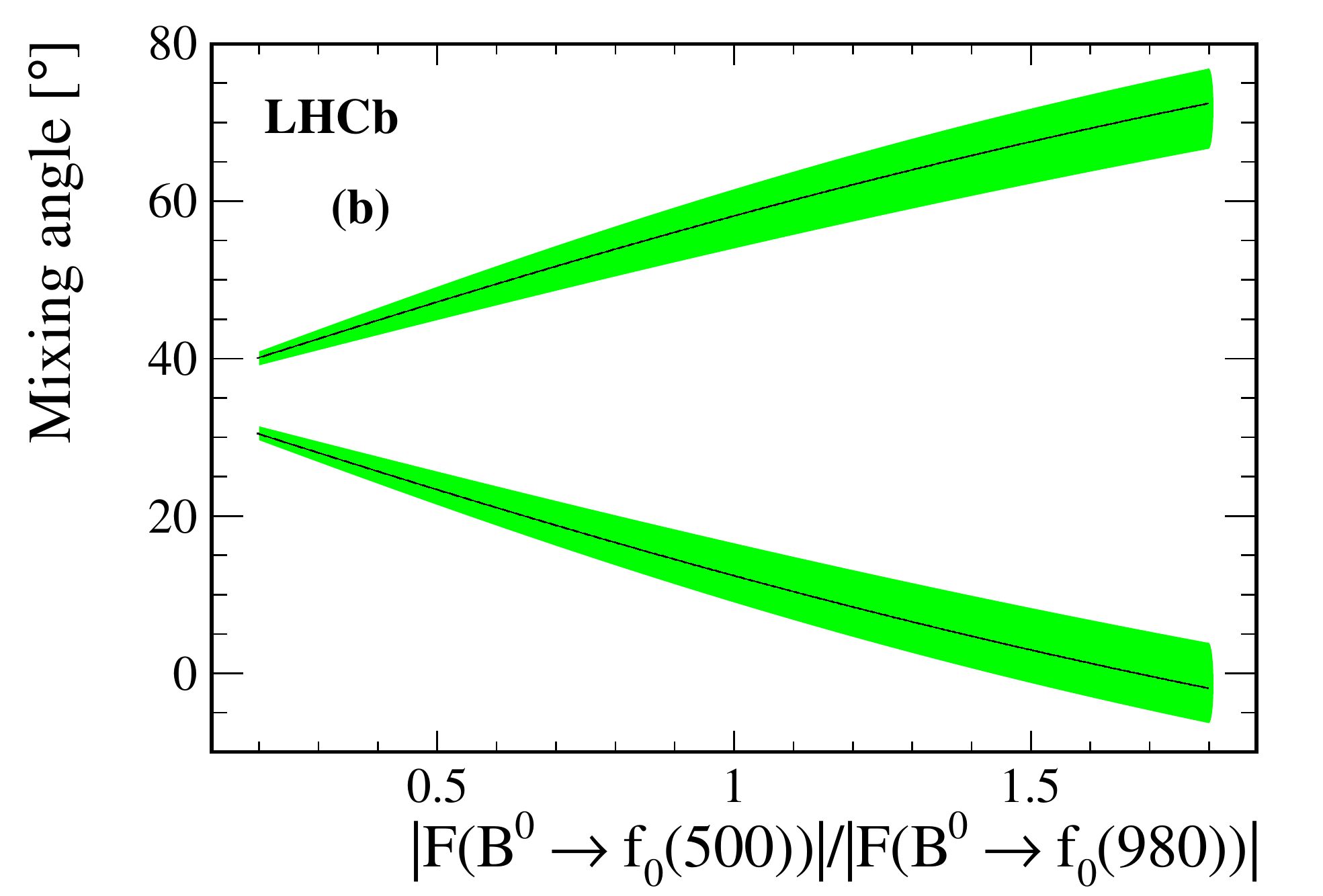}}
\end{minipage}
\caption{
Mixing angle as a function of form factor ratio for the (a) $q\bar{q}$ model and (b) $[qq'][\bar{q}\bar{q'}]$ tetraquark model.
Green band gives 1$\sigma$ interval around central values (black solid line).
}
\label{fig: mixangle}
\end{figure}

The parameter $r^f$ is related to the mixing angle by the equation
\begin{equation}
 r^f = \tan^2\varphi_{\textrm{mix}} \times \left| \frac{F(B^0 \to f_0(980))}{F(B^0 \to f_0(500))} \right|^2
 \end{equation}
 in the $q\bar{q}$ model and by
 \begin{equation}
 r^f = \left|\frac{1-\sqrt{2}\tan\omega_{\textrm{mix}}}{\tan\omega_{\textrm{mix}} + \sqrt{2}} \right|^2 \times \left| \frac{F(B^0 \to f_0(980))}{F(B^0 \to f_0(500))} \right|^2
 \end{equation}
 in the $[qq'][\bar{q}\bar{q'}]$ tetraquark model~\cite{Wang:2009azc,Li:2012sw}.
The form factors $F(B^0 \to f_0(980))$ and $F(B^0 \to f_0(500))$ are evaluated at the four-momentum transfer squared equal to the square of the $\Dzb$ mass.
 Finally, values of the mixing angles as a function of form factor ratio are obtained in Fig.~\ref{fig: mixangle} for the $q\bar{q}$ model and the $[qq'][\bar{q}\bar{q'}]$ tetraquark model.
Such angles have also been computed by LHCb for the decays $B^0_{(s)} \to J/\psi\pi^+\pi^-$~\cite{LHCb-PAPER-2012-045,LHCb-PAPER-2013-069,LHCb-PAPER-2014-012}.

The expectation is that the ratio of form factors should be close to unity.  However, LHCb has recently performed a search for the decay $B^0_s \to \Dzb f_0(980)$~\cite{LHCb-PAPER-2015-012}.  The limit set on this decay is below the value expected in a simple model based on our measured value of $\BR(B^0 \to \Dzb f_0(500))$ and assuming equal form factors.  More complicated models may be needed in order to explain all results.

The above discussion is one possible interpretation of the results. Another possible mechanism~\cite{rescattering1, rescattering2} involves the generation of pseudo-scalar resonances through the interactions of $\pi^+\pi^-$ mesons.

\subsection{Isospin analysis of the {\boldmath $B \to D \rho$} system}

The measured branching fraction of the $B^0 \to \Dzb \rho(770)^0$ decay, presented in Table~\ref{tab: finalBr}, can be used to perform an isospin analysis of the $B \to D \rho$ system. Isospin symmetry relates the amplitudes of the decays $B^+ \to \Dzb \rho(770)^+$, $B^0 \to D^- \rho(770)^+$, and   $B^0 \to \Dzb \rho(770)^0$, which can be written as linear combinations of the isospin eigenstates $A_{I}$ with  $I=1/2$ and $3/2$~\cite{Rosner:1999zm,Neubert:2001sj}
\begin{eqnarray}
A(\Dzb\rho^+) &=& \sqrt{3}A_{3/2}, \\ \nonumber
A(D^-\rho^+) &=& \sqrt{1/3}A_{3/2} + \sqrt{2/3}A_{1/2}, \\ \nonumber
A(\Dzb \rho^0) &=& \sqrt{2/3}A_{3/2} - \sqrt{1/3}A_{1/2},  \nonumber
\end{eqnarray}
leading to
\begin{equation}
A(\Dzb\rho^+)= A(D^-\rho^+)+\sqrt{2}A(\Dzb\rho^0).
\end{equation}
The strong phase difference between the amplitudes $A_{1/2}$ and $A_{3/2}$ is denoted by $\delta_{D\rho}$. Final-state interactions between the states $\Dzb \rho^0$  and  $D^-\rho^+$ may lead to a value of $\delta_{D\rho}$ different from zero and through constructive interference, to a larger value of $\BR(B^0 \to \Dzb \rho^0)$ than the prediction obtained within the factorisation approximation.  In the heavy-quark limit, the factorisation model predicts~\cite{Beneke:2000ry,Cheng:1998kd} $\delta_{D\rho}= {\cal O}(\Lambda_{\rm QCD}/m_b)$ and the amplitude ratio ${R_{D\rho} \equiv \frac{|A_{1/2}|}{\sqrt{2}|A_{3/2}|} = 1 +{\cal O}(\Lambda_{\rm QCD}/m_b)}$, where $m_b$ represents the $b$ quark mass and $\Lambda_{\rm QCD}$ the QCD scale. 

Using our measurement of $\BR(B^0 \to \Dzb \rho^0)$ together with the world average values of $\BR(B^0 \to D^- \rho^+)$, $\BR(B^+ \to \Dzb \rho^+)$, and the ratio of lifetimes $\tau(B^+)/\tau(B^0)$~\cite{PDG}, we obtain
\begin{equation}
R_{D\rho} = \sqrt{\frac{1}{2}}\left( \frac{3\,(\BR(D^-\rho^+) + \BR(\Dzb \rho^0))}{\BR(\Dzb \rho^+)}\times \frac{\tau_{B^+}}{\tau_{B^0}} - 1 \right)^{1/2}
\end{equation}
and
\begin{equation}
\cos\delta_{D\rho} = \frac{1}{4R_{D\rho}} \times \left( \frac{\tau_{B^+}}{\tau_{B^0}}\times \frac{3\,(\BR(D^-\rho^+) - 2\,\BR(\Dzb\rho^0))}{\BR(\Dzb\rho^+)} + 1  \right).
\end{equation}
With a frequentist statistical approach~\cite{Charles:2004jd}, $R_{D\rho}$ and $\cos\delta_{D\rho}$ are calculated for the Isobar and K-matrix models in Table~\ref{tab: Isospin}.
\begin{table}
\centering
\caption{Results of $R_{D\rho}$ and $\cos\delta_{D\rho}$.}
\label{tab: Isospin}
\begin{tabular}{lcc}
\hline
Model  & $R_{D\rho}$ & $\cos\delta_{D\rho}$\\
\hline
\vspace{0.1cm}
Isobar  & $0.69 \pm 0.15$ & $0.984^{+0.113}_{-0.048}$ \\
K-matrix & $0.69 \pm 0.15$ & $0.987^{+0.114}_{-0.048}$\\
[+0.3ex]
\hline
\end{tabular}
\end{table}
These results are not  significantly different from the predictions of factorisation models. As opposed to the theoretical expectations~\cite{Rosner:1999zm,Neubert:2001sj} and in contrast to the $B \to D^{(*)}\pi$ system~\cite{BaBarColSup}, non-factorisable final-state interaction effects do not introduce a sizeable phase difference between the isospin amplitudes in the $B \to D \rho$ system . The precision on $R_{D\rho}$ and $\cos\delta_{D\rho}$ is dominated by that of the branching fractions of the decays  $B^+ \to \Dzb \rho(770)^+$ (14\%) and $B^0 \to D^- \rho(770)^+$ (17\%)~\cite{PDG}.
The precision of the branching fraction of the $B^0 \to \Dzb \rho(770)^0$ decay is 7.3\% (9.2\%)  for the Isobar (K-matrix) model (see Table~\ref{tab: finalBr}).

\section{Conclusion}
\label{sec:conclusion}

A Dalitz plot analysis of the $\btodpipi$  decay is presented. The decay model contains four components from $\Dzb \pi^-$ resonances, four P-wave $\pi^+\pi^-$ resonances and one D-wave $\pi^+\pi^-$ resonance. Two models are used to describe the S-wave $\pi^+\pi^-$ resonances. The Isobar model uses four components, including the $f_0(500)$, $f_0(980)$, $f_0(2020)$ resonances and a nonresonant contribution. The K-matrix approach describes the $\pi^+\pi^-$ S-wave using a $5 \times 5$ scattering matrix with a production vector. The overall branching fraction of $\btodpipi$  and quasi-two-body decays are measured. Significant contributions from the $f_0(500)$, $f_0(980)$, $\rho(1450)$ and $D_3^*(2760)^-$ mesons are observed for the first time. For the latter, this is a confirmation of the observation from previous inclusive measurements, and the spin-parity of this resonance is determined for the first time to be $J^P=3^-$. This suggests a spectroscopic assignment of $^3D_3$, and shows that the $1D$ family of charm resonances can be explored in Dalitz plot analysis of $B$-meson decays in the same way as recently seen for the charm-strange resonances~\cite{LHCb-PAPER-2014-035,LHCb-PAPER-2014-036}. Evidence for the $f_0(2020)$ meson is also seen for the first time. The measured branching fractions of two-body decays are more precise than the existing world averages and there is good agreement between values from the Isobar and K-matrix models.

The masses and widths of the $\Dzb \pi^-$ resonances are also determined. The measured masses and widths of the $D_0^*(2400)^-$ and $D_3^*(2760)^-$ states are consistent with the previous measurements. 
The precision on the $D_0^*(2400)^-$ meson is much improved.
For the measurement on the mass and width of the $D_2^*(2460)^-$ meson,  the broad $D_0^*(2400)^-$ component was excluded  from the fit model in the former LHCb inclusive analysis~\cite{LHCb-PAPER-2013-026}, due to a high correlation with the background lineshape parameters, while here it is included. The present result therefore supersedes the former measurement. 

The significant contributions found for both the $f_0(500)$ and $f_0(980)$ allow us to constrain on  the mixing angle between the $f_0(500)$ and $f_0(980)$ resonances. An isospin analysis in the $\btodrho$ decays using our improved measurement of the branching fraction of the decay $\btodrhoz$ is performed, indicating that non-factorisable effects from final-state interactions are limited in the $D\rho$ system.

\section*{Acknowledgements}
\noindent We express our gratitude to our colleagues in the CERN
accelerator departments for the excellent performance of the LHC. We
thank the technical and administrative staff at the LHCb
institutes. We acknowledge support from CERN and from the national
agencies: CAPES, CNPq, FAPERJ and FINEP (Brazil); NSFC (China);
CNRS/IN2P3 (France); BMBF, DFG, HGF and MPG (Germany); INFN (Italy);
FOM and NWO (The Netherlands); MNiSW and NCN (Poland); MEN/IFA (Romania);
MinES and FANO (Russia); MinECo (Spain); SNSF and SER (Switzerland);
NASU (Ukraine); STFC (United Kingdom); NSF (USA).
The Tier1 computing centres are supported by IN2P3 (France), KIT and BMBF
(Germany), INFN (Italy), NWO and SURF (The Netherlands), PIC (Spain), GridPP
(United Kingdom).
We are indebted to the communities behind the multiple open
source software packages on which we depend. We are also thankful for the
computing resources and the access to software R\&D tools provided by Yandex LLC (Russia).
Individual groups or members have received support from
EPLANET, Marie Sk\l{}odowska-Curie Actions and ERC (European Union),
Conseil g\'{e}n\'{e}ral de Haute-Savoie, Labex ENIGMASS and OCEVU,
R\'{e}gion Auvergne (France), RFBR (Russia), XuntaGal and GENCAT (Spain), Royal Society and Royal
Commission for the Exhibition of 1851 (United Kingdom).

\newpage
{\noindent\bf\Large Appendices}

\appendix
\section{Unnormalised Legendre polynomial weighted moments}
\label{sec:legendre}
\begin{figure}[!tbh]
\centerline{\includegraphics[width = 1.05\linewidth]{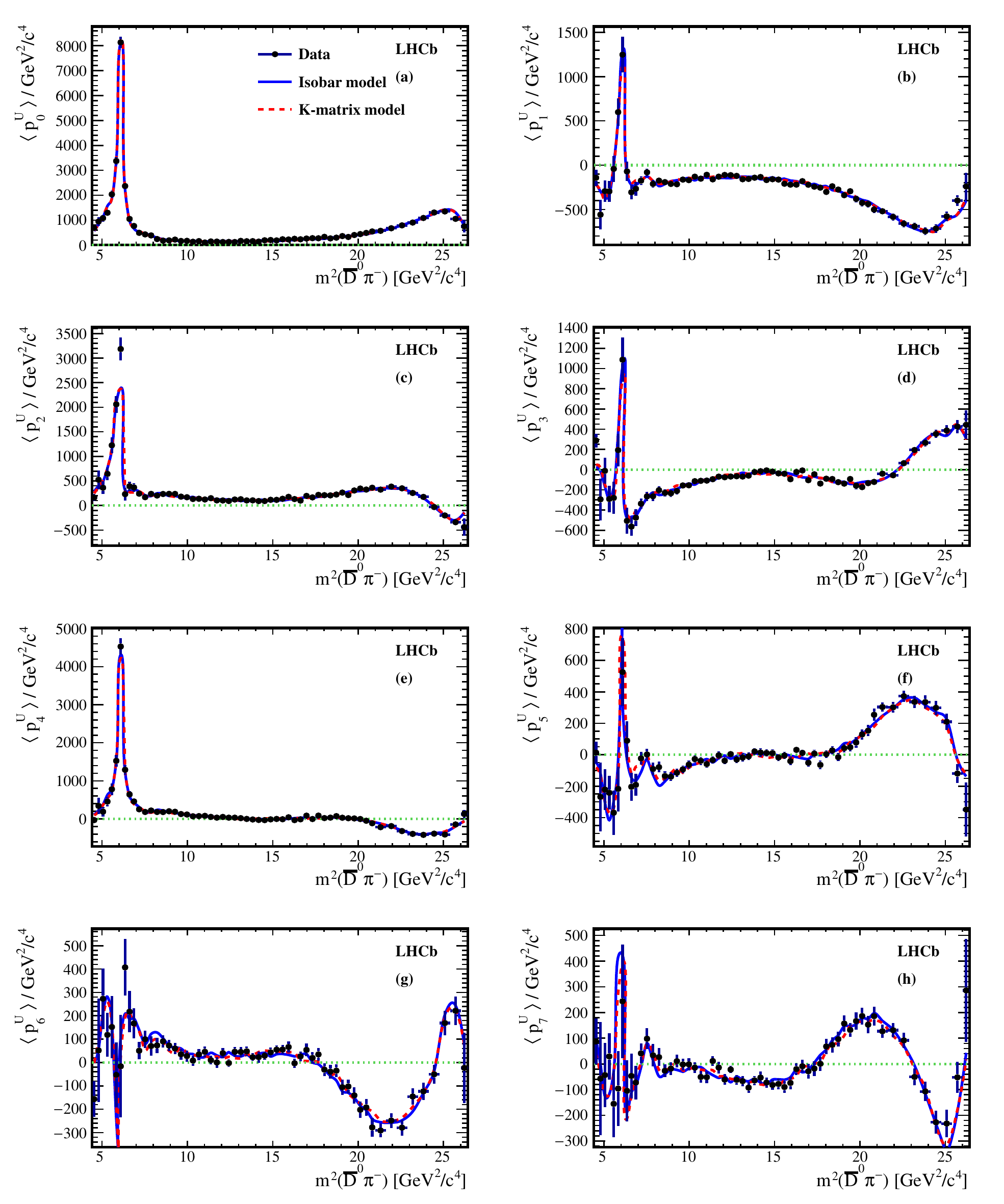}}
\caption{
The first eight unnormalised Legendre polynomial weighted moments (0 to 7 correspond to (a) to (h)) for background-subtracted and efficiency-corrected $B^0 \to \Dzb \pi^+\pi^-$ data and the Dalitz plot fit results as a function of $m^2(\Dzb\pi^-)$. }
\label{fig: Poly13}
\end{figure}
\begin{figure}
\centerline{\includegraphics[width = 1.05\linewidth]{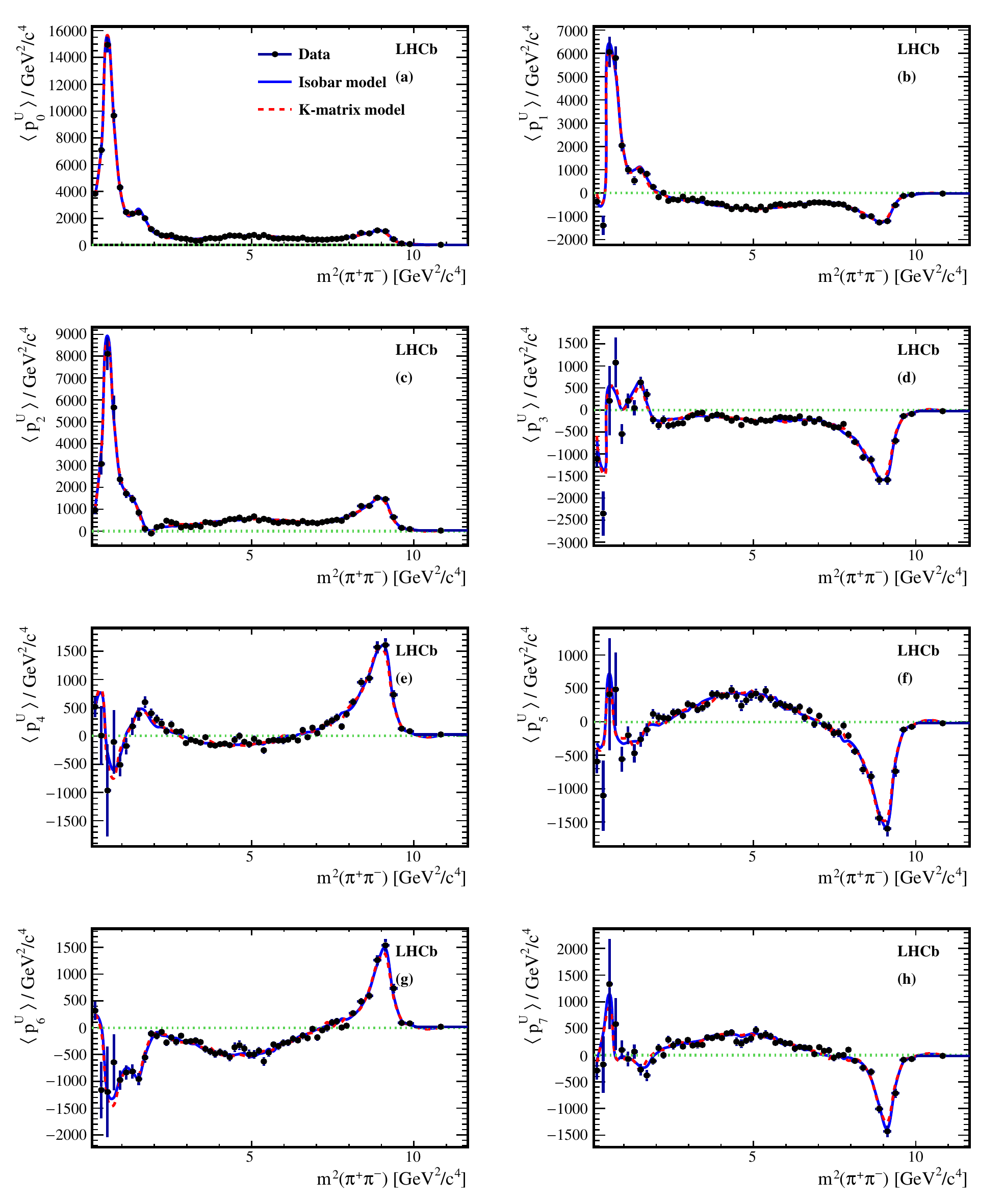}}
\caption{
The first eight unnormalised Legendre polynomial weighted moments (0 to 7 correspond to (a) to (h)) for background-subtracted and efficiency-corrected $B^0 \to \Dzb \pi^+\pi^-$ data and the Dalitz plot fit results as a function of $m^2(\pi^+\pi^-)$. }
\label{fig: Poly23}
\end{figure}

\begin{figure}[!tbh]
\centerline{\includegraphics[width = 1.05\linewidth]{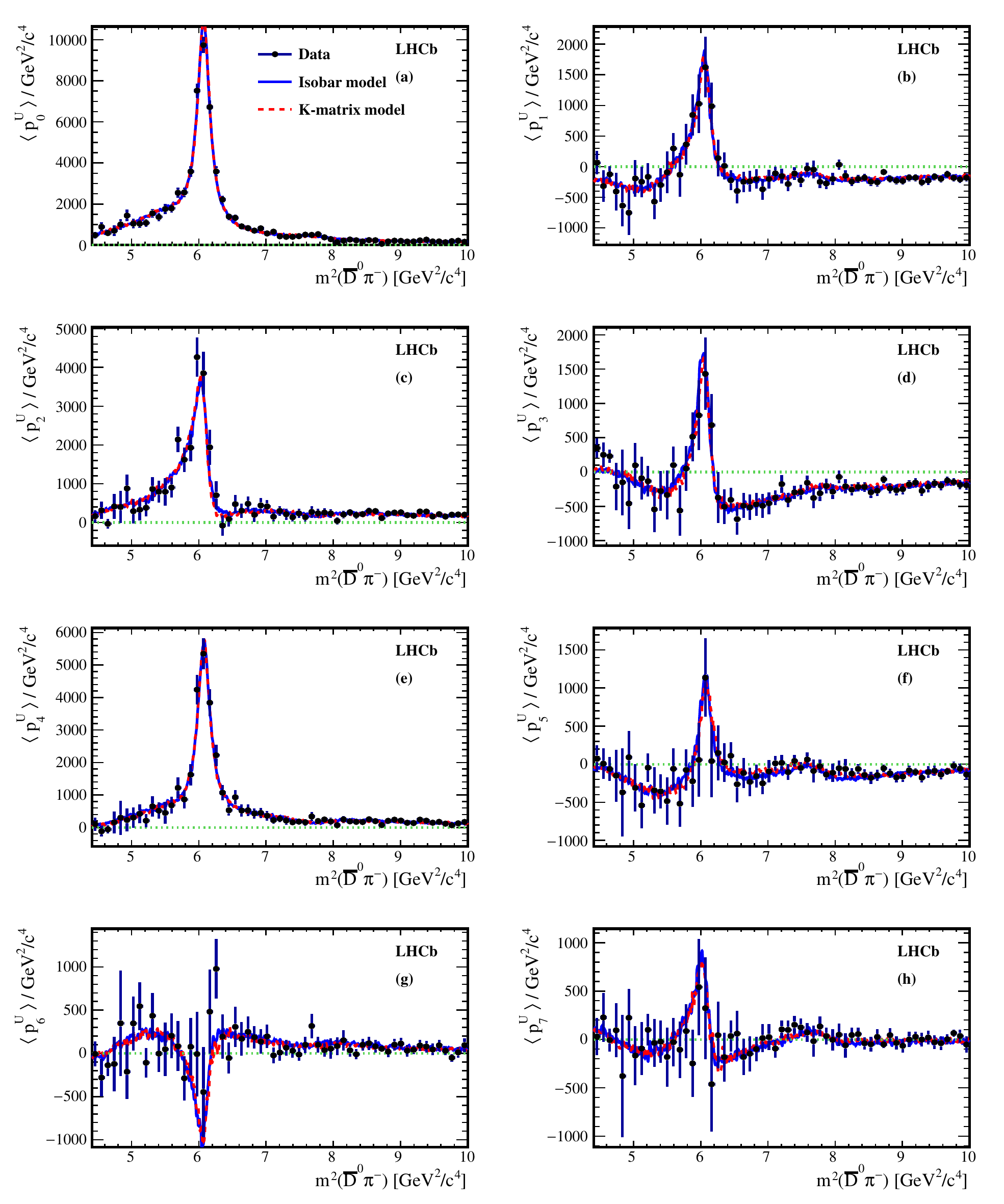}}
\caption{
The first eight unnormalised Legendre polynomial weighted moments (0 to 7 correspond to (a) to (h)) for background-subtracted and efficiency-corrected $B^0 \to \Dzb \pi^+\pi^-$ data and the Dalitz plot fit results as a function of $m^2(\Dzb\pi^-)$.
Only results in the region $m^2(\Dzb\pi^-)< 10$ GeV$^2$/$c^4$ are shown.
}
\label{fig: Poly13Sub}
\end{figure}
\begin{figure}
\centerline{\includegraphics[width = 1.05\linewidth]{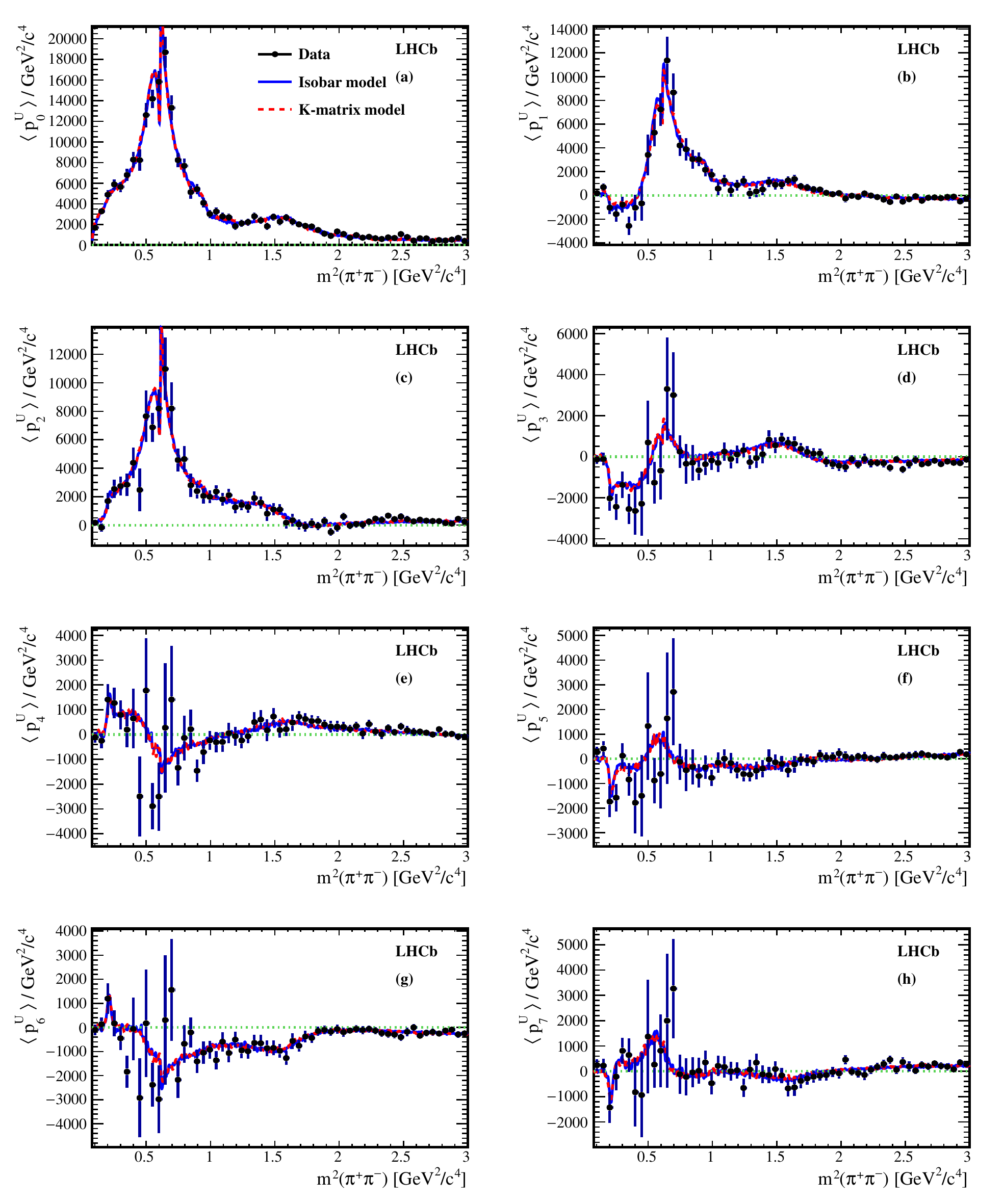}}
\caption{
The first eight unnormalised Legendre polynomial weighted moments (0 to 7 correspond to (a) to (h)) for background-subtracted and efficiency-corrected $B^0 \to \Dzb \pi^+\pi^-$ data and the Dalitz plot fit results as a function of $m^2(\pi^+\pi^-)$.
Only results in the region $m^2(\pi^+\pi^-)< 3$ GeV$^2$/$c^4$ are shown.
}
\label{fig: Poly23Sub}
\end{figure}

Figures~\ref{fig: Poly13} and~\ref{fig: Poly23} show the distributions of the unnormalised Legendre polynomial weighted moments $<p_L^U>$ which display the contributions of resonances with spin larger than $L/2$.
The $\rho(770)$ resonance can clearly be seen in the distributions with $L \leq 2$ and the $D^*_2(2460)^-$ resonance in the distributions with $L\leq 4$.
Figures~\ref{fig: Poly13Sub} and~\ref{fig: Poly23Sub} display an expanded version in low mass regions.
The distributions from the Isobar and the K-matrix models are compatible with those from data.

\clearpage

\section{Systematic uncertainties on the parameters in the Dalitz plot analysis}
\label{sec:Systematicuncertainties}

\subsection{Systematic uncertainties for the Isobar model}
\begin{table}[!htb]
\centering
\footnotesize
\caption{Systematic uncertainties on the $\dbpim$ resonant masses (MeV/$c^2$) and widths (MeV) for the Isobar model. \label{tab:IsoMW}}
\begin{tabular}{l......}
\hline
\multirow{2}{2cm}{\mbox{Source}}   &  \multicolumn{2}{c}{$D^*_0(2400)^-$}   &  \multicolumn{2}{c}{$D^*_2(2460)^-$}  &  \multicolumn{2}{c}{$D^*_3(2760)^-$}  \\\cline{2-7}
 &  \multicolumn{1}{c}{$\Gamma_0$}  &  \multicolumn{1}{c}{$m_0$}  &  \multicolumn{1}{c}{$\Gamma_0$}  &  \multicolumn{1}{c}{$m_0$}  &  \multicolumn{1}{c}{$\Gamma_0$}  &  \multicolumn{1}{c}{$m_0$} \\
\hline
PID                   &  1.9  &  0.5  &  <0.1  &  <0.1  &  1.2  &  0.3  \\
Trigger               &  0.5  &  0.2  &   0.2  &  <0.1  &  1.9  &  0.7  \\
Reconstruction        &  0.2  &  0.1  &  <0.1  &  <0.1  &  0.5  &  0.1  \\
Simulation statistic  &  0.6  &  0.1  &  0.1  &  <0.1  &  0.6  &  0.1  \\
Background model      &  1.5  &  0.6  &  0.1  &  <0.1  &  3.4  &  0.4 \\
$D^*(2010)^-$ veto    &  4.4  &  0.8  &  <0.1  &  <0.1  &  4.6  &  0.4 \\ \hline
Total (experiment)    &  5.1  &  1.1  &  0.3  &  <0.1  &  6.2  &  1.0 \\
\hline
Additional resonances  &  10.7  &  0.4  &  0.1   &  0.1  &  21.0  &  5.1 \\
RBW parameters         &  0.1  &  1.9  &  0.1   &  <0.1  &  1.0  &  1.5 \\
$\pi\pi$ res. mass, width  &  3.4  &  0.1  &  0.1   & 0.1  &  2.9  &  1.3 \\
$B^0$, $D^0$ mass  &  0.2   &  0.2  &  0.2  &  0.2   &  0.2   &  0.2  \\
$f_0(500)$ model  &  2.3  &  3.6  &  0.5  &  <0.1  &  9.0  &  3.5  \\
$f_0(980)$ model  &  2.8  &  0.7  &  0.2  &  0.1  &  3.5  &  1.2  \\
\hline
Total (model)  &  11.8  &  4.2  &  0.6  &  0.3  &  23.3  &  6.6 \\
\hline
Total (all)  &  12.9  &  4.3  &  0.7  &  0.3  &  24.1  &  6.7 \\
\hline
\end{tabular}
\end{table}

\begin{table}
\centering
\footnotesize
\caption{Systematic uncertainties on the moduli of the complex coefficients of the resonant contributions for the Isobar model.
The moduli are normalised to that of $\rho(770)$. \label{tab:IsoAmp}}
\begin{tabular}{l......} 
\hline
Source   &   \multicolumn{1}{c}{Nonres.}   &   \multicolumn{1}{c}{$f_0(500)$}   &   \multicolumn{1}{c}{$\omega(782)$}  &  \multicolumn{1}{c}{$f_0(980)$} &  \multicolumn{1}{c}{$f_2(1270)$}  &  \multicolumn{1}{c}{$\rho(1450)$}  \\
\hline
PID  &  0.02  &  0.15  & <0.01 &  0.03  &  <0.001  &  0.01 \\
Trigger  &  0.02  &  0.03  & <0.01 &  0.01  &  <0.001  & <0.01 \\
Reconstruction  & <0.01 &  0.01  & <0.01 & <0.01 &  <0.001  & <0.01 \\
Simulation statistic  & <0.01 &  0.11  & <0.01 &  0.01  &  <0.001  & <0.01 \\
Background model  & <0.01 &  0.07  & <0.01 &  0.02  &  <0.001  & <0.01 \\
$D^*(2010)^-$ veto  &  0.03  &  0.20  & <0.01 &  0.08  &  0.002  &  0.01 \\
\hline
Total (experiment)  &  0.04  &  0.29  & <0.01 &  0.09  &  <0.001  &  0.01 \\
\hline
Additional resonances  &  0.34 &  0.61 & <0.01&  0.03 &  0.003 & <0.01 \\
RBW parameters  &  0.11 &  0.10 &  0.01 &  0.04 &  0.001 &  0.01 \\
$\pi\pi$ res. mass, width  &  0.06 &  0.41 & <0.01&  0.03 &  0.001 & <0.01 \\
$f_0(500)$ model  &  0.36 &   \multicolumn{1}{c}{\textrm{\al\all n/a}}   & <0.01&  0.46 &  0.004 &  0.01 \\
$f_0(980)$ model  &  0.01 &  0.30 & <0.01& \multicolumn{1}{c}{\textrm{\al\all n/a}} &  <0.001 &  0.01 \\
\hline
Total (model)  &  0.51  &  0.80  &  0.01  &  0.46  &  0.005  &  0.02 \\
\hline
Total  &  0.51  &  0.85  &  0.01  &  0.47  &  0.006  &  0.02 \\
\hline
 &  &  &  &  &  & \\
\hline
Source  &  \multicolumn{1}{c}{$\rho(1700)$}  &  \multicolumn{1}{c}{$f_0(2020)$}  &  \multicolumn{1}{c}{$\Dzb\pi^-$ P-wave}  &  \multicolumn{1}{c}{$D^*_0(2400)^-$}  &  \multicolumn{1}{c}{$D^*_2(2460)^-$}  &  \multicolumn{1}{c}{$D^*_3(2760)^-$} \\
\hline
PID  &  <0.001 &  0.06 &  0.14 &  0.21 & <0.01&  <0.001 \\
Trigger   &  0.001 &  0.02 &  0.17 &  0.08 &  0.02 &  <0.001 \\
Reconstruction  &  <0.001 & <0.01&  0.01 &  0.02 & <0.01&  <0.001 \\
Simulation statistic  &  0.001 &  0.02 &  0.10 &  0.03 &  0.01 &  <0.001 \\
Background model  &  0.002 &  0.01 &  0.03 &  0.06 & <0.01&  0.002 \\
$D^*(2010)^-$ veto  &  0.006 &  0.20 &  \multicolumn{1}{c}{\textrm{\al\all n/a}}  &  0.20 & <0.01&  0.002 \\
\hline
Total (experiment)  &  0.006  &  0.21  &  0.25  &  0.31  &  0.02  &  0.003 \\
\hline
Additional resonances  &  0.005 &  0.37 &  1.74 &  0.47 &  0.01 &  0.007 \\
RBW parameters  &  0.002 &  0.27 &  0.05 &  0.12 &  0.01 &  <0.001 \\
$\pi\pi$ res. mass, width  &  0.002 &  0.73 &  0.16 &  0.19 & <0.01 &  0.001 \\
$f_0(500)$ model  &  0.004 &  1.56 &  0.62 &  0.20 &  0.01 &  0.003 \\
$f_0(980)$ model  &  0.003 &  0.06 &  0.06 &  0.18 &  0.01 &  0.002 \\
\hline
Total (model)  &  0.008  &  1.78  &  1.86  &  0.59  &  0.02  &  0.008\\
\hline
Total (all) &  0.010  &  1.80  &  1.87  &  0.66  &  0.03  &  0.008\\
\hline
\end{tabular}
\end{table}

\begin{table}
\centering
\footnotesize
\caption{Systematic uncertainties on the phases ($^{\circ}$) of the complex coefficients of the resonant contributions for the Isobar model.
The phase of $\rho(700)$ is set to $0^{\circ}$ as the reference. \label{tab:IsoPha}}
\begin{tabular}{l......} 
\hline
Source  &  \multicolumn{1}{c}{Nonres.}  &  \multicolumn{1}{c}{$f_0(500)$}  &  \multicolumn{1}{c}{$\omega(782)$}  &  \multicolumn{1}{c}{$f_0(980)$} &  \multicolumn{1}{c}{$f_2(1270)$}  &  \multicolumn{1}{c}{$\rho(1450)$}  \\
\hline
PID  &  0.2 &  0.1 &  0.4 &  0.1 &  0.1 &  0.1 \\
Trigger  &  1.1 &  0.3 &  0.1 &  0.1 &  0.6 &  0.2 \\
Reconstruction  & <0.1& <0.1& <0.1& <0.1&  0.1 &  0.1 \\
Simulation statistic  &  0.3 &  0.1 &  0.3 &  0.2 &  0.3 &  0.3 \\
Background model  &  0.7 &  0.1 & <0.1&  0.4 &  0.4 &  0.4 \\
$D^*(2010)^-$ veto  &  1.8 &  1.2 &  0.2 &  1.4 &  1.4 &  4.8 \\
\hline
Total (experiment)  &  2.3  &  1.3  &  0.6  &  1.5  &  1.6  &  4.8 \\
\hline
Additional resonances  &  3.0 &  2.8 & <0.1&  2.7 &  1.6 &  1.1 \\
RBW parameters  &  0.3 &  1.1 &  0.1 &  1.3 & <0.1&  0.3\\
$\pi\pi$ res. mass, width  &  2.3 &  0.1 &  0.3 &  0.4 &  1.2 &  2.2 \\
$f_0(500)$ model  &  3.7 &  \multicolumn{1}{c}{\textrm{\all n/a}}  &  0.3 &  10.5 &  3.0 &  3.6 \\
$f_0(980)$ model  &  1.0 &  2.1 &  0.2 &  \multicolumn{1}{c}{\textrm{\all n/a}}  &  1.1 &  1.1 \\
\hline
Total (model)  &  5.4  &  3.7  &  0.5  &  10.9  &  3.8  &  4.5 \\
\hline
Total (all)  &  5.8  &  3.9  &  0.7  &  11.0  &  4.1  &  6.6 \\
\hline
 &  &  &  &  &  & \\
\hline
Source   &  \multicolumn{1}{c}{$\rho(1700)$}  &  \multicolumn{1}{c}{$f_0(2020)$}  &  \multicolumn{1}{c}{$\Dzb\pi^-$ P-wave}  &  \multicolumn{1}{c}{$D^*_0(2400)^-$}  &  \multicolumn{1}{c}{$D^*_2(2460)^-$}  &  \multicolumn{1}{c}{$D^*_3(2760)^-$} \\
\hline
PID  &  0.4 & <0.1&  0.2 &  0.3 &  0.1 &  0.2 \\
Trigger   &  0.5 &  0.2 &  0.2 &  0.7 & <0.1&  0.6 \\
Reconstruction  & <0.1&  0.1 &  0.1 &  0.1 &  0.1 &  0.1 \\
Simulation statistic  &  0.7 &  0.2 &  0.1 &  0.2 &  0.1 &  0.1 \\
Background model  &  0.6 &  0.9 &  0.1 & <0.1&  0.2 &  0.3 \\
$D^*(2010)^-$ veto  &  4.4 &  0.6 &  \multicolumn{1}{c}{\textrm{\all n/a}}  &  2.7 &  0.8 &  1.2 \\
\hline
Total (experiment)  &  4.5  &  1.1  &  0.3  &  2.8  &  0.8  &  1.4 \\
\hline
Additional resonances  &  0.3 &  3.4 &  5.7 &  1.8 &  1.2 &  3.9 \\
RBW parameters  &  1.0 &  1.1 &  0.6 &  1.6 &  0.7 &  1.8 \\
$\pi\pi$ res. mass, width  &  0.5 &  1.3 &  0.5 &  0.3 &  0.7 &  2.0 \\
$f_0(500)$ model  &  0.3 &  26.5 &  4.2 &  3.9 &  2.4 &  1.9 \\
$f_0(980)$ model  &  2.1 &  1.8 &  0.4 &  0.1 &  0.9 &  0.2 \\
\hline
Total (model)  &  2.4  &  26.8  &  7.1  &  4.6  &  3.0  &  5.1 \\
\hline
Total (all)  &  5.1  &  26.9  &  7.1  &  5.4  &  3.1  &  5.3 \\
\hline
\end{tabular}
\end{table}

\begin{table}
\centering
\scriptsize
\caption{Systematic uncertainties on the fit fractions (\%) of the resonant contributions for the Isobar model. \label{tab:IsoFraction}}
\begin{tabular}{l.......} 
\hline
Source  &  \multicolumn{1}{c}{Nonres.}  &  \multicolumn{1}{c}{$f_0(500)$}  &  \multicolumn{1}{c}{$\rho(770)$}  &  \multicolumn{1}{c}{$\omega(782)$}  &  \multicolumn{1}{c}{$f_0(980)$} &  \multicolumn{1}{c}{$f_2(1270)$}  &  \multicolumn{1}{c}{$\rho(1450)$}  \\
\hline
PID  &  0.02 &  0.23 &  0.37 &  0.01 &  0.04 &  0.07 &  0.06 \\
Trigger  &  0.01 &  0.03 &  0.09 & <0.01& <0.01&  0.05 &  0.03 \\
Reconstruction  &  0.01 &  0.09 &  0.39 &  0.01 &  0.01 &  0.01 & <0.01 \\
Simulation statistic  & <0.01&  0.09 &  0.26 & <0.01&  0.01 &  0.04 & <0.01 \\
Background model  &  0.01 &  0.08 &  0.07 & <0.01&  0.01 &  0.05 &  0.04 \\
$D^*(2010)^-$ veto  & 0.06 & 0.15 & 0.01 &<0.01& 0.10 & 0.29 & 0.03 \\
\hline
Total (experiment)  &  0.07  &  0.31  &  0.61  &  0.01  &  0.11  &  0.31  &  0.08 \\
\hline
Additional resonances  &  0.51 &  0.57 &  0.81 &  0.02 & <0.01&  0.53 &  0.01 \\
RBW parameters  &  0.21 &  0.03 &  0.50 &  0.01 &  0.07 &  0.10 &  0.15 \\
$\pi\pi$ res. mass, width  &  0.09 &  0.60 &  0.08 & <0.01&  0.03 &  0.16 & <0.01 \\
$f_0(500)$ model  &  0.57 &  2.25 &  0.04 &  0.01 &  0.50 &  0.94 &  0.14 \\
$f_0(980)$ model  &  0.03 &  0.48 &  0.19 &  0.01 &  0.20 &  0.11 &  0.09 \\
\hline
Total (model)  &  0.80  &  2.45  &  0.98  &  0.03  &  0.54  &  1.10  &  0.22 \\
\hline
Total (all)  &  0.80  &  2.47  &  1.15  &  0.03  &  0.56  &  1.14  &  0.24 \\
\hline
 &  &  &  &  &  &  & \\
\hline
Source   &  \multicolumn{1}{c}{$\rho(1700)$}  &  \multicolumn{1}{c}{$f_0(2020)$}  &  \multicolumn{1}{c}{S-wave}  &  \multicolumn{1}{c}{$\Dzb\pi^-$ P-wave}  &  \multicolumn{1}{c}{$D^*_0(2400)^-$}  &  \multicolumn{1}{c}{$D^*_2(2460)^-$}  &  \multicolumn{1}{c}{$D^*_3(2760)^-$} \\
\hline
PID  &  0.01 &  0.01 &  0.31 &  0.19 &  0.12 &  0.43 &  0.02 \\
Trigger  & <0.01&  0.03 &  0.04 &  0.02 &  0.06 &  0.28 &  0.01 \\
Reconstruction  & <0.01&  0.01 &  0.12 &  0.13 &  0.14 &  0.18 &  0.02 \\
Simulation statistic  & <0.01&  0.01 &  0.05 &  0.06 &  0.04 &  0.09 &  0.01 \\
Background model   &  0.02 &  0.01 &  0.04 &  0.05 &  0.03 &  0.01 &  0.02 \\
$D^*(2010)^-$ veto  & 0.07 & 0.14 & 0.26 &  \multicolumn{1}{c}{\textrm{\al\all n/a}}  & 0.01 & 0.50 & 0.06 \\
\hline
Total (experiment)  &  0.07  &  0.15  &  0.43  &  0.24  &  0.20  &  0.74  &  0.07 \\
\hline
Additional resonances  &  0.04 &  0.24 &  1.43 &  1.60 &  0.01 &  0.01 &  0.07 \\
RBW parameters  &  0.02 &  0.21 &  0.07 &  0.17 &  0.28 &  0.44 &  0.02 \\
$\pi\pi$ res. mass, width  &  0.02 &  0.22 &  0.05 &  0.17 &  0.13 &  0.02 & 0.02 \\
$f_0(500)$ model  &  0.03 &  0.92 &  0.25 &  0.60 &  0.13 &  0.22 &  0.03 \\
$f_0(980)$ model  &  0.02 &  0.04 &  0.14 &  0.01 &  0.10 &  0.09 &  0.03 \\
\hline
Total (model)  &  0.06  &  1.00  &  1.46  &  1.73  &  0.35  &  0.50  &  0.09 \\
\hline
Total (all)  &  0.10  &  1.01  &  1.52  &  1.74  &  0.40  &  0.90  &  0.11 \\
\hline
\end{tabular}
\end{table}

\clearpage

\subsection{Systematic uncertainties for the K-matrix model}

\begin{table}[!htb]
\centering
\footnotesize
\caption{Systematic uncertainties on the $\dbpim$ resonant masses (MeV/$c^2$) and widths (MeV) for the K-matrix model. \label{tab:KMaMW}}
\begin{tabular}{l......} 
\hline
\multirow{2}{2cm}{\mbox{Source}}   &  \multicolumn{2}{c}{$D^*_0(2400)^-$}   &  \multicolumn{2}{c}{$D^*_2(2460)^-$}  &  \multicolumn{2}{c}{$D^*_3(2760)^-$}  \\\cline{2-7}
 &  \multicolumn{1}{c}{$\Gamma_0$}  &  \multicolumn{1}{c}{$m_0$}  &  \multicolumn{1}{c}{$\Gamma_0$}  &  \multicolumn{1}{c}{$m_0$}  &  \multicolumn{1}{c}{$\Gamma_0$}  &  \multicolumn{1}{c}{$m_0$} \\
\hline
PID  &  8.9  &  5.9   &  1.0  &  0.3  &  1.1  &  8.6 \\
Trigger  &  1.1  &  0.6   &  0.1  &  <0.1  &  0.1  &  0.1 \\
Reconstruction  &  <0.1  &  <0.1  &  <0.1  &  <0.1  &  <0.1  &  <0.1   \\
Simulation statistic  &  4.6  &  0.5   &  0.2  &  0.1  &  2.4  &  0.4 \\
Background model  &  2.3  &  0.9   &  0.1  &  <0.1  &  2.7  &  0.8 \\
$D^*(2010)^-$ veto  &  14.9  &  9.0   &  1.4  &  0.3  &  12.1  &  5.4 \\
\hline
Total (experiment)  &  18.1  &  10.8  &  1.7  &  0.4  &  12.7  &  10.2\\
\hline
Additional resonances  &  8.5  &  1.0   &  0.3  &  0.2  &  1.4  &  1.1 \\
RBW parameters  &  6.0  &  1.5   &  <0.1  &  0.1  &  8.9  &  2.3  \\
$\pi\pi$ res. mass, width  &  1.4  &  0.9  &  0.1  &  0.1  &  1.3  &  0.1  \\
$B^0$, $D^0$ mass  &  0.2   &  0.2   &  0.2   &  0.2   &  0.2   &  0.2 \\
\hline
Total (model)  &  10.5  &  2.0  &  0.4  &  0.3  &  9.1  &  2.6 \\
\hline
Total (all)  &  21.0   &  11.0   &  1.8   &  0.5   &  15.6   &  10.5 \\
\hline
\end{tabular}
\end{table}

\begin{table}
\centering
\footnotesize
\caption{Systematic uncertainties on the moduli of the complex coefficients of the resonant contributions for the K-matrix model. The moduli are normalised to that of $\rho(770)$.}
\label{tab:KMaAmp}
\begin{tabular}{l....} 
\hline
Source  &  \multicolumn{1}{c}{$\omega(782)$}  &  \multicolumn{1}{c}{$f_2(1270)$}  &  \multicolumn{1}{c}{$\rho(1450)$}  &  \multicolumn{1}{c}{$\rho(1700)$} \\
\hline
PID  & <0.01&  0.001 &  0.04 &  0.037 \\
Trigger  & <0.01&  <0.001 & <0.01&  0.002 \\
Reconstruction  & <0.01&  <0.001 & <0.01&  <0.001 \\
Simulation statistic  & <0.01&  0.003 &  0.02 &  0.004 \\
Background model  & <0.01&  <0.001 & <0.01&  0.001 \\
$D^*(2010)^-$ veto  &  0.01 &  0.005 &  0.07 &  0.067 \\
\hline
Total (experiment)  &  0.01  &  0.006  &  0.08  &  0.077 \\
\hline
Additional resonances   & <0.01&  0.003 & <0.01&  0.003 \\
RBW parameters  &  0.01 &  <0.001 &  0.01 &  0.003 \\
$\pi\pi$ res. mass, width  & <0.01&  <0.001 & <0.01&  0.010 \\
\hline
Total (model)  &  0.01  &  0.003  &  0.01  &  0.011 \\
\hline
Total (all)  &  0.01  &  0.007  &  0.08  &  0.077 \\
\hline
 &  &  &  & \\
\hline
Source   &  \multicolumn{1}{c}{$\Dzb\pi^-$ P-wave}  &  \multicolumn{1}{c}{$D^*_0(2400)^-$}  &  \multicolumn{1}{c}{$D^*_2(2460)^-$}  &  \multicolumn{1}{c}{$D^*_3(2760)^-$} \\
\hline
PID  &  0.21 &  0.10 & <0.01&  0.003 \\
Trigger   &  0.12 &  0.01 &  0.02 &  <0.001 \\
Reconstruction  & <0.01& <0.01& <0.01&  0.001 \\
Simulation statistic   &  0.62 &  0.72 &  0.03 &  0.003 \\
Background model  &  0.13 &  0.13 & <0.01&  0.002 \\
$D^*(2010)^-$ veto  & <0.01 &  0.39 & <0.01&  0.001 \\
\hline
Total (experiment)  &  0.68  &  0.84  &  0.04  &  0.005 \\
\hline
Additional resonances  &  0.52 &  0.22 & <0.01&  0.002 \\
RBW parameters  &  0.30 &  0.46 & <0.01&  0.002 \\
$\pi\pi$ res. mass, width  &  0.05 &  0.04 & <0.01&  0.001 \\
\hline
Total (model)  & 0.60  &  0.51  & <0.01 &  0.003 \\
\hline
Total (all)  &  0.91  &  0.98  &  0.04  &  0.006 \\
\hline
\end{tabular}
\end{table}

\begin{table}
\centering
\footnotesize
\caption{Systematic uncertainties on the phases ($^{\circ}$) of the complex coefficients of the resonant contributions for the K-matrix model. The phase of $\rho(700)$ is set to $0^{\circ}$ as reference. \label{tab:KMaPha}}
\begin{tabular}{l....} 
\hline
Source   &  \multicolumn{1}{c}{$\omega(782)$}  &  \multicolumn{1}{c}{$f_2(1270)$}  &  \multicolumn{1}{c}{$\rho(1450)$}  &  \multicolumn{1}{c}{$\rho(1700)$} \\
\hline
PID  &  0.8 &  5.6 &  4.1 &  12.8 \\
Trigger  & <0.1&  0.2 & <0.1&  0.3 \\
Reconstruction  &  0.1 & <0.1& <0.1&  0.0 \\
Simulation statistic  &  0.5 &  1.1 &  1.6 &  1.5 \\
Background model  &  0.2 &  1.3 &  0.3 &  0.1 \\
$D^*(2010)^-$ veto  &  1.1 &  6.2 &  7.1 &  19.2 \\
\hline
Total (experiment)  &  1.5  &  8.5  &  8.4  &  23.1 \\
\hline
Additional resonances  &  0.2 &  1.9 &  2.3 &  3.3 \\
RBW parameters  &  0.4 &  1.3 &  0.4 &  0.6 \\
$\pi\pi$ res. mass, width  &  0.1 &  1.1 &  5.0 &  3.0 \\
\hline
Total (model)  &  0.5  &  2.6  &  5.5  &  4.5 \\
\hline
Total (all)  &  1.5  &  8.9  &  10.0  &  23.6 \\
\hline
 &  &  &  & \\
\hline
Source  &  \multicolumn{1}{c}{$\Dzb\pi^-$ P-wave}  &  \multicolumn{1}{c}{$D^*_0(2400)^-$}  &  \multicolumn{1}{c}{$D^*_2(2460)^-$}  &  \multicolumn{1}{c}{$D^*_3(2760)^-$} \\
\hline
PID  &  3.2 &  7.0 &  0.2 &  11.8 \\
Trigger  &  0.4 &  0.9 &  0.3 &  0.1 \\
Reconstruction  & <0.1& <0.1& <0.1&  0.1 \\
Simulation statistic  &  0.1 &  0.8 &  0.1 &  1.3 \\
Background model  &  0.7 &  0.3 &  0.4 &  0.1 \\
$D^*(2010)^-$ veto  &  \multicolumn{1}{c}{\textrm{\all n/a}}  &  9.1 &  0.4 &  9.5 \\
\hline
Total (experiment)  &  3.3  &  11.5  &  0.7  &  15.2 \\
\hline
Additional resonances  &  6.6 &  0.2 &  1.8 &  0.4 \\
RBW parameters   &  1.0 &  1.5 &  0.4 &  2.2 \\
$\pi\pi$ res. mass, width   &  0.4 &  0.7 &  0.2 &  0.3 \\
\hline
Total (model)  &  6.7  &  1.7  &  1.9  &  2.3 \\
\hline
Total (all)  &  7.5  &  11.7  &  2.0  &  15.4 \\
\hline
\end{tabular}
\end{table}

\begin{table}
\centering
\footnotesize
\caption{Systematic uncertainties on the fit fractions (\%) of the resonant contributions for the K-matrix model. \label{tab:KMaFraction}}
\begin{tabular}{l.....} 
\hline
Source   &  \multicolumn{1}{c}{$\rho(770)$}  &  \multicolumn{1}{c}{$\omega(782)$}  &  \multicolumn{1}{c}{$f_2(1270)$}  &  \multicolumn{1}{c}{$\rho(1450)$}  &  \multicolumn{1}{c}{$\rho(1700)$} \\
\hline
PID  &  1.36 & <0.01&  0.06 &  0.50 &  0.37 \\
Trigger  &  0.33 &  0.01 &  0.04 &  0.01 &  0.01 \\
Reconstruction  &  0.02 & <0.01& <0.01& <0.01& <0.01 \\
Simulation statistic  &  1.17 &  0.01 &  0.45 &  0.22 &  0.02 \\
Background model  &  0.07 & <0.01&  0.15 &  0.04 &  0.01 \\
$D^*(2010)^-$ veto  & 1.09 &<0.01& 0.68 & 0.61 & 0.48\\
\hline
Total (experiment)  &  2.13  &  0.01  &  0.83  &  0.82  &  0.61 \\
\hline
Additional resonances  &  0.61 &  0.02 &  0.56 &  0.01 &  0.05 \\
RBW parameters  &  0.49 &  0.01 &  0.15 &  0.21 &  0.04 \\
$\pi\pi$ res. mass, width  &  0.12 &  0.01 &  0.04 &  0.04 &  0.10 \\
\hline
Total (model)  &  0.79  &  0.02  &  0.58  &  0.21  &  0.12 \\
\hline
Total (all)  &  2.28  &  0.03  &  1.02  &  0.85  &  0.62\\
\hline
 &  &  &  &  & \\
\hline
Source   &  \multicolumn{1}{c}{S-wave}  &  \multicolumn{1}{c}{$\Dzb \pi^-$ P-wave}  &  \multicolumn{1}{c}{$D^*_0(2400)^-$}  &  \multicolumn{1}{c}{$D^*_2(2460)^-$}  &  \multicolumn{1}{c}{$D^*_3(2760)^-$} \\
\hline
PID  &  0.77 &  0.59 &  0.13 &  0.28 &  0.13 \\
Trigger  &  0.09 &  0.03 &  0.01 &  0.39 &  0.03 \\
Reconstruction  &  0.01 & <0.01& <0.01& <0.01& <0.01 \\
Simulation statistic  &  0.31 &  0.28 &  0.49 &  0.38 &  0.01 \\
Background model  &  0.23 &  0.13 &  0.08 &  0.03 &  0.04 \\
$D^*(2010)^-$ veto  & 1.44 &  \multicolumn{1}{c}{\textrm{\al\all n/a}}  & 0.69 & 0.86 & 0.12 \\
\hline
Total (experiment)  &  1.68  &  0.67  &  0.86  &  1.06  &  0.18 \\
\hline
Additional resonances  &  1.08 &  0.63 &  0.23 &  0.28 &  0.03 \\
RBW parameters  &  0.18 &  0.40 &  0.47 &  0.46 &  0.06 \\
$\pi\pi$ res. mass, width  &  0.07 &  0.07 &  0.02 &  0.01 &  0.01 \\
\hline
Total (model)  &  1.10  &  0.75  &  0.52  &  0.54  &  0.07\\
\hline
Total (all)  &  2.01  &  1.00  &  1.01  &  1.19  &  0.20\\
\hline
\end{tabular}
\end{table}

\clearpage

\section{Results for the interference fit fractions}
\label{sec:interferencefitfractions}

The central values of the interference fit fractions for the Isobar (K-matrix) model are given in Table~\ref{tab:IFF_Iso} (Table~\ref{tab:IFF_Kma}). The statistical, experimental systematic and model-dependent uncertainties on these quantities are given in Tables~\ref{tab:IFF_stat_Iso},~\ref{tab:IFF_exp_Iso} and \ref{tab:IFF_the_Iso} (Tables~\ref{tab:IFF_stat_Kma},~\ref{tab:IFF_exp_Kma} and \ref{tab:IFF_the_Kma}).

\begin{table}[!htb]
\centering
\scriptsize
\caption{\label{tab:IFF_Iso}
Interference fit fractions (\%) of the resonant contributions for the Isobar model with $m(\Dzb\pi^{\pm})>2.1$ \gevcc.
The resonances are:
($A_0$) nonresonant S-wave, ($A_1$) $f_0(500)$, ($A_2$) $f_0(980)$, ($A_3$) $f_0(2020)$, ($A_4$) $\rho(770)$, ($A_5$) $\omega(782)$,
($A_6$) $\rho(1450)$, $(A_7)$ $\rho(1700)$, $(A_8)$ $f_2(1270)$, $(A_9)$ $\Dzb\pi^-$ P-wave, $(A_{10})$ $D_0^*(2400)^-$, $(A_{11})$ $D_2^*(2460)^-$, $(A_{12})$ $D_3^*(2760)^-$.
The diagonal elements correspond to the fit fractions given in Table~\ref{tab: finalfraction}.
}
\begin{tabular}{cccccccccccccc}
\hline
 & $A_0$ & $A_1$ & $A_2$ & $A_3$& $A_4$ & $A_5$ & $A_6$ & $A_7$ & $A_8$ & $A_9$ & $A_{10}$ & $A_{11}$ & $A_{12}$ \\
\hline
$A_{0}$ & 2.82  & \al 2.70  & $-0.37$  & $-0.25$  & \al0.00  & \al\all 0.00  & \all \al0.00  & \all\al0.00  & \all\al0.13  & \all\al0.79  & $-1.70$  & $-2.12$  & \al\all0.06   \\
$A_{1}$ & $-$ & 13.23  & $-1.02$  & $-4.53$  & \al0.00  & \al\all0.00  & \all\al0.00  & \all\al0.00  & \all\al0.14  & \all\al3.37  & \all\al0.97  & \al\all3.81  & \al\all0.57   \\
$A_{2}$ & $-$  & $-$ & \all\al1.56  & \al\all0.79  & \al0.00  & \all\al0.00  & \all\al0.00  & \all\al0.00  & $-0.21$  & $-0.60$  & \all\al0.63  & $-0.90$  & $-$0.14   \\
$A_{3}$ & $-$  & $-$  & $-$ & \all\al 1.58  & \al0.00  & \all\al 0.00  & \all\al0.00  & \all\al0.00  & $-0.17$  & $-1.39$  & $-1.27$  & $-1.76$  & $-0.16$   \\
$A_{4}$ & $-$  & $-$  & $-$  & $-$ & 37.54  & $-0.78$  & \all\al2.43  & \all\al1.53  & \all\al0.00  & $-5.71$  & $-1.54$  & $-3.26$  & $-0.78$   \\
$A_{5}$ & $-$  & $-$  & $-$  & $-$  & $-$ & \all\al0.49  & $-0.01$  & \all\al0.00  & \all\al0.00  & \all\al0.00  & \all\al0.01  & $-$0.01  & \al\all0.00   \\
$A_{6}$ & $-$  & $-$  & $-$  & $-$  & $-$  & $-$ & \all\al1.54  & $-0.06$  & \al\all0.00  & \all\al0.26  & $-0.74$  & \al\all0.94  & \al\all0.04   \\
$A_{7}$ & $-$  & $-$  & $-$  & $-$  & $-$  & $-$  & $-$ & \all\al0.38  & \al\all0.00  & $-0.89$  & $-0.66$  & $-0.58$  & $-0.13$   \\
$A_{8}$ & $-$  & $-$  & $-$  & $-$  & $-$  & $-$  & $-$  & $-$ & \all10.28  & $-2.29$  & $-0.89$  & $-1.43$  & $-0.27$   \\
$A_{9}$ & $-$  & $-$  & $-$  & $-$  & $-$  & $-$  & $-$  & $-$  & $-$ & \all\al9.21  & \all\al0.00  & $-0.01$  & \al\all0.00   \\
$A_{10}$ & $-$  & $-$  & $-$  & $-$  & $-$  & $-$  & $-$  & $-$  & $-$  & $-$ & \al\all9.00  & \all\al0.01  & \al\all0.00   \\
$A_{11}$ & $-$  & $-$  & $-$  & $-$  & $-$  & $-$  & $-$  & $-$  & $-$  & $-$  & $-$ & \all28.83  & \al\all0.00   \\
$A_{12}$ & $-$  & $-$  & $-$  & $-$  & $-$  & $-$  & $-$  & $-$  & $-$  & $-$  & $-$  & $-$ & \al\all1.22   \\
\hline
\end{tabular}
\end{table}

\begin{table}
\centering
\footnotesize
\caption{\label{tab:IFF_Kma}
Interference fit fractions (\%) of the resonant contributions for the K-matrix model with $m(\Dzb\pi^{\pm})>2.1$ \gevcc.
The resonances are:
($A_0$) K-matrix S-wave, ($A_1$) $\rho(770)$, ($A_2$) $\omega(782)$,
($A_3$) $\rho(1450)$, $(A_4)$ $\rho(1700)$, $(A_5)$ $f_2(1270)$, $(A_6)$ $\Dzb\pi^-$ P-wave, $(A_{7})$ $D_0^*(2400)^-$, $(A_{8})$ $D_2^*(2460)^-$, $(A_{9})$ $D_3^*(2760)^-$
The diagonal elements correspond to the fit fractions given in Table~\ref{tab: finalfraction}.
}
\begin{tabular}{ccccccccccc}
\hline
 & $A_0$ & $A_1$ & $A_2$ & $A_3$& $A_4$ & $A_5$ & $A_6$ & $A_7$ & $A_8$ & $A_9$  \\
\hline
 $A_{0}$& 16.51  & \al0.00  & \al\all0.00  & \al\all0.00  & \al\all0.00  & $-0.06$  & \al\all2.37  & $-1.45$  & $-0.10$  & \al\all0.01   \\
$A_{1}$ & $-$ & 36.15  & $-0.84$  & \al\all4.20  & \al\all2.10  & \al\all0.00  & $-5.39$  & $-1.88$  & $-2.81$  & $-0.90$   \\
$A_{2}$ & $-$  & $-$ & \al\all0.50  & $-0.01$  & \al\all0.00  & \al\all0.00  & \al\all0.00  & \al\all0.01  & $-0.01$  & \al\all0.00   \\
$A_{3}$ & $-$  & $-$  & $-$ & \al\all2.16  & $-0.43$  & \al\all0.00  & $-0.15$  & $-1.14$  & \al\all0.73  & $-0.04$   \\
$A_{4}$ & $-$  & $-$  & $-$  & $-$ & \al\all0.83  & \al\all0.00  & $-1.49$  & $-0.99$  & $-1.12$  & $-0.24$   \\
$A_{5}$ & $-$  & $-$  & $-$  & $-$  & $-$ & \al\all9.88  & $-2.03$  & $-0.73$  & $-1.50$  & $-0.35$   \\
$A_{6}$ & $-$  & $-$  & $-$  & $-$  & $-$  & $-$ & \al\all9.22  & \al\all0.00  & $-$0.01  & \al\all0.00   \\
$A_{7}$ & $-$  & $-$  & $-$  & $-$  & $-$  & $-$  & $-$ & \al\all9.27  & \al\all0.01  & \al\all0.00   \\
$A_{8}$ & $-$  & $-$  & $-$  & $-$  & $-$  & $-$  & $-$  & $-$ & \all28.13  & \al\all0.00   \\
$A_{9}$ & $-$  & $-$  & $-$  & $-$  & $-$  & $-$  & $-$  & $-$  & $-$ & \al\all1.58   \\
\hline
\end{tabular}
\end{table}

\begin{table}
\centering
\footnotesize
\caption{\label{tab:IFF_stat_Iso}
Statistical uncertainties on the interference fit fractions (\%) of the resonant contributions for the Isobar model with $m(\Dzb\pi^{\pm})>2.1$ \gevcc.
The resonances are:
($A_0$) nonresonant S-wave, ($A_1$) $f_0(500)$, ($A_2$) $f_0(980)$, ($A_3$) $f_0(2020)$, ($A_4$) $\rho(770)$, ($A_5$) $\omega(782)$,
($A_6$) $\rho(1450)$, $(A_7)$ $\rho(1700)$, $(A_8)$ $f_2(1270)$, $(A_9)$ $\Dzb\pi^-$ P-wave, $(A_{10})$ $D_0^*(2400)^-$, $(A_{11})$ $D_2^*(2460)^-$, $(A_{12})$ $D_3^*(2760)^-$.
The diagonal elements correspond to the statistical uncertainties on the fit fractions given in Table~\ref{tab: finalfraction}.
}
\begin{tabular}{cccccccccccccc}
\hline
 & $A_0$ & $A_1$ & $A_2$ & $A_3$& $A_4$ & $A_5$ & $A_6$ & $A_7$ & $A_8$ & $A_9$ & $A_{10}$ & $A_{11}$ & $A_{12}$ \\
\hline
$A_{0}$& 0.34 & 0.29 & 0.11 & 0.07 & 0.00 & 0.00 & 0.00 & 0.00 & 0.02 & 0.13 & 0.36 & 0.26 & 0.03  \\
$A_{1}$& $-$ & 0.89 & 0.54 & 0.64 & 0.00 & 0.00 & 0.00 & 0.00 & 0.04 & 0.22 & 0.45 & 0.20 & 0.08  \\
$A_{2}$& $-$ & $-$ & 0.29 & 0.15 & 0.00 & 0.00 & 0.00 & 0.00 & 0.03 & 0.09 & 0.09 & 0.11 & 0.04  \\
$A_{3}$& $-$ & $-$ & $-$ & 0.36 & 0.00 & 0.00 & 0.00 & 0.00 & 0.02 & 0.18 & 0.20 & 0.22 & 0.05  \\
$A_{4}$& $-$ & $-$ & $-$ & $-$ & 1.00 & 0.33 & 0.65 & 0.32 & 0.00 & 0.35 & 0.21 & 0.22 & 0.12  \\
$A_{5}$& $-$ & $-$ & $-$ & $-$ & $-$ & 0.13 & 0.00 & 0.00 & 0.00 & 0.00 & 0.00 & 0.00 & 0.00  \\
$A_{6}$& $-$ & $-$ & $-$ & $-$ & $-$ & $-$ & 0.32 & 0.24 & 0.00 & 0.24 & 0.18 & 0.19 & 0.04  \\
$A_{7}$& $-$ & $-$ & $-$ & $-$ & $-$ & $-$ & $-$ & $^{+0.25}_{-0.12}$ & 0.00 & 0.23 & 0.15 & 0.20 & 0.03  \\
$A_{8}$& $-$ & $-$ & $-$ & $-$ & $-$ & $-$ & $-$ & $-$ & 0.49 & 0.20 & 0.14 & 0.07 & 0.04  \\
$A_{9}$& $-$ & $-$ & $-$ & $-$ & $-$ & $-$ & $-$ & $-$ & $-$ & 0.56 & 0.00 & 0.00 & 0.00  \\
$A_{10}$& $-$ & $-$ & $-$ & $-$ & $-$ & $-$ & $-$ & $-$ & $-$ & $-$ & 0.60 & 0.00 & 0.00  \\
$A_{11}$& $-$ & $-$ & $-$ & $-$ & $-$ & $-$ & $-$ & $-$ & $-$ & $-$ & $-$ & 0.69 & 0.00  \\
$A_{12}$& $-$ & $-$ & $-$ & $-$ & $-$ & $-$ & $-$ & $-$ & $-$ & $-$ & $-$ & $-$ & 0.19  \\
\hline
\end{tabular}
\end{table}

\begin{table}
\centering
\footnotesize
\caption{\label{tab:IFF_exp_Iso}
Experimental systematic uncertainties on the interference fit fractions (\%) of the resonant contributions for the Isobar model with $m(\Dzb\pi^{\pm})>2.1$ \gevcc.
The resonances are:
($A_0$) nonresonant S-wave, ($A_1$) $f_0(500)$, ($A_2$) $f_0(980)$, ($A_3$) $f_0(2020)$, ($A_4$) $\rho(770)$, ($A_5$) $\omega(782)$,
($A_6$) $\rho(1450)$, $(A_7)$ $\rho(1700)$, $(A_8)$ $f_2(1270)$, $(A_9)$ $\Dzb\pi^-$ P-wave, $(A_{10})$ $D_0^*(2400)^-$, $(A_{11})$ $D_2^*(2460)^-$, $(A_{12})$ $D_3^*(2760)^-$.
The diagonal elements correspond to the statistical uncertainties on the fit fractions given in Table~\ref{tab: finalfraction}.
}
\begin{tabular}{cccccccccccccc}
\hline
 & $A_0$ & $A_1$ & $A_2$ & $A_3$& $A_4$ & $A_5$ & $A_6$ & $A_7$ & $A_8$ & $A_9$ & $A_{10}$ & $A_{11}$ & $A_{12}$ \\
\hline
$A_{0}$& 0.07 & 0.06 & 0.03 & 0.03 & 0.00 & 0.00 & 0.00 & 0.00 & 0.07 & 0.02 & 0.11 & 0.09 & 0.01  \\
$A_{1}$& $-$ & 0.31 & 0.22 & 0.20 & 0.00 & 0.00 & 0.00 & 0.00 & 0.07 & 0.06 & 0.26 & 0.04 & 0.01  \\
$A_{2}$& $-$ & $-$ & 0.11 & 0.05 & 0.00 & 0.00 & 0.00 & 0.00 & 0.14 & 0.02 & 0.04 & 0.04 & 0.02  \\
$A_{3}$& $-$ & $-$ & $-$ & 0.15 & 0.00 & 0.00 & 0.00 & 0.00 & 0.11 & 0.08 & 0.06 & 0.11 & 0.02  \\
$A_{4}$& $-$ & $-$ & $-$ & $-$ & 0.61 & 0.05 & 0.31 & 0.16 & 0.00 & 0.18 & 0.05 & 0.03 & 0.02  \\
$A_{5}$& $-$ & $-$ & $-$ & $-$ & $-$ & 0.01 & 0.00 & 0.00 & 0.00 & 0.00 & 0.00 & 0.00 & 0.00  \\
$A_{6}$& $-$ & $-$ & $-$ & $-$ & $-$ & $-$ & 0.08 & 0.03 & 0.00 & 0.11 & 0.12 & 0.06 & 0.01  \\
$A_{7}$& $-$ & $-$ & $-$ & $-$ & $-$ & $-$ & $-$ & 0.07 & 0.00 & 0.10 & 0.07 & 0.03 & 0.01  \\
$A_{8}$& $-$ & $-$ & $-$ & $-$ & $-$ & $-$ & $-$ & $-$ & 0.31 & 0.16 & 0.11 & 0.07 & 0.02  \\
$A_{9}$& $-$ & $-$ & $-$ & $-$ & $-$ & $-$ & $-$ & $-$ & $-$ & 0.24 & 0.00 & 0.00 & 0.00  \\
$A_{10}$& $-$ & $-$ & $-$ & $-$ & $-$ & $-$ & $-$ & $-$ & $-$ & $-$ & 0.20 & 0.00 & 0.00  \\
$A_{11}$& $-$ & $-$ & $-$ & $-$ & $-$ & $-$ & $-$ & $-$ & $-$ & $-$ & $-$ & 0.74 & 0.00  \\
$A_{12}$& $-$ & $-$ & $-$ & $-$ & $-$ & $-$ & $-$ & $-$ & $-$ & $-$ & $-$ & $-$ & 0.07  \\
\hline
\end{tabular}
\end{table}

\begin{table}
\centering
\footnotesize
\caption{\label{tab:IFF_the_Iso}
Model-dependent systematic uncertainties on the interference fit fractions (\%) of the resonant contributions for the Isobar model with $m(\Dzb\pi^{\pm})>2.1$ \gevcc.
The resonances are:
($A_0$) nonresonant S-wave, ($A_1$) $f_0(500)$, ($A_2$) $f_0(980)$, ($A_3$) $f_0(2020)$, ($A_4$) $\rho(770)$, ($A_5$) $\omega(782)$,
($A_6$) $\rho(1450)$, $(A_7)$ $\rho(1700)$, $(A_8)$ $f_2(1270)$, $(A_9)$ $\Dzb\pi^-$ P-wave, $(A_{10})$ $D_0^*(2400)^-$, $(A_{11})$ $D_2^*(2460)^-$, $(A_{12})$ $D_3^*(2760)^-$.
The diagonal elements correspond to the statistical uncertainties on the fit fractions given in Table~\ref{tab: finalfraction}.
}
\begin{tabular}{cccccccccccccc}
\hline
 & $A_0$ & $A_1$ & $A_2$ & $A_3$& $A_4$ & $A_5$ & $A_6$ & $A_7$ & $A_8$ & $A_9$ & $A_{10}$ & $A_{11}$ & $A_{12}$ \\
\hline
$A_{0}$& 0.80 & 0.61 & 0.31 & 0.17 & 0.00 & 0.00 & 0.00 & 0.00 & 0.03 & 0.28 & 0.56 & 0.45 & 0.01  \\
$A_{1}$& $-$ & 2.45 & 2.00 & 3.03 & 0.00 & 0.00 & 0.00 & 0.00 & 0.16 & 0.72 & 0.79 & 0.98 & 0.08  \\
$A_{2}$& $-$ & $-$ & 0.54 & 0.67 & 0.00 & 0.00 & 0.00 & 0.00 & 0.02 & 0.15 & 0.13 & 0.28 & 0.08  \\
$A_{3}$& $-$ & $-$ & $-$ & 1.00 & 0.00 & 0.00 & 0.00 & 0.00 & 0.08 & 0.51 & 0.34 & 0.69 & 0.06  \\
$A_{4}$& $-$ & $-$ & $-$ & $-$ & 0.98 & 0.03 & 0.47 & 0.12 & 0.00 & 0.54 & 0.33 & 0.27 & 0.09  \\
$A_{5}$& $-$ & $-$ & $-$ & $-$ & $-$ & 0.03 & 0.00 & 0.00 & 0.00 & 0.00 & 0.00 & 0.00 & 0.00  \\
$A_{6}$& $-$ & $-$ & $-$ & $-$ & $-$ & $-$ & 0.22 & 0.08 & 0.00 & 0.31 & 0.18 & 0.14 & 0.03  \\
$A_{7}$& $-$ & $-$ & $-$ & $-$ & $-$ & $-$ & $-$ & 0.06 & 0.00 & 0.12 & 0.07 & 0.04 & 0.02  \\
$A_{8}$& $-$ & $-$ & $-$ & $-$ & $-$ & $-$ & $-$ & $-$ & 1.10 & 0.49 & 0.33 & 0.09 & 0.05  \\
$A_{9}$& $-$ & $-$ & $-$ & $-$ & $-$ & $-$ & $-$ & $-$ & $-$ & 1.73 & 0.00 & 0.00 & 0.00  \\
$A_{10}$& $-$ & $-$ & $-$ & $-$ & $-$ & $-$ & $-$ & $-$ & $-$ & $-$ & 0.35 & 0.00 & 0.00  \\
$A_{11}$& $-$ & $-$ & $-$ & $-$ & $-$ & $-$ & $-$ & $-$ & $-$ & $-$ & $-$ & 0.50 & 0.00  \\
$A_{12}$& $-$ & $-$ & $-$ & $-$ & $-$ & $-$ & $-$ & $-$ & $-$ & $-$ & $-$ & $-$ & 0.09  \\
\hline
\end{tabular}
\end{table}

\begin{table}
\centering
\footnotesize
\caption{\label{tab:IFF_stat_Kma}
Statistical uncertainties on the interference fit fractions (\%) of the resonant contributions for the K-matrix model with $m(\Dzb\pi^{\pm})>2.1$ \gevcc.
The resonances are:
($A_0$)K-matrix S-wave, ($A_1$) $\rho(770)$, ($A_2$) $\omega(782)$,
($A_3$) $\rho(1450)$, $(A_4)$ $\rho(1700)$, $(A_5)$ $f_2(1270)$, $(A_6)$ $\Dzb\pi^-$ P-wave, $(A_{7})$ $D_0^*(2400)^-$, $(A_{8})$ $D_2^*(2460)^-$, $(A_{9})$ $D_3^*(2760)^-$
The diagonal elements correspond to the statistical uncertainties on the fit fractions shown in Table~\ref{tab: finalfraction}.
}
\begin{tabular}{ccccccccccc}
\hline
 & $A_0$ & $A_1$ & $A_2$ & $A_3$& $A_4$ & $A_5$ & $A_6$ & $A_7$ & $A_8$ & $A_9$  \\
\hline
 $A_{0}$& 0.70 & 0.00 & 0.00 & 0.00 & 0.00 & 0.04 & 0.28 & 0.49 & 0.43 & 0.16  \\
$A_{1}$& $-$ & 1.00 & 0.34 & 0.71 & 0.31 & 0.00 & 0.41 & 0.22 & 0.25 & 0.14  \\
$A_{2}$& $-$ & $-$ & 0.13 & 0.00 & 0.00 & 0.00 & 0.01 & 0.00 & 0.00 & 0.00  \\
$A_{3}$& $-$ & $-$ & $-$ & 0.42 & 0.29 & 0.00 & 0.29 & 0.23 & 0.20 & 0.07  \\
$A_{4}$& $-$ & $-$ & $-$ & $-$ & 0.21 & 0.00 & 0.23 & 0.16 & 0.19 & 0.04  \\
$A_{5}$& $-$ & $-$ & $-$ & $-$ & $-$ & 0.58 & 0.22 & 0.16 & 0.08 & 0.05  \\
$A_{6}$& $-$ & $-$ & $-$ & $-$ & $-$ & $-$ & 0.58 & 0.00 & 0.00 & 0.00  \\
$A_{7}$& $-$ & $-$ & $-$ & $-$ & $-$ & $-$ & $-$ & 0.60 & 0.00 & 0.00  \\
$A_{8}$& $-$ & $-$ & $-$ & $-$ & $-$ & $-$ & $-$ & $-$ & 0.72 & 0.00  \\
$A_{9}$& $-$ & $-$ & $-$ & $-$ & $-$ & $-$ & $-$ & $-$ & $-$ & 0.22  \\
\hline
\end{tabular}
\end{table}

\begin{table}
\centering
\footnotesize
\caption{\label{tab:IFF_exp_Kma}
Experimental systematic uncertainties on the interference fit fractions (\%) of the resonant contributions for the K-matrix model with $m(\Dzb\pi^{\pm})>2.1$ \gevcc.
The resonances are:
($A_0$) K-matrix S-wave, ($A_1$) $\rho(770)$, ($A_2$) $\omega(782)$,
($A_3$) $\rho(1450)$, $(A_4)$ $\rho(1700)$, $(A_5)$ $f_2(1270)$, $(A_6)$ $\Dzb\pi^-$ P-wave, $(A_{7})$ $D_0^*(2400)^-$, $(A_{8})$ $D_2^*(2460)^-$, $(A_{9})$ $D_3^*(2760)^-$
The diagonal elements correspond to the statistical uncertainties on the fit fractions shown in Table~\ref{tab: finalfraction}.
}
\begin{tabular}{ccccccccccc}
\hline
 & $A_0$ & $A_1$ & $A_2$ & $A_3$& $A_4$ & $A_5$ & $A_6$ & $A_7$ & $A_8$ & $A_9$  \\
\hline
$A_{0}$& 1.68 & 0.00 & 0.00 & 0.00 & 0.00 & 0.10 & 0.84 & 1.88 & 1.21 & 0.36  \\
$A_{1}$& $-$ & 2.13 & 0.06 & 1.42 & 1.02 & 0.00 & 0.87 & 0.37 & 0.14 & 0.29  \\
$A_{2}$& $-$ & $-$ & 0.01 & 0.00 & 0.00 & 0.00 & 0.01 & 0.00 & 0.00 & 0.00  \\
$A_{3}$& $-$ & $-$ & $-$ & 0.82 & 0.73 & 0.00 & 0.13 & 0.30 & 0.10 & 0.15  \\
$A_{4}$& $-$ & $-$ & $-$ & $-$ & 0.61 & 0.00 & 0.88 & 0.51 & 0.89 & 0.14  \\
$A_{5}$& $-$ & $-$ & $-$ & $-$ & $-$ & 0.83 & 0.27 & 0.16 & 0.18 & 0.16  \\
$A_{6}$& $-$ & $-$ & $-$ & $-$ & $-$ & $-$ & 0.67 & 0.00 & 0.00 & 0.00  \\
$A_{7}$& $-$ & $-$ & $-$ & $-$ & $-$ & $-$ & $-$ & 0.86 & 0.00 & 0.00  \\
$A_{8}$& $-$ & $-$ & $-$ & $-$ & $-$ & $-$ & $-$ & $-$ & 1.06 & 0.00  \\
$A_{9}$& $-$ & $-$ & $-$ & $-$ & $-$ & $-$ & $-$ & $-$ & $-$ & 0.18  \\
\hline
\end{tabular}
\end{table}

\begin{table}
\centering
\footnotesize
\caption{\label{tab:IFF_the_Kma}
Model-dependent systematic uncertainties on the interference fit fractions (\%) of the resonant contributions for the K-matrix model with $m(\Dzb\pi^{\pm})>2.1$ \gevcc.
The resonances are:
($A_0$) K-matrix S-wave, ($A_1$) $\rho(770)$, ($A_2$) $\omega(782)$,
($A_3$) $\rho(1450)$, $(A_4)$ $\rho(1700)$, $(A_5)$ $f_2(1270)$, $(A_6)$ $\Dzb\pi^-$ P-wave, $(A_{7})$ $D_0^*(2400)^-$, $(A_{8})$ $D_2^*(2460)^-$, $(A_{9})$ $D_3^*(2760)^-$
The diagonal elements correspond to the statistical uncertainties on the fit fractions shown in Table~\ref{tab: finalfraction}.
}
\begin{tabular}{ccccccccccc}
\hline
 & $A_0$ & $A_1$ & $A_2$ & $A_3$& $A_4$ & $A_5$ & $A_6$ & $A_7$ & $A_8$ & $A_9$  \\
\hline
$A_{0}$& 1.10 & 0.00 & 0.00 & 0.00 & 0.00 & 0.02 & 0.24 & 0.25 & 0.40 & 0.15  \\
$A_{1}$& $-$ & 0.79 & 0.02 & 0.41 & 0.25 & 0.00 & 0.26 & 0.29 & 0.19 & 0.05  \\
$A_{2}$& $-$ & $-$ & 0.02 & 0.00 & 0.00 & 0.00 & 0.00 & 0.00 & 0.00 & 0.00  \\
$A_{3}$& $-$ & $-$ & $-$ & 0.21 & 0.08 & 0.00 & 0.20 & 0.09 & 0.12 & 0.03  \\
$A_{4}$& $-$ & $-$ & $-$ & $-$ & 0.12 & 0.00 & 0.14 & 0.12 & 0.08 & 0.02  \\
$A_{5}$& $-$ & $-$ & $-$ & $-$ & $-$ & 0.58 & 0.14 & 0.19 & 0.08 & 0.08  \\
$A_{6}$& $-$ & $-$ & $-$ & $-$ & $-$ & $-$ & 0.75 & 0.00 & 0.00 & 0.00  \\
$A_{7}$& $-$ & $-$ & $-$ & $-$ & $-$ & $-$ & $-$ & 0.52 & 0.00 & 0.00  \\
$A_{8}$& $-$ & $-$ & $-$ & $-$ & $-$ & $-$ & $-$ & $-$ & 0.54 & 0.00  \\
$A_{9}$& $-$ & $-$ & $-$ & $-$ & $-$ & $-$ & $-$ & $-$ & $-$ & 0.07  \\
\hline
\end{tabular}
\end{table}

\clearpage

\section{Results of the K-matrix parameters}
The moduli and phases of the K-matrix parameters in Eq.~(\ref{eq:kmatrix}) are listed in Table~\ref{tab: finalKMa}.
The break-down of systematic uncertainties are shown in Tables~\ref{tab:KMaKMaAmp} and~\ref{tab:KMaKMaPha}.

\begin{table}[!tbh]
\centering
\small
\caption{The moduli and phases of the K-matrix parameters.
The first uncertainty is statistical, the second the experimental systematic, and the third the model-dependent systematic.
The moduli are normalised to that of the $\rho(770)$ contribution and the phase of $\rho(770)$ is set to 0$^{\circ}$.
}
\label{tab: finalKMa}
\begin{tabular}{lcc}
\hline
Parameter  & Modulus & Phase ($^{\circ}$) \\\hline
$f_{10}$ & $\al17.0 \pm \al3.3 \pm \al9.5 \pm \al3.7$& $347.3 \pm 13.7 \pm 18.7 \pm \al3.2$ \\
$f_{11}$ & $\al14.9 \pm 17.1 \pm 20.3 \pm \al8.0$ & $160.0 \pm 70.1 \pm 39.6 \pm 26.2$ \\
$f_{12}$ & $111.3 \pm 23.1 \pm 23.8 \pm 12.8$ & $226.1 \pm 12.0 \pm 11.2 \pm \al4.9$ \\
$f_{13}$ & $\al28.7 \pm 14.2 \pm \al8.1 \pm \al5.3$ & $186.5 \pm 30.0 \pm 30.4 \pm \al8.6$ \\
$f_{14}$ & $\al31.0 \pm 12.8 \pm 13.4 \pm \al8.8$ & $10.61 \pm 25.9 \pm 15.2 \pm \al2.9$ \\
$\beta_0$ & $\al\al9.5 \pm \al1.8 \pm \al2.9 \pm \al1.1$& $\al20.7 \pm 15.2 \pm 13.5 \pm 10.4$ \\
$\beta_1$ & $\al17.2 \pm \al6.4 \pm \al6.2 \pm \al4.8$& $\al19.6 \pm 19.4 \pm 14.4 \pm \al3.7$ \\
$\beta_2$ & $\al34.9 \pm \al7.6 \pm 14.3 \pm \al3.1$& $128.3 \pm 12.1 \pm \al2.1 \pm \al1.9$ \\
$\beta_3$ & $\al53.5 \pm 14.3 \pm \al9.2 \pm \al4.2$& $138.7 \pm 15.5 \pm \al7.2 \pm\al3.9$ \\
$\beta_4$ & $\al52.5 \pm 10.2 \pm 22.4 \pm \al5.9$& $305.0 \pm 10.5 \pm 13.5 \pm \al2.2$ \\
\hline
\end{tabular}
\end{table}
\begin{table}
\centering
\footnotesize
\caption{Systematic uncertainties on the moduli of the K-matrix parameters. The moduli are normalised to that of $\rho(770)$. \label{tab:KMaKMaAmp}}
\begin{tabular}{l..........} 
\hline
Source   &  \multicolumn{1}{c}{$f_{10}$}  &  \multicolumn{1}{c}{$f_{11}$}  &  \multicolumn{1}{c}{$f_{12}$}  &  \multicolumn{1}{c}{$f_{13}$}  &  \multicolumn{1}{c}{$f_{14}$}  &  \multicolumn{1}{c}{$\beta_0$}  &  \multicolumn{1}{c}{$\beta_1$}  &  \multicolumn{1}{c}{$\beta_2$}  &  \multicolumn{1}{c}{$\beta_3$}  &  \multicolumn{1}{c}{$\beta_4$} \\
\hline
PID  &  8.1 &  5.5 &  22.5 &  3.2 &  13.2 &  2.5 &  3.6 &  10.7 &  8.0 &  18.0 \\
Trigger  &  0.4 &  1.1 &  2.5 &  1.0 &  1.2 &  0.2 &  0.7 &  0.6 &  1.1 &  1.0 \\
Reconstruction   &  0.1 &  0.9 &  0.4 &  0.5 &  0.7 &  0.1 &  0.4 &  0.4 &  0.6 &  0.5 \\
Simulation statistic  &  0.3 &  4.6 &  2.8 &  2.7 &  1.6 &  0.1 &  1.8 &  1.0 &  0.6 &  0.7 \\
Background model  &  0.6 &  0.5 &  6.5 & <0.1&  0.2 &  0.1 &  0.7 &  1.7 &  2.4 &  2.7 \\
$D^*(2010)^-$ veto  &  4.9 &  18.9 &  1.9 &  6.8 &  0.6 &  1.4 &  4.6 &  9.2 &  3.7 &  13.1 \\
\hline
Total (experiment)  &  9.5  &  20.3  &  23.8  &  8.1  &  13.4  &  2.9  &  6.2  &  14.3  &  9.2  &  22.4 \\
\hline
Additional resonances  &  3.5 &  8.0 &  12.6 &  3.8 &  7.7 &  1.1 &  4.6 &  3.0 &  3.5 &  5.6 \\
RBW parameters  &  1.2 &  0.4 &  2.3 &  3.3 &  4.0 &  0.2 &  1.1 &  0.5 &  2.2 &  1.9 \\
$\pi\pi$ res. mass, width  &  0.3 &  0.7 &  0.1 &  1.5 &  1.7 &  0.2 &  0.6 &  0.4 &  0.7 &  0.3 \\
\hline
Total (model)  &  3.7  &  8.0  &  12.8  &  5.3  &  8.8  &  1.1  &  4.8  &  3.1  &  4.2  &  5.9 \\
\hline
Total (all)  &  10.2  &  21.8  &  27.0  &  9.6  &  16.0  &  3.1  &  7.8  &  14.6  &  10.1  &  23.2 \\
\hline
\end{tabular}
\end{table}
\begin{table}
\centering
\footnotesize
\caption{Systematic uncertainties on the phases ($^{\circ}$) of the K-matrix parameters.
The phase of $\rho(700)$ is set to 0$^{\circ}$ as the reference. \label{tab:KMaKMaPha}}
\begin{tabular}{l..........} 
\hline
Source   &  \multicolumn{1}{c}{$f_{10}$}  &  \multicolumn{1}{c}{$f_{11}$}  &  \multicolumn{1}{c}{$f_{12}$}  &  \multicolumn{1}{c}{$f_{13}$}  &  \multicolumn{1}{c}{$f_{14}$}  &  \multicolumn{1}{c}{$\beta_0$}  &  \multicolumn{1}{c}{$\beta_1$}  &  \multicolumn{1}{c}{$\beta_2$}  &  \multicolumn{1}{c}{$\beta_3$}  &  \multicolumn{1}{c}{$\beta_4$} \\
\hline
PID  &  18.0  &  35.0 &  9.6 &  25.9 &  5.8 &  10.4 &  10.4 &  1.7 &  1.4 &  4.3 \\
Trigger  &  0.5 &  0.2 &  0.3 &  0.2 & <0.1&  0.5 &  0.3 &  0.1 &  0.5 &  0.1 \\
Reconstruction  &  0.3 &  2.4 &  0.6 &  0.5 &  0.6 &  0.6 &  0.4 &  0.1 &  0.6 &  0.2 \\
Simulation statistic  &  4.0 &  1.0 &  0.2 &  2.0 &  7.1 &  5.2 &  4.6 &  0.8 &  1.9 &  1.3 \\
Background model  &  3.1 &  8.2 &  0.4 &  3.9 &  0.4 &  1.9 &  2.5 &  0.1 &  0.3 &  0.4 \\
$D^*(2010)^-$ veto  &  0.3 &  16.4 &  5.8 &  15.2 &  12.1 &  6.5 &  8.5 &  1.0 &  6.7 &  12.7 \\
\hline
Total (experiment)  &  18.7  &  39.6  &  11.2  &  30.4  &  15.2  &  13.5  &  14.4  &  2.1  &  7.2  &  13.5 \\
\hline
Additional resonances  &  0.4 &  5.0 &  4.1 &  2.1 &  1.0 &  3.6 &  1.9 &  0.2 &  3.7 &  1.9 \\
RBW parameters   &  3.2 &  25.2 &  2.6 &  8.3 &  0.6 &  9.2 &  3.2 &  1.9 &  1.0 &  1.1 \\
$\pi\pi$ res. mass, width   &  0.3 &  5.0 &  0.5 &  1.0 &  2.6 &  3.2 &  0.4 &  0.2 &  0.7 &  0.3 \\
\hline
Total (model)  &  3.2  &  26.2  &  4.9  &  8.6  &  2.9  &  10.4  &  3.7  &  1.9  &  3.9  &  2.2 \\
\hline
Total (all)  &  19.0  &  47.5  &  12.3  &  31.6  &  15.5  &  17.0  &  14.9  &  2.9  &  8.1  &  13.7 \\
\hline
\end{tabular}
\end{table}
\clearpage

\newpage

\addcontentsline{toc}{section}{References}
\ifx\mcitethebibliography\mciteundefinedmacro
\PackageError{LHCb.bst}{mciteplus.sty has not been loaded}
{This bibstyle requires the use of the mciteplus package.}\fi
\providecommand{\href}[2]{#2}

\clearpage
\newpage

\centerline{\large\bf LHCb collaboration}
\begin{flushleft}
\small
R.~Aaij$^{41}$, 
B.~Adeva$^{37}$, 
M.~Adinolfi$^{46}$, 
A.~Affolder$^{52}$, 
Z.~Ajaltouni$^{5}$, 
S.~Akar$^{6}$, 
J.~Albrecht$^{9}$, 
F.~Alessio$^{38}$, 
M.~Alexander$^{51}$, 
S.~Ali$^{41}$
% ...

{\footnotesize \it
% ...
$ ^{5}$Clermont Universit\'{e}, Universit\'{e} Blaise Pascal, CNRS/IN2P3, LPC, Clermont-Ferrand, France\\
$ ^{6}$CPPM, Aix-Marseille Universit\'{e}, CNRS/IN2P3, Marseille, France\\
% ...
$ ^{9}$Fakult\"{a}t Physik, Technische Universit\"{a}t Dortmund, Dortmund, Germany\\
% ...
$ ^{37}$Universidad de Santiago de Compostela, Santiago de Compostela, Spain\\
$ ^{38}$European Organization for Nuclear Research (CERN), Geneva, Switzerland\\
% ...
$ ^{41}$Nikhef National Institute for Subatomic Physics, Amsterdam, The Netherlands\\
% ...
$ ^{46}$H.H. Wills Physics Laboratory, University of Bristol, Bristol, United Kingdom\\
% ...
$ ^{51}$School of Physics and Astronomy, University of Glasgow, Glasgow, United Kingdom\\
$ ^{52}$Oliver Lodge Laboratory, University of Liverpool, Liverpool, United Kingdom\\
% ...
}
\end{flushleft}

\end{document}